\documentclass[10pt,a4paper]{article}
\usepackage{graphics,epsfig}
\usepackage{palatino}
\usepackage{amstext,amsfonts,amssymb,amsthm,amsmath,mathrsfs,epic,eepic}
\usepackage{subfigure}  
\usepackage{color}
\usepackage[normalem]{ulem}
\usepackage[left]{lineno}
\usepackage{setspace}
\usepackage{float}
\usepackage{graphics,epsfig}
\usepackage{color}
\usepackage{caption}
\usepackage{capt-of}
\DeclareMathOperator*{\argmax}{arg\,max}

\begin{document}

\newcommand{\figref}[1]{Fig~\ref{#1}}
\newcommand{\figsref}[1]{Figs~\ref{#1}}
\newcommand{\figssref}[2]{Figs~\ref{#1}-\ref{#2}}
\newcommand{\etal}{\emph{et al}}
\newcommand{\be}{\begin{equation}}
\newcommand{\ee}{\end{equation}}
\newcommand{\secref}[1]{Section \ref{#1}}
\definecolor{darkgreen}{rgb}{0.0,0.33,0.0}
\definecolor{midgreen}{rgb}{0.0,0.5,0.0}
\newcommand{\TODO}[1]{{\bf \color{midgreen}{($\bigstar$ #1)}}}
\newcommand{\TF}[1]{{\color{midgreen}{#1}}}
\title{Evolution of new regulatory functions on biophysically realistic fitness landscapes}

\author{Tamar Friedlander*, Roshan Prizak*, Nicholas H. Barton \and Ga\v{s}per Tka\v{c}ik \\
{\normalsize * - equal contribution}
\\
Institute of Science and Technology Austria, Am Campus 1, A-3400 \\ Klosterneuburg, Austria}

\maketitle


\begin{abstract}
Gene expression is controlled by networks of regulatory proteins that interact specifically with external signals and DNA regulatory sequences. These interactions force the network components to co-evolve so as to continually maintain function. Yet, existing models of evolution mostly focus on isolated genetic elements. In contrast, we study the essential process by which regulatory networks grow: the duplication and subsequent specialization of network components. We synthesize a biophysical model of molecular interactions with the evolutionary framework to find the conditions and pathways by which new regulatory functions emerge. We show that specialization of new network components is usually slow, but can be drastically accelerated in the presence of regulatory crosstalk and mutations that promote promiscuous interactions between network components.
\end{abstract}

\clearpage

\section*{Introduction}
Phenotypes evolve largely through changes in gene regulation~\cite{king_evolution_1975,gilad_expression_2006,wray_evolutionary_2007, carroll_evolution_2005}, and such evolution may be flexible and rapid~\cite{yona_relay_2015,madan_babu_evolutionary_2006}.
Of particular importance are mutations affecting affinity and specificity of transcription factors (TFs) for their upstream signals or for their binding sites, short fragments of DNA that TFs interact with to activate or repress transcription of specific target genes. Mutations in these binding sites or at sites that alter TF specificity are crucial because of their ability to ``rewire'' the regulatory network---to weaken or completely remove existing interactions and add new ones, either functional or spurious.
Emergence of novel functions in such a network will usually be constrained to evolutionary trajectories that maintain a viable pattern of existing interactions.
This raises a fundamental question about the effects of such constraints on the accessibility of different regulatory architectures and the timescales needed to reach them.

The case that we focus on here is the divergence of  gene regulation, which can give rise to a variety of new phenotypes, e.g., via expansion in TF families. A regulatory function previously accomplished by a single (or several) TF(s) is now carried out by a larger number of TFs, allowing for additional fine-tuning and precision, or, alternatively, for an expansion of the regulatory scope~\cite{kacser_evolution_1984,simionato_origin_2007,larroux_genesis_2008,hobert_molecular_2010,achim_structural_2014,mckeown_evolution_2014,baker_extensive_2011,sayou_promiscuous_2014,pougach_duplication_2014, Nadimpalli_pervasive_2015,arendt_evolution_2008}. The main avenue for such expansions are gene duplications~\cite{ohno_evolution_2013,magadum_gene_2013,andersson_gene_2009,yona_chromosomal_2012}, which generate copies of the TFs and thus provide the ``raw material'' for evolutionary diversification. Subsequent specialization of TFs often involves divergence in both their inputs (e.g., ligands) and outputs (regulated genes)~\cite{wittkopp_cis-regulatory_2012,wray_evolutionary_2007}. Examples range from repressors involved in bacterial carbon metabolism that arose from the same ancestor via a series of duplication-divergence events~\cite{nguyen_phylogenetic_1995}, and ancestral TF Lys14  in the metabolism of \emph{S. cerevisiae}, which diverged into 3 different TFs regulating different subsets of genes in \emph{C. albicans}~\cite{perez_how_2014}, to many variants of Lim and Pou-homeobox genes involved in neural development across different organisms~\cite{hobert_functions_2000}. In some systems the ligand sensing and gene regulatory functions are distributed across two or more molecules, as  for bacterial two-component pathways~\cite{parkinson_signal_1993} and eukaryotic signaling cascades~\cite{bowler_emerging_1994}; here, too, specialization can occur by a series of mutations in multiple relevant components.

Immediately following a duplication event, molecular recognition between TFs, their input signals, and their binding sites is specific but undifferentiated between the two TF copies. Under selection to specialize, recognition sequences and ligand preferences of the two TFs can diverge, but only if some degree of matching between TFs and their binding sites is continually retained  to ensure network function. Binding sites are thus forced to coevolve in tandem with the TFs, yet little is known about the resulting limits to evolutionary outcomes and their dependence on important parameters: the number of regulated genes, the length and specificity of the binding sites, the correlations between the input signals, and so on.

Theoretical understanding of TF duplication is still incomplete, with existing models  predominantly belonging to two categories.
The first category of gene duplication-differentiation models studies subfunctionalization of isolated proteins (e.g., enzymes) that do not have any regulatory role~\cite{innan_evolution_2010}. When cis-regulatory mutations that control the expression of the duplicated gene are included~\cite{force_preservation_1999,lynch_probability_2000,lynch_probability_2001,force_origin_2005,proulx_multiple_2012}, this is done in a simplified fashion, e.g., by a small number of discrete alleles that represent TF binding sites appearing and disappearing at fixed rates~\cite{force_origin_2005,proulx_multiple_2012}. Because this approach ignores the essentials of molecular recognition, it cannot model co-evolution between TFs  and their binding sites---the topic of our interest.

The second category of studies tracks regulatory sequences explicitly and uses a biophysical description of TF-BS (binding site) interactions, properly accounting for the fact that TFs can bind a variety of DNA sequences with different affinities~\cite{maerkl_systems_2007,wunderlich_different_2009,payne_robustness_2014}. In conjunction with thermodynamic models of gene regulation~\cite{shea_or_1985,kinney_using_2010,sherman_thermodynamic_2012,he_thermodynamics-based_2010}, this approach has been used to study the evolution of binding sites given a single TF~\cite{berg_adaptive_2004,lassig_biophysics_2007,payne_robustness_2014,lynch_evolutionary_2015,tugrul_dynamics_2015}, while mostly overlooking the issue of TF duplication and subfunctionalization (but see~\cite{poelwijk_evolutionary_2006,burda_distribution_2010}).

Here we synthesize these two frameworks---the biophysical description of gene regulation and the evolutionary modeling of TF specialization---to construct a realistic description of the fundamental step by which regulatory networks have evolved.
A biophysical model of this setup gives rise to complex fitness landscapes that are markedly different from simple forms considered previously; in what follows, we show that realistic landscapes exert a major influence over the evolutionary outcomes and dynamics.

\section*{Results}
\subsection*{A biophysically realistic fitness landscape}
In our model, $n_{\rm TF}$ transcription factors regulate $n_G$ genes by binding to sites of length $L$ base pairs; for simplicity, we consider each gene to have one such binding site. The specificity of a TF for any sequence is determined by the TF's preferred (consensus) sequence; sequences matching consensus are assigned lowest energy, $E=0$, which corresponds to tightest binding, and every mismatch between the consensus and the binding site increases the energy by $\epsilon$; this additive ``mismatch'' model has a long history in gene regulation literature~\cite{von_hippel_specificity_1986,gerland_physical_2002,lassig_biophysics_2007,maerkl_systems_2007}.

The equilibrium probability that the binding site of gene $j$  ($j=1,\dots,n_G$) is bound by active TFs of any type $i$ ($i=1,\dots,n_{\rm TF}$) is a proxy for the gene expression level and is given by the thermodynamic model of gene regulation~\cite{shea_or_1985,bintu_transcriptional_2005}:
\be
\label{eq:p_ON_t}
p_{jm}(\{k_{ij}\}, \{C_i(m)\}) = \frac{\sum_i C_{i}(m) e^{-\epsilon k_{ij}}}{1 + \sum_i C_{i}(m) e^{-\epsilon k_{ij}}},
\ee
where $C_{i}(m)$ is dimensionless concentration of active TFs of type $i$ in condition $m$, $k_{ij}$ is the number of mismatches between the consensus sequence of the $i$-th TF species and the binding site of the $j$-th gene, and $\epsilon$ is the energy per mismatch in units of $k_BT$. Concentration $C_i(m)$ of active TFs depends on condition $m$, which can represent either time or space (e.g., during developmental gene expression programs) or a discrete external environment (e.g., the presence/absence of particular chemical signals). The simplest case considered here assumes the existence of two such signals that can be either present or absent, in any combination, for a total number of 4 possible environments ($m=00,01,10,11$), occurring with probabilities $\alpha_m$; an important parameter will be the correlation, $-1\leq\rho\leq 1$, between the two signals. Each TF has two binary alleles, $\sigma_i \in [00,01,10,11]$, determining its specificity for the two signals. If the TF $i$ is responsive to a signal and that signal is present in environment $m$, then its active concentration $C_i(m)=C_0$; otherwise, $C_i(m)=0$. Given constants $C_0$, $\epsilon$, and the genotype $\mathcal{D}$---comprising TF consensus and binding site sequences as well as TF sensitivity alleles $\sigma_i$---the thermodynamic model of Eq.~(\ref{eq:p_ON_t}) fully specifies expression levels for all genes in all environments (Supplementary Notes Section 1).

\begin{figure}
\centering
\includegraphics[width=\linewidth]{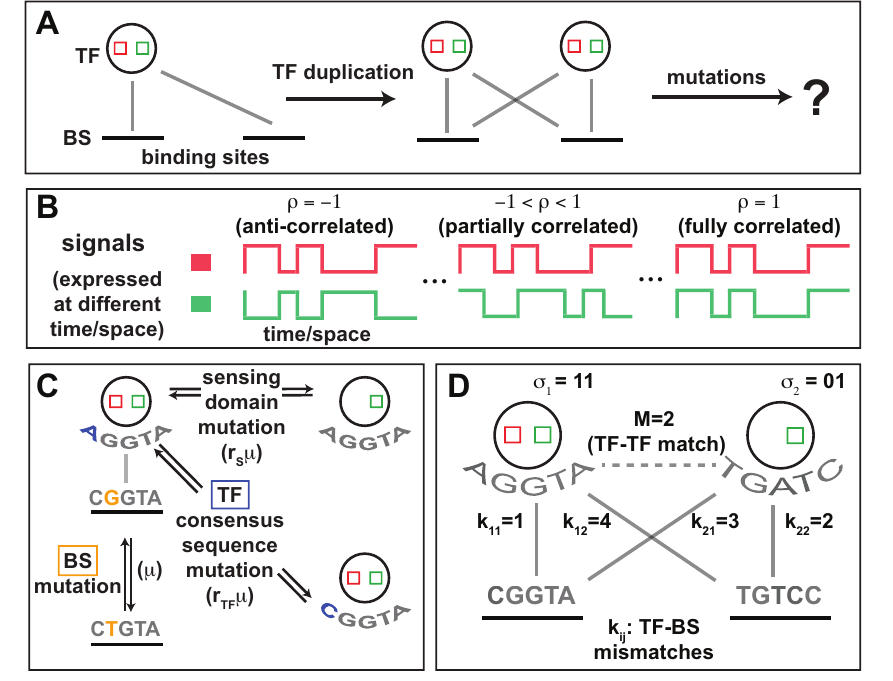}
  \caption[]
 { \label{fig:model} \textbf{Schematic of the model.}
 {\bf (A)} TF, initially responsive to two external signals (red and green ``slots'')
 and regulating two genes, duplicates and the additional copy fixes in the population. Immediately after duplication, the two copies are undifferentiated.
 {\bf (B)} A crucial parameter that will determine the fate of the duplicate is the correlation, $\rho$, of the two signals that activate or induce expression of  the TFs. The signals can correspond to different time periods in development, spatial regions in the organism or tissue, or external conditions / ligands.
 {\bf (C)} Various mutation types that can occur post-duplication with their associated rates.
{\bf (D)} After accumulating several mutations, the pattern of mismatches between TF consensus sequences and the binding sites is reflected in new values of $\{k_{ij}\}$, which determine the activation levels of the two genes according to Eq.~(\ref{eq:p_ON_t}). $M$, the number of matches between the consensus sequences of the two TFs (with a value between $0$ and $L$), keeps track of the overall divergence of the TF specificities. For a list of model parameters and baseline values see Supplementary Notes Table 1.
}
  \end{figure}

\figref{fig:model}A illustrates this setup for a simple case $n_{\rm TF}=n_G=2$, assuming that the two copies of the TF emerged through an initial gene duplication event and are fixed in the population. The original TF regulates two downstream genes by binding to their binding sites. It is sensitive to both external signals, which can be present with a varying degree of correlation (\figref{fig:model}B). After duplication, three types of mutation can occur, as shown in \figref{fig:model}C: point mutations in the binding sites (rate $\mu$), mutations in the TF coding sequence that change TF's preferred (consensus) specificity (rate $r_{\rm TF}\mu$) and mutations in the two signal-sensing alleles (rate $r_{\rm S}\mu$), which can give each TF specificity to both signals, to one of them, or to neither. An example in \figref{fig:model}D shows the state of the system after several mutations have affected the degree of (mis)match between the TFs and the binding sites, $k_{ij}$; an especially important quantity that tracks the overall divergence of the TF specificity is denoted as $M$, the match between the two TF consensus sequences.

\begin{center}
\includegraphics[width=\linewidth]{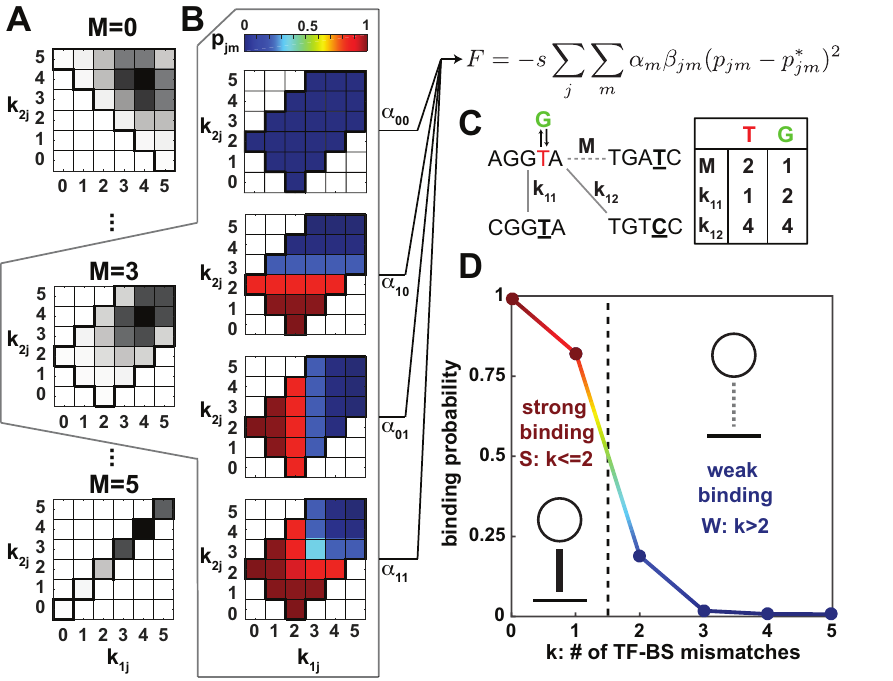}
    \captionof{figure}
 { \label{fig:k1k2_given_M} \textbf{Biophysical and evolutionary constraints shape the genotype-phenotype-fitness map after TF duplication.} {\bf (A)} Match, $M$, between transcription factor consensus sequences (here, of length $L=5$), constrains the possible mismatch values, $k_{1j},k_{2j}$, between the gene's binding site and either TF. For example, when the two TFs are identical ($M=L=5$, bottom left), they must have equal mismatches with all genes ($k_{1j}=k_{2j}$). Some combinations of mismatches are impossible given $M$ (white), while others are realized by different numbers of genotypes (grayscale).
{\bf (B)} Expression level (color) for a regulated gene given all mismatch combinations, $k_{1j},k_{2j}$, at $M=3$. Impossible mismatch combinations are white. Each of the four panels shows expression levels in four possible environments, $m=00,10,01,11$. Fitness $F$ depends on the structure of mismatches (A), the biophysics of binding (B), and the frequencies of different environments, $\alpha_m$. Here we choose $\alpha$ so that the marginal probability of each input signal is always $\frac{1}{2}$ but the correlation can be varied, and assign weight $\beta_{jm}=1$ whenever the gene should be induced but is not, and $\beta_{jm}=\frac{1}{2}$ when it should not be induced but is.
{\bf (C)} A single point mutation, e.g. a change in one TF's binding specificity from {\tt T} to {\tt G}, can simultaneously affect the match, $M$, and either increase, decrease, or leave intact the mismatches, $k_{11}$ and $k_{12}$, that determine fitness.
{\bf (D)} TF-BS interactions with mismatch $k$ that is low enough to ensure a high binding probability ($p > 2/3$) are assigned to a ``strong binding'' phenotype (solid link); conversely, $p < 1/3$ is a ``weak binding'' phenotype (dotted link).
}
\end{center}

To complete the evolutionary model, a fitness function is required. We assume selection for the genes to acquire distinct expression patterns in response to external signals, and thus define this fully specialized state as having the highest fitness in our model. Specifically, we penalize the deviations in actual gene expression, $p_{jm}$, from the ideal expression levels, $p^*_{jm}$:
\be
\label{eq:F}
F = -s \sum_j \sum_m \alpha_m \beta_{jm} (p_{jm} - p^*_{jm})^2,
\ee
where the ideal expression level $p^*_{jm}$ is 1 (fully induced) for the first gene if signal 1 is present and the expression is 0 (not induced) otherwise, and similarly for the second gene; $\beta_{jm}$ can be used to vary the relative weight of different errors (e.g. of a gene being uninduced when it should be induced and vice versa, see Supplementary Notes Section 5), and $s$ is the selection intensity. Importantly, selection does not directly depend on the TFs, but only on the  expression state of the genes they regulate; genes, however, can only be expressed when TFs bind to proper binding sites, implicitly selecting on TFs.

We consider mutation rates to be low enough  that a beneficial mutation fixes before another beneficial mutation arises~\cite{desai_beneficial_2007}, allowing us to assume that the population is almost always fixed. The probability that the population occupies a particular genotypic state, $P(\mathcal{D},t)$, evolves according to a continuous-time discrete-space Markov chain that specifies the rate of transition between any two genotypes. The transition rates are a product between the mutation rates between different states and the fixation probability that depends on the fitness advantage a mutant has over the ancestral genotypes~\cite{kimura_probability_1962,lassig_biophysics_2007}.
The size of genotype space is high-dimensional but still tractable, because our model only requires us to keep track of mismatches and not full sequences, i.e., to write out the dynamical equations for the reduced-genotypes, $\mathcal{G}=\{M,k_{ij},\sigma_i\}$. Standard Markov chain techniques can then be used to compute the evolutionary steady state, first hitting times to reach specific evolutionary outcomes, or to perform stochastic simulations (Supplementary Notes Section 2).

\figref{fig:k1k2_given_M} shows the interplay of biophysical constraints that give rise to a realistic fitness landscape for our problem. Given a  match, $M$, between two TF consensus sequences, only certain combinations of mismatches, ($k_{1j},k_{2j}$), of the TFs with each of the two binding sites are possible. A particular allowed combination can be realized by different numbers of genotypes, as shown in \figref{fig:k1k2_given_M}A, providing a detailed account of the entropy of the neutral distribution. For each of the four environments, Eq.~(\ref{eq:p_ON_t}) predicts gene expression at every pair of mismatch values (\figref{fig:k1k2_given_M}B); together with the probabilities of different environments occurring, the gene expression pattern determines the genotypes's fitness, $F$. TF specialization then unfolds on this landscape by different types of mutations (e.g., \figref{fig:k1k2_given_M}C). Although the landscape is complex and high-dimensional, it is highly structured and ultimately fully specified by only a handful of biophysical parameters. Furthermore, because of the sigmoidal shape of binding probability as a function of mismatch $k$ [Eq.~(\ref{eq:p_ON_t})], it is possible to assign phenotypes of ``strong'' and ``weak'' binding to every TF-BS interaction, allowing us to depict network interactions graphically, as shown in \figref{fig:k1k2_given_M}D, and to classify the possible macroscopic evolutionary outcomes, as we will show next.

\subsection*{Evolutionary outcomes in steady state}
Evolutionary outcomes in steady state are determined by a balance between selection and drift. The steady state distribution over reduced-genotypes is~\cite{gillespie_population_2004}
\be
P_{\rm SS}(\mathcal{G}) = P(\mathcal{G},t\rightarrow\infty) = P_0(\mathcal{G}) \exp(2 N F(\mathcal{G})),
\label{eq:balance}
\ee
where $P_0$ is the neutral distribution of genotypes and $N$ is the population size. Eq.~(\ref{eq:balance}) is similar to the energy/entropy balance of statistical physics~\cite{sella_application_2005}, with fitness $F$ playing the role of energy and $\log P_0$ the role of entropy; in our model, both of these quantities are explicitly computable, as is the resulting steady state distribution.

Understanding the high dimensional distribution over genotypes is difficult, but classification of individual TF-BS interactions into ``strong'' and ``weak'' ones, as described above, allows us to systematically and uniquely assign every genotype to one of a few possible macroscopic outcomes, or ``macrostates,'' graphically depicted in \figref{fig:ss_dist}A and defined precisely in Supplementary Notes Section 1. Thus, in the {\tt No Regulation} state, input signals are not transduced to the target genes, either because TF-BS mismatches are high and there is no binding or because TFs themselves lose responsiveness to the input signals; in the {\tt One TF Lost} state, a single TF regulates both genes (as before duplication), while the other TF is lost, i.e., its specificity has diverged so far that it does not bind any of the sites;  the {\tt Specialize Binding} state corresponds to each TF regulating its own gene without cross-regulating the other but the signal sensing domains are not yet signal specific, as they are in the {\tt Specialize Both}, the state which we have defined to have the highest fitness. Finally, the {\tt Partial} macrostate predominantly features configurations where each of the TFs binds at least one binding site, but one of the TFs still binds both sites or retains responsiveness for both input signals; functionally, these configurations lead to large ``crosstalk,'' where input signals are non-selectively transmitted to both target genes.

Ultimately, these macrostates are the functional network phenotypes that we care about. The number of genotypes in each macrostate, however, can vary by orders of magnitude; for example, the {\tt No Regulation} state is larger by $\sim 10^4$ relative to the high-fitness {\tt Specialize Both} state, for our baseline choice of parameters ($L=5, \epsilon = 3$). Selection can act against this strong entropic bias, and the distribution of fitness values across genotypes within each macrostate is shown in \figref{fig:ss_dist}B. Clearly, the mean or median fitness within each macrostate is a poor substitute for the detailed structure of fitness levels that depend nonlinearly on TF-BS mismatches and the degeneracy of the sequence space. Unlike the entropic term in \figref{fig:ss_dist}A, fitness also depends on the statistics of the environment, $\alpha_m$, and in particular, the correlation $\rho$ between the two signals. For example, when the signals are strongly correlated, the {\tt Initial} state right after duplication or the {\tt One TF Lost} state can achieve quite high fitnesses, since responding to the wrong signal or having a high degree of crosstalk will still ensure largely appropriate gene expression pattern in all likely environments. In contrast, at strong negative correlation, many genotypes in {\tt Specialize Binding} and {\tt Initial} states will suffer a large fitness penalty because their sensing domains are not specialized for the correct signals, while the {\tt Specialize Both} state will have high fitness regardless of the environmental signal correlation.
%
\begin{center}
\includegraphics[width=\linewidth]{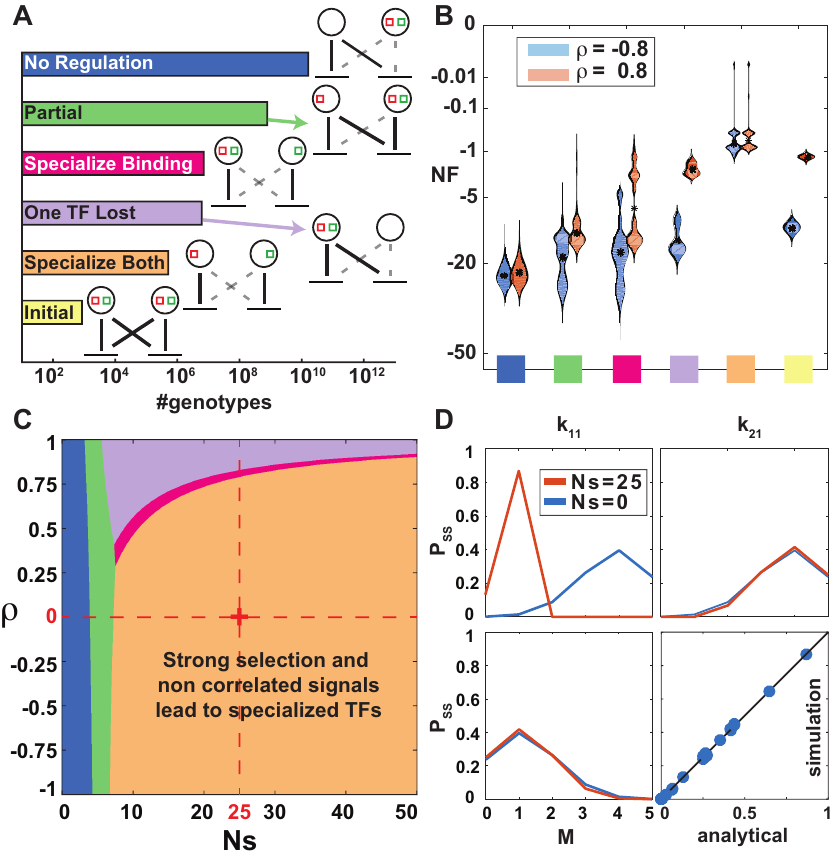}
    \captionof{figure}
 { \label{fig:ss_dist}
  \textbf{Steady state evolutionary outcomes of TF duplication.}
  {\bf (A)}  Evolutionary macrostates (see text) depicted graphically as network phenotypes with solid (dashed) lines indicating strong (weak) TF-BS interactions. Logarithmic scale indicates the number of genotypes in each macrostate.
  {\bf (B)} Distribution of fitness values across genotypes in each macrostate (color-coded as in A), shown as violin plots, for two values of signal correlation, $\rho$. Black dots = median fitness in the macrostate.
  {\bf (C)} Most probable outcome of gene duplication in steady state (color-coded as in A), as a function of selection strength, $Ns$, and the correlation between two external signals, $\rho$.
{\bf (D)} Steady state distributions for mismatches ($P_{\rm SS}(k_{ij}|\sigma_1=10,\sigma_2=01)$, upper row) and the match between the two TF consensus sequences ($P_{\rm SS}(M|\sigma_1=10,\sigma_2=01)$, lower left), under strong selection (red; at baseline parameters denoted by the red cross in C) and neutrality (blue; Bernoulli distributions). Comparison between analytical calculation and 400 replicates of the stochastic simulation (lower right). Here and in subsequent figures, baseline parameter values are $L=5$, $\epsilon=3$, $r_{\rm S}=r_{\rm TF}=1$.
}
\end{center}

%
How do fitness and entropy combine to determine macroscopic evolutionary outcomes? \figref{fig:ss_dist}C shows the most probable macrostate as a function of selection strength and signal correlation (Supplementary Notes Section 3). At weak selection, specific TF-BS interactions cannot be maintained against mutational entropy and the system settles into the most numerous, {\tt No Regulation} state. Higher selection strengths can maintain a limited number of TF-BS interactions in {\tt Partial} states. Beyond a threshold value for $Ns$, the evolutionary outcome depends on the signal correlation: when signals are anti-correlated or weakly correlated, the TFs reach the fully specialized state, whereas high positive correlation favors losing one TF and having the remaining TF regulate both genes and respond to both signals. As signal correlation increases, so does the selection strength required to support full specialization.

The map of evolutionary outcomes is very robust to parameter variations. The energy scale of TF-DNA interactions is that of hydrogen bonds: $\epsilon\sim 3$ (in $k_B T$ units), consistent with direct measurements. The scale of $C_0$ is set to ensure that consensus sites are occupied at saturation while fully mismatching sites are essentially empty.  The only remaining important biophysical parameter is $L$, the length of the binding sites. As expected, increasing $L$ expands the regions of {\tt No Regulation} and {\tt Partial} at low $Ns$, due to entropic effects. Surprisingly, however, one can demonstrate that the important boundary between the {\tt Specialize} and {\tt One TF Lost} states is independent of $L$; furthermore, the map in~\figref{fig:ss_dist}C is exactly robust to the overall rescaling of the mutation rate, $\mu$, and even to separate rescaling of individual rates $r_{\rm S}, r_{\rm TF}$. 

We compare the steady-state marginal distributions of TF-BS mismatches and the match, $M$, between the two TFs, under strong selection to specialize ($Ns=25$) vs neutral evolution ($Ns=0$). Mismatch distributions for $k_{11}$ and $k_{21}$ in \figref{fig:ss_dist}D display a clear difference in the two  regimes: strong selection favors a small mismatch of the BS with the cognate TF, sufficient to ensure strong binding but nonzero due to entropy, and a large mismatch with the noncognate TF, to reduce crosstalk.
Surprisingly, however, the distribution of matches $M$ between two TF consensus sequences shows only a tiny signature of selection, with both distributions peaking around $1$ match.
As a consequence, inferring selection to specialize from measured binding preferences of real TFs might not be feasible with realistic amounts of data.
%
\begin{center}
\includegraphics[width=\linewidth]{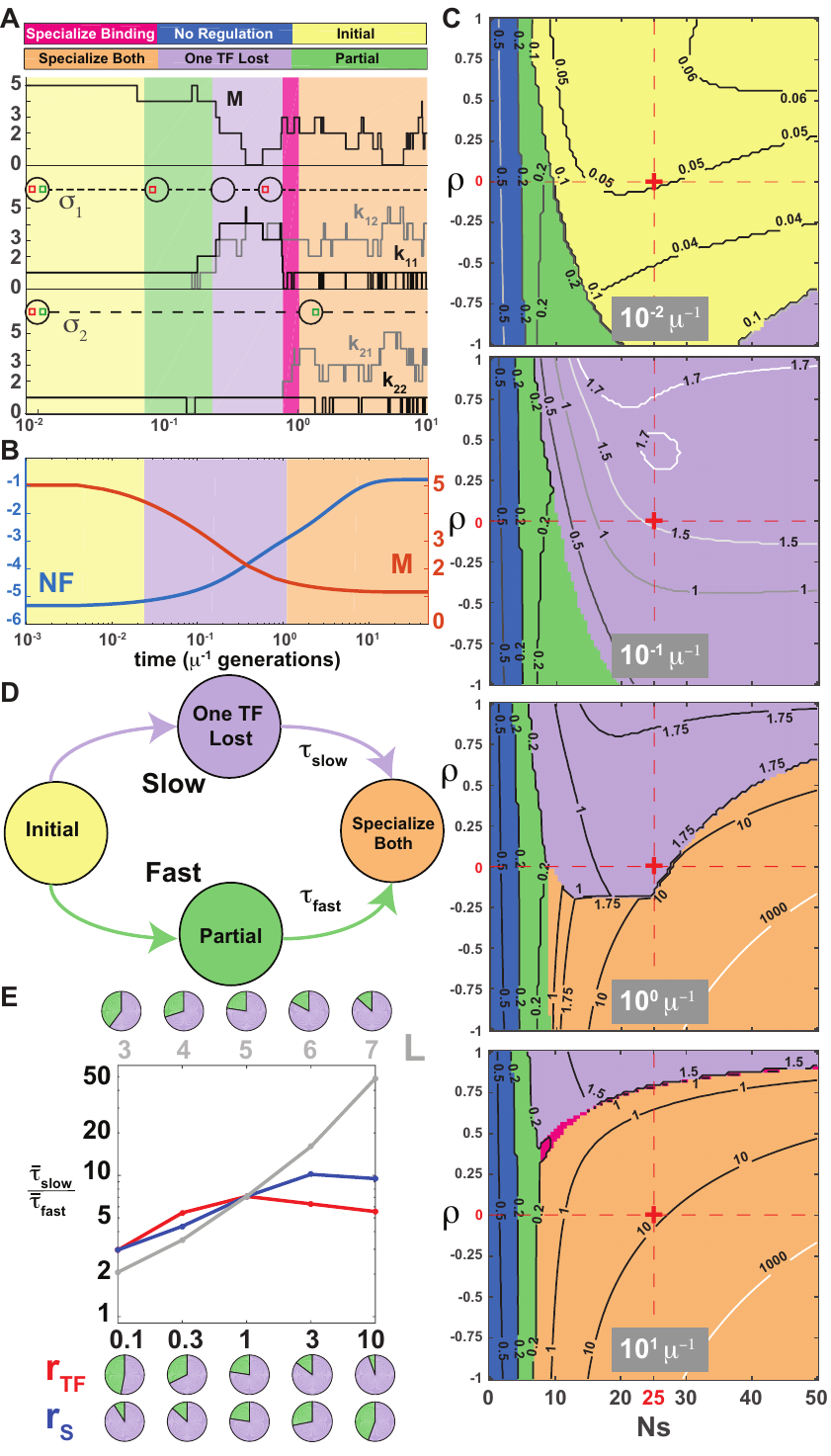}
    \captionof{figure}
 { \label{fig:dyn}
\textbf{Slow and fast pathways to TF specialization.}
 {\bf (A)} Temporal traces of TF-TF match $M$ (top), and TF-BS mismatches $k_{ij}$ (middle: TF1, bottom: TF2) with the corresponding signal specificity mutations denoted on dashed lines, for one example evolutionary trajectory at baseline parameters.  Macrostates are color-coded as in the top legend and~\figref{fig:ss_dist}.
 {\bf (B)} Average dynamics of fitness $NF$ (blue, left scale) and TF-TF match $M$ (red, right scale). For every timepoint, the dominant macrostate is denoted in color.
 {\bf (C)} Snapshots of dominant macrostates (at increasing time post-duplication as indicated in the panels), shown for different combinations of selection strength $Ns$ and signal correlation $\rho$ as in \figref{fig:ss_dist}.
 Contours mark dwell times in the dominant macrostates (in units of $\mu^{-1}$). Red cross = baseline parameters.
 {\bf (D)} Schematic of the two alternative pathways to specialization. $\tau_{\rm slow}$ and $\tau_{\rm fast}$ are the  total times to specialization for the ``slow'' and the ``fast'' pathway, respectively.
 {\bf (E)} Relative duration of the two pathways, as a function of binding site length $L$ (gray line, top axis), TF consensus sequence mutation rate $r_{\rm TF}$ (red), and signal domain mutation rate $r_{\rm S}$ (blue, bottom axis). Pie charts indicate the fraction of slow (pink) and fast (green) pathways at each parameter value.
 }
\end{center}

\subsection*{Evolutionary dynamics and fast pathways towards specialization}

Next, we focus on evolutionary trajectories and the timescales to reach the fully specialized state after gene duplication.
 An example trajectory is shown in \figref{fig:dyn}A: the two TFs start off identical (with maximal match, $M=L=5$) until, as a result of the loss of specificity for both signals, TF1 starts to drift, diverging from TF2 (sharply decreasing $M$ in {\tt One TF Lost} state) and losing interactions with both binding sites. Subsequently TF1 reacquires preference to the red signal, which drives the reestablishment of TF1 specificity for one binding site during a short {\tt Specialize Binding} epoch,  followed quickly by the specialization of TF2 for the green signal at the start of {\tt Specialize Both} epoch of maximal fitness.

Dynamics of the TF-TF match, $M$, and the scaled fitness, $NF$, become smooth and gradual when discrete transitions and the consequent large jumps in fitness are averaged over individual realizations, as in~\figref{fig:dyn}B. Importantly, we learn that the sequence of dominant macrostates leading towards the final (and steady) state, {\tt Specialize Both}, involves a long intermediate epoch when the system is in the {\tt One TF Lost} state. We examine this sequence of most likely macrostates in detail in~\figref{fig:dyn}C, and visualize it analogously to the map of evolutionary outcomes in steady state shown in~\figref{fig:ss_dist}C. High $Ns$ and correlation ($\rho$) values favor trajectories passing through the {\tt One TF Lost} state, while intermediate $Ns$  ($5\lesssim Ns \lesssim 20$) and low correlation values enable transitions through {\tt Partial} macrostate; along the latter trajectory, the binding of neither TF is completely abolished. Typical dwell times in dominant states, indicated as contours in~\figref{fig:dyn}C, suggest that specialization via the {\tt One TF Lost} state should be slower than through the {\tt Partial} state, which is best seen at $t=1/\mu$, 
where specialization has already occurred at intermediate $Ns$ and low, but not high, $\rho$ values.

It is easy to understand why pathways towards specialization via the {\tt One TF Lost} state are slow. As the example in~\figref{fig:dyn}A illustrates, so long as one TF maintains binding to both sites and thus network function (especially when signals are strongly correlated), the other TF's specificity will be unconstrained to neutrally drift and lose binding to both sites, an outcome which is entropically highly favored. After the TF's sensory domain specializes, however, the binding has to re-evolve essentially from scratch in a process that is known to be slow~\cite{tugrul_dynamics_2015} unless selection strength is very high. In contrast to this ``Slow'' pathway, the ``Fast'' pathway via the {\tt Partial} state relies on sequential loss of ``crosstalk'' TF-BS interactions, with the divergence of TF consensus sequences followed in lock-step by mutations in cognate binding sites. Specifically, the likely intermediary of the fast pathway is a {\tt Partial} configuration in which the first TF responds to both signals but only regulates one gene, whereas the second TF is already specialized for one signal, but still regulates both genes.

The fast and the slow pathways are summarized in~\figsref{fig:dyn}D. A detailed analysis (Supplementary Notes Section 4)
reveals how different biophysical and evolutionary parameters change the relative probability and the average duration (\figref{fig:dyn}E
of both pathways. For example, increasing the length, $L$, of the binding sites favors the slow pathway as well as drastically increases its duration, leading to very slow evolutionary dynamics. In contrast, time to specialize via the fast pathway is unaffected by an increase in $L$.
 Increasing the rate of TF-specificity-affecting mutations, $r_{\rm TF}$, has a qualitatively similar effect, while increasing the mutation rate affecting the sensory domain, $r_{\rm S}$, favors the fast pathway. Indeed, in the limit when $r_{\rm S}$ is much larger than the other two mutation rates, 
the sensing domain specializes almost instantaneously, making the complete loss of binding by either TF very deleterious and thus avoiding the {\tt One TF Lost} state; the adaptation dynamics is initially rapid, with binding sites responding to diverging TF consensus sequences, and subsequently slow, when TF consensus sequences further minimize their match, $M$, in a nearly neutral process.
%
%
%

%
\subsection*{Promiscuity-promoting mutations}
Typically, each TF must regulate more than one target gene. As the number of regulated genes per TF ($n_G/n_{\rm TF}$)
increases, intuition suggests that the evolution of the TF's consensus sequence should become more and more constrained: while a mutation in an individual binding site can lower the total fitness by increasing mismatch and thereby impeding TF-BS binding, a single mutation in the TF's consensus has the ability to simultaneously weaken the interaction with many binding sites, leading to a high fitness penalty. Our analysis of the biophysical fitness landscape confirmed that the landscape gets progressively more frustrated as the number of regulated genes per TF increases, due to the explosion of constraints that TFs have to satisfy to ensure the maintenance of functional regulation (Supplementary Notes Section 7). 
Consequently, one can expect extremely long times to specialization. How can it nevertheless proceed at observable rates?

%
\begin{center}
\includegraphics[width=\linewidth]{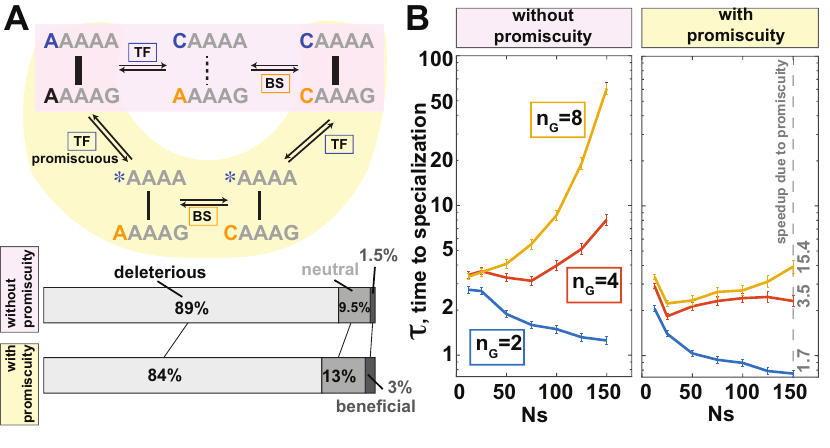}
  \captionof{figure}
 { \label{fig:prom}
 {\bf Promiscuity-promoting mutations speed up specialization with multiple regulated genes per TF.}
 {\bf (A)} In the absence of promiscuity-promoting mutations, a compensatory series of point mutations in the TF's consensus (upper sequence) and its binding site (lower sequence) is needed to maintain TF-BS specificity (top; light red). Alternatively, in the presence of promiscuity-promoting mutations in the TF consensus, a position in the TF's recognition sequence (marked by a star) can lose and later regain sequence specificity (middle; light yellow). Promiscuity decreases the fraction of deleterious mutations along typical pathways to specialization (bottom, computed using baseline parameters).
 {\bf (B)} Time to specialization as a function of selection strength, $Ns$, without (left) and with (right) promiscuity promoting mutations in the TF, for different numbers of regulated genes per TF, $n_G$ (color).
}
\end{center}
%

Energy matrices for many real TFs display  
``promiscuous'' specificity where, at a particular position within the binding site, binding to multiple nucleotides is equally preferable. We wondered how our findings would be affected if consensus sequence specificity of the TFs could pass through such intermediate promiscuous states.   \figref{fig:prom}A shows how TF consensus sequence and the corresponding binding site can co-evolve using point mutations, or using the new ``promiscuity-promoting'' mutation type for the TF: promiscuity-promoting mutation renders one position in the recognition sequence of the TF insensitive to the corresponding DNA base in the binding site (Supplementary Notes Section 8). Evolutionary pressure on the binding sites is therefore temporarily relieved, until the specificity of the TF is reestablished by a back mutation. Without promiscuity-promoting mutations, TF-BS co-evolution must proceed in a tight sequence of compensatory mutations; with promiscuity-promoting mutations, such a precise sequence is no longer required, although one extra mutation is needed to reestablish high TF-BS specificity. With promiscuity, the fraction of deleterious mutations along the evolutionary path towards specialization is reduced, an effect that grows stronger with increasing $L$. As shown in~\figref{fig:prom}B, this has drastic effects on the time to specialization. Without promiscuity, increasing the selection strength, $Ns$, decreases the required time when each TF regulates one gene, as expected for a landscape with large neutral plateaus but with no fitness barriers. For $n_G>2$, however, the landscape develops barriers that need to be crossed, and evolutionary time starts increasing with $Ns$. In contrast, promiscuity enables fast emergence of TF specialization even with multiple regulated genes in a broad range of evolutionary parameters (although there are also costs due to high promiscuity).

\section*{Discussion}
The role that the shape of a fitness landscape plays for the dynamics and the final outcomes of evolution has been appreciated in population genetics for a long time. This has stimulated a large body of theoretical research into evolution on toy model landscapes~\cite{kauffman_towards_1987,kryazhimskiy_dynamics_2009}, as well as motivated efforts to map out real, small-scale landscapes experimentally. For limited classes of problems, mostly those involving molecular recognition,  biophysical constraints are informative enough to permit computational exploration of complex landscapes. Such is the case for the secondary structure of RNA~\cite{schuster_sequences_1994}, antibody-antigen interactions, protein-protein interactions, and transcription factor-DNA binding, explored here. We exploit this prior knowledge to construct a fitness landscape for a more complicated evolutionary event, the specialization of two TFs after duplication, a key evolutionary step by which gene regulatory networks expand. The biophysical model naturally captures a number of essential features, without having to introduce them ``by hand'': the fact that specialization is driven by avoidance of regulatory crosstalk; the importance of the mutational entropy; the dependence on number of downstream genes; the existence of transient network configurations preceding specialization, which crucially impact dynamics; and the importance for evolutionary outcomes of the statistical properties of the signals that TFs respond to. Importantly, the expressive power of our framework does not come at increased modeling cost: while complex, the fitness landscape is still determined only by a few, mostly known, parameters, and an exponentially large space of genotypes can be systematically coarse grained to a small set of functional network phenotypes. This combination of biophysical and co-evolutionary approaches is applicable generally to the evolution of molecular interactions, e.g., in protein interaction networks.

In steady state, our results robustly identify correlation between the environmental signals that drive TFs as a key determinant for specialization, as shown in \figref{fig:ss_dist}C. Unless the new signal, for which a post-duplication TF can specialize, is sufficiently independent (uncorrelated) from the existing signals
that the regulatory network processes,  one TF copy will be lost due to drift.
As a consequence, the \emph{effective dimensionality of environmental signals} dictates the complexity of genetic regulatory networks~\cite{friedlander_evolution_2015}, reminiscent of information-theoretic tradeoffs in sensory neuroscience; in evolutionary terms, selection to maintain complex regulation needs to withstand the mutational flux into vastly more numerous but less functional network phenotypes. Recently, it has been shown that finite biochemical specificity also limits the complexity of genetic regulatory networks~\cite{friedlander_intrinsic_2016}; an interesting direction for future research is to understand how the balance between regulatory crosstalk, environmental signal statistics, and evolutionary constraints ultimately determines the number of TFs that can be stably maintained.  A related question concerns the expected match between pairs of TFs in a large network as a signature of selection for specialized function; for an isolated pair of TFs, our results in~\figref{fig:ss_dist}D predict only a tiny deviation from neutrality.

Timescales and pathways to specialization are completely shaped by the properties of the biophysical fitness landscape, and thus cannot be captured by simple allelic models that ignore the topology of the sequence space (Supplementary Notes Section 6). We show that the fast pathway to specialization transitions through {\tt Partial} states where neither of the two TFs completely loses binding. Interestingly, it is exactly the existence of crosstalk interactions that permits fast adaptation via these transient states, by maintaining the network function through one TF, while the other is free to diverge in a series of mutations to the TF and its future binding site~\cite{shultzaberger_probing_2012}. Crosstalk thus enables some amount of network plasticity during early adaptation, yet is ultimately selected against, when TFs become fully specialized~\cite{rowland_crosstalk_2014,eldar_social_2011}. In the protein-protein-interaction literature, {\tt Partial} states are sometimes referred to as promiscuous states, and they have been suggested as evolutionarily accessible intermediaries that relieve the two interacting molecules of the need to evolve in a tight (and likely very slow) series of compensatory mutations \cite{aakre_evolving_2015}. 
In contrast to the fast pathway, the slow pathway involves a complete loss of TF-BS binding interactions; the long timescale emerges from long dwell times while the TF  and the binding sites evolve in a nearly neutral landscape before TF-BS specificity is reacquired. Long binding sites and (perhaps counter-intuitively) fast TF mutation rates favor the slow pathway, while fast sensing domain mutation rates favor the fast pathway.

The situation changes qualitatively when each TF regulates more genes~\cite{sengupta_specificity_2002}. On the one hand, entropy makes pathways that pass through the {\tt One TF Lost} state dynamically uncompetitive, as multiple binding sites would have to emerge {\emph de novo} to reestablish interactions with a diverged TF. This would favor fast pathways through {\tt Partial} states. On the other hand, the biophysical fitness landscape develops frustration (or sign epistasis) as $n_G>2$ and the timescales to specialization lengthen with increasing selection strength when passing through {\tt Partial} states. We demonstrate that frustration is relieved by promiscuity-promoting mutations in the transcription factor, enabling fast emergence of specialization even with multiple regulated genes.

Taken together, our results paint a picture of TF specialization that most likely proceeds through intermediate states with high crosstalk, in which one TF has already specialized for its input signals but not yet for the target genes, while the other TF is not yet specialized for the input signals but only regulates one gene. In addition, these intermediate states are likely to be more promiscuous, binding different sites with the same affinity, with the promiscuity reverting to specific binding towards the end of specialization. This picture is qualitatively different from the paradigmatic idea of a simple and sequential progression of compensatory mutations in the TF and its binding sites~\cite{de_vos_breaking_2015,poelwijk_evolutionary_2006}. It depends fundamentally on the biophysical model of TF-BS interactions, predicts significantly faster specialization times, as well as the existence of promiscuous TF variants that are starting to be observed in genomic analyses of duplication-specialization events~\cite{sayou_promiscuous_2014,pougach_duplication_2014}.

%
%
\textbf{Acknowledgments}
We thank the People Programme (Marie Curie Actions) of the European Union's Seventh Framework Programme (FP7/2007-2013) under REA grant agreement Nr. 291734 (T.F.), ERC grant Nr. 250152 (N.B.), and Austrian Science Fund grant FWF P28844 (G.T.).

\newpage
\baselineskip = 14 pt \leftline{\large \bf Evolution of new regulatory functions on biophysically realistic fitness landscapes}
\smallskip
\leftline{\large \it Supporting Information}
\bigskip
\baselineskip = 14 pt \leftline{Tamar Friedlander, Roshan Prizak, Nicholas H. Barton and Ga\v{s}per Tka\v{c}ik}
\bigskip
\leftline{\today}
\smallskip
\hrule\bigskip\bigskip

\tableofcontents

\newpage

\section{Model description and parameters}
\label{sec:Model}
\subsection{Biophysical model}
    Consider a transcription factor (TF) that activates $n_G$ ($\geq 2$) downstream genes. The starting point of our evolutionary model is a duplication event of the TF, where the duplicate is fixed in the population.
Gene regulation is accomplished by the binding of either TF (original or duplicate) to a short DNA sequence of length $L$ associated with the gene (abbreviated below as 'BS': binding site). For simplicity we assume each gene has only a single BS. We describe the DNA-binding preference of each TF by its (unique) consensus sequence -  the $L$-base-pair sequence to which it binds with highest affinity. We begin by assuming that each TF has only a unique consensus sequence and later on relax this assumption (see Section \ref{sec:Promiscuity}).
In our simple model, a gene is activated when its BS is bound by an activating TF. The probability that the binding site of gene $j$ is bound by either TF is calculated using the thermodynamic model of gene regulation~\cite{shea_or_1985,gerland_physical_2002}:

\be
\label{eq:p_ON_t}
p_{jm}(\{k_{ij}\}, \{C_i(m)\}) = \frac{\sum_i C_{i}(m) e^{-\epsilon k_{ij}}}{1 + \sum_i C_{i}(m) e^{-\epsilon k_{ij}}},
\ee

where $\{k_{ij}\}_{i=1}^2$ is the number of sequence mismatches between the consensus sequence of the $i$-th TF species and the binding site of the $j$-th gene and $\epsilon$ is the energy per mismatch. We consider multiple environments $m$ that differ in TF concentrations: $C_{i}(m)$ is the dimensionless concentration of the $i$-th TF in environment $m$. Associated with each TF $i$ is an associated (complex) allele $\sigma_i$ that determines the TF concentration $C_i(m)$ in different environments. Eq.~(\ref{eq:p_ON_t}) assumes that all base pairs have equal and additive contributions to the binding energy, such that the binding probability only depends on the number of mismatches $k_{ij}$~\cite{von_hippel_specificity_1986,gerland_physical_2002,lassig_biophysics_2007,maerkl_systems_2007}.

Together, the TF consensus sequences, the BS sequences and the complex alleles $\sigma_i$ compose the genotype. Genotypes come from the space of all possible genotypes $\mathcal{D}$, and they completely describe the regulatory activity of the system in different environments.

\noindent We study two variants of the model, depending on whether $\sigma_i$ is evolvable or not.
\subsubsection*{Main model}
In this model variant, which is described in the main text, transcription factors are equipped with an evolvable signal sensing domain (captured by $\sigma_i$). The original TF senses two distinct external signals. Each of the downstream genes is suitable to respond to only one of the two signals. Before duplication the genes are constrained to follow the only TF available which responds to both signals. The extra TF formed in the duplication event offers an additional degree of freedom in regulating these genes, if the TFs specialize such that each of them senses only one of the two signals and regulates only a subset of the genes.

This model variant is applicable to more general pathway architecture than a TF that implements both signal sensing and gene regulation in the same molecule. Often these two functions are split between different components of the same pathway; for example, a separate upstream component senses the signal(s) and consequently activates the TF (e.g. by phosphorylation or another modification). 
Additionally, TF production is also regulated. One can also think of the evolution of the regulatory sequences of the gene coding for the TF in terms of our model. Since our model is defined in very general terms, it can capture such situations as well.

\subsubsection*{Alternative model}
In the alternative model, which we explore in the SI, transcription factors have no explicit evolvable signal sensing domain (no complex allele $\sigma_i$ associated with them), but can be expressed at different time or location as determined by $C_i(m)$. Before duplication the genes are constrained to follow the only TF available, and are thus expressed at the same time or location. After TF duplication, the two copies immediately specialize to be active at different time slots (different parts of the cell cycle, different phases of developmental process) or space (different tissues), and as such enable distinct expression patterns for the downstream genes. This variant is a limiting case of the main model, with the main difference being the lack of an evolvable TF signal sensing domain. It also acts as an approximation when the signal sensing domain evolves very quickly, resulting in a quick divergence of TF expression patterns.

Gene birth can occur via different biological mechanisms, some of them allowing for the emergence of slightly modified copies of original genes or allowing for different regulation of the same coding sequence.
One such mechanism is called 'retroposition': creation of duplicate gene copies in new genomic positions through the reverse transcription of mRNAs from source genes (also known as RNA-based duplication or retroduplication)~\cite{kaessmann_RNA-based_2009}. These newly formed genes often lack regulatory elements of the parental gene and may also be slightly modified due to transcription errors (that are significantly more common than DNA-duplication errors). It was shown that transcription of these so-called 'retrogenes' is very common and often relies on regulatory elements of neighboring genes~\cite{vinckenbosch_evolutionary_2006}.

\subsection{Evolutionary model}
We define fitness such that the specialized genotypes have higher fitness compared to the initial non-specialized genotypes. The fitness of a genotype equals the squared deviation of the actual expression $p_{jm}$ from the ideal one $p^*_{jm}$, summed over all genes $j$ and averaged over all environments $m$:

\be
\label{eq:F}
F = -s \sum_j \sum_m \alpha_m \beta_{jm} (p_{jm} - p^*_{jm})^2,
\ee

where $s$ denotes the selection intensity and $\alpha_m$ is the frequency of the $m$-th environment. We define environments by the presence or absence of the signals, which result in different active TF concentrations depending on their signal responsiveness. $\beta_{jm}$ is the penalty for each type of deviation from the ideal expression level, allowing for diverse penalties for different genes or at different environments. For example, a gene which is not expressed when needed can incur a higher penalty than the expression of a gene that is not necessary in a given environment. To capture these latter interactions, which we call crosstalk interactions, we exploited $\beta_{jm}$ to tune the fitness penalty in \secref{sec:beta}. Expression levels $p_{jm}$ for a genotype are calculated using \eqref{eq:p_ON_t} by obtaining the dimensionless concentrations of the TFs, $C_i(m)$, from their signal sensing alleles $\sigma_i$, and the mismatches, $k_{ij}$, from the TF consensus sequences and the BS sequences.

Note that the fixation probability in \eqref{eq:P_fix} below, depends, via the fitness, and in turn via the binding probabilities, directly on the TFs' signal sensing alleles $\sigma_i$, and the mismatches $k_{ij}$ of the BS sequences with the TF consensus sequences, but not on $M$, the match between the TF consensus sequences. But, as shown in Fig. 2A of the main text, the set of possible $k_{ij}$'s is constrained by $M$, and hence, there is implicit selection on $M$. Also, importantly, selection does not directly depend on the TFs and BSs, but only via their biophysical interaction to result in appropriate gene regulation, thereby requiring concerted evolution of TFs and BSs.

The evolutionary process proceeds via three types of mutations:
The BS of each downstream gene can acquire point-mutations at rate $\mu$; the consensus sequence of each TF can have point-mutations at rate $r_{\mbox{\tiny{TF}}}\mu$. These two mutation types can modify the (mis)match values $M$ and $k_{ij}$.
A third type of mutation exists in the first model variant: the signal-sensing domain of each TF has two components, each of them can alternate between two alleles (sensitive/ non-sensitive to one of the two signals) at rate  $r_s\mu$.
 Owing to the faster time-scales over which gene regulation evolves, we consider only these types of mutations on the BSs and TFs. In particular, we assume no change in the coding regions of the downstream genes themselves, only in their regulation.

\subsection{Putting the pieces together}
In our main model, we consider $n_G=2$ downstream genes (models considering larger sets of downstream genes are explored in \secref{sec:MultipleGenes}), each of which is equipped with a binding site of length $L$, and two signals, with the presence/absence of the first (second) signal requiring the expression/silencing of the first (second) gene. In other words, information should be passed from the first signal to the first gene and from the second signal to the second gene.

The presence ('$1$') or absence ('$0$') of these two signals defines the different environments $m \in \{00,01,10,11\}$ that are possible, with $\alpha_m$ denoting the frequency of environment $m$. The frequency of each signal can be obtained as $f_1=\alpha_{10}+\alpha_{11}$ and $f_2=\alpha_{01}+\alpha_{11}$. Assuming that both signals appear at equal frequencies, $f_1=f_2$, and that each signal is present (or absent) half of the time, $f_1=f_2=0.5$, we obtain the following relations between $\rho$, the correlation between the signals, and $\alpha_m$:
\[\alpha_{00}=\alpha_{11} = \frac{1}{4}(1+\rho)\]
\[\alpha_{10}=\alpha_{01} = \frac{1}{4}(1-\rho).\]

Thus when the signals are uncorrelated ($\rho=0$), we have $\alpha_{00}=\alpha_{10}=\alpha_{01}=\alpha_{11}=1/4$. When the signals are fully correlated ($\rho=1$) we obtain $\alpha_{00}=\alpha_{11}=0.5$ and $\alpha_{10}=\alpha_{01}=0$ and vice versa for anti-correlation ($\rho=-1$). We explore asymmetric environments in \secref{subsec:asymm}.

The information transmission between signals and genes is mediated by TFs which contain a signal-sensing domain and a DNA-binding domain. TFs, on sensing a signal, become active and can induce the expression of a gene by binding to its binding site. We define each TF $i$ by its consensus sequence,  the sequence of length $L$ for which the TF has the highest affinity, 
and its signal sensing allele $\sigma_i \in \{ 00,01,10,11\}$, which describes its responsiveness to the two signals. If a TF $i$ is responsive to a signal and that signal is present in environment $m$, then its active dimensionless concentration $C_i(m)=C_0$, and $C_i(m)=0$ otherwise. For simplicity, we assume only these two concentration levels.

The regulatory network is described by its genotype, $\mathcal{D}$, consisting of the consensus sequences and the signal sensing alleles of the two TFs, and the BS sequences of the (two) genes. As described in Eq.~(\ref{eq:p_ON_t}) and Eq.~(1) of the main text, the probability $p_{jm}$ that the binding site of gene $j$ is bound in environment $m$ depends on, apart from $\epsilon$, the mismatches $k_{ij}$ (which can be obtained from the genotype sequences) between the consensus sequence of TF $i$ and the BS of gene $j$, and the signal sensing alleles $\sigma_i$ which determine the active concentrations $C_i(m)$.

In Eq.~(\ref{eq:F}) and Eq.~(2) of the main text, we define the fitness of a genotype by considering the deviation of the actual expression levels $p_{jm}$ from the ideal expression levels $p_{jm}^*$. We define the ideal expression level of gene $j$ in environment $m$, $p_{jm}^*$, such that $p_{jm}^*=1$ if signal $j$ is present in environment $m$ and $p_{jm}^*=0$ if signal $j$ is absent in environment $m$. We consider the penalty $\beta_{jm}=1$ if gene $j$ is required in environment $m$ and $\beta_{jm}=\beta_X$ ($\beta_X \in [0,1]$) if gene $j$ is not required in environment $m$. $\beta_{X}$ quantifies the relative penalty on crosstalk interactions between signals and genes, compared to functional interactions. We explore the role of $\beta_X$ in \secref{sec:beta}. In Table 1 we list the model parameters and their baseline values used in calculations (unless stated otherwise).

With the fitness of genotypes and the mutations between them defined, we consider an evolutionary framework to study the evolutionary dynamics of this regulatory system. We assume mutation rates to be low enough such that a beneficial mutation fixes before an additional mutation (beneficial or not) arises. The condition under which this assumption is valid was found by Desai and Fisher~\cite{desai_beneficial_2007} and reads $\frac{\log(4N\Delta F)}{\Delta F}\ll \frac{1}{4N\mu_b \Delta F}$. $\Delta F$ is the fitness advantage of the beneficial mutant, $N$ is the population size and $\mu_b$ is the rate of beneficial mutations.

Under this condition the population is almost always fixed (monomorphic), and its evolutionary trajectory is captured by a series of discrete transitions between different genotypes. Consequently, when a new mutation emerges, it competes with only one other genotype.
The fixation probability of a new mutation that alters the genotype from $y$ to $x$ equals

\be
\label{eq:P_fix}
\Phi_{y\rightarrow x} = \frac{1 - \exp(-(F(x)-F(y)))}{1 - \exp(-2N(F(x)-F(y)))},
\ee
where the fitness $F$ is defined by \eqref{eq:F} given the frequencies of the various environments $\alpha_m$ and the desired expression pattern of the genes $p^*_{jm}$ at each.
\eqref{eq:P_fix} applies to a diploid population in which the mutant $x$ appears in a single copy over a uniform background of the other genotype $y$. For diploids, the fitness difference $\Delta F=F(x)-F(y)$ refers to the fitness difference between the two homozygotes or to twice the selective advantage of the heterozygote (one copy of the mutant) over the prevailing homozygote genotype \cite{gillespie_population_2004}. The overall rate of substitution from genotype $y$ to $x$ is given by~\cite{lassig_biophysics_2007}:

\be
\label{eq:subs_rate}
r_{xy} = 2N\mu_{xy}\Phi_{y\rightarrow x},
\ee

where $\mu_{xy}$ denotes the mutation rate from genotype $y$ to $x$. We illustrate the evolutionary model further in \secref{sec:Methods}.

\subsection{Space of reduced-genotypes}

The size of the genotype space is huge, $|\mathcal{D}| = 4^{4L+2}\approx 10^{13.25}$ for $L=5$, which makes it hard to analytically track the evolutionary model. Since the fitnesses of genotypes depend only on the mismatches $k_{ij}$ and the signal sensing alleles $\sigma_i$, and the mutations only alter $k_{ij},\sigma_i$ and the TF consensus sequences' match $M$, we consider the space of "reduced-genotypes", $\mathcal{G}=\{M,k_{ij},\sigma_i\}$, keeping track of only these reduced features of the genotype. The size of the reduced-genotype space is $|\mathcal{G}|<16(L+1)^5\approx10^{5.09}$ for $L=5$, which is tractable. Hence, for analytical calculations, we treat the regulatory network in the reduced-genotype space $\mathcal{G}$, and for simulations, we treat the regulatory network in the full genotypic space $\mathcal{D}$. Note that the reduced genotype representation in our model framework is \emph{not} an approximation, but is an exact solution of the full genotype model, with the tractability gained due to clever bookkeeping of states in the sequence space.

\subsection{Classification of genotypes into ``macrostates''}
\label{sec:MacroState}
Since our interest is in the biological function implemented by the network, we further coarse-grain the space of reduced-genotypes $\mathcal{G}$, and classify these reduced-genotypes into six possible macro-states, $\mathcal{M}=\{\texttt{No Regulation},$
$\texttt{Initial},$ $\texttt{One TF Lost},$ $\texttt{Specialize Both},$
$\texttt{Specialize Binding},$ $\texttt{Partial}\}$, by distinguishing only between "strong" and "weak" interactions. 
We set a threshold $k_T$ and consider an interaction as weak, $k_{ij}\in \mathcal{W}$, if $k_{ij}>k_{T}$, and strong, $k_{ij}\in \mathcal{S}$, if $k_{ij}\le k_{T}$.
In the basic version of the model where both TFs have same biophysical properties (in particular same $L$) $k_T$ is the same for all TF-BS interactions (but see the extension in Section \ref{sec:Promiscuity}). The threshold $k_T$ for each TF-BS pair $ij$ is set such that for mismatches $k<k_T$, $p_{jm_{i}}\ge0.5$ and for $k>k_T$, $p_{jm_{i}}<0.5$ when only TF $i$ is present and other TF(s) are absent, $C_{i}(m_{i})=C_{0}$.

Tje full genotypic space $\mathcal{D}$ is a union of sequences belonging to different macrostates $z$:

\be \mathcal{D}=\bigcup\limits_{z \in \mathcal{M}}S_z, \ee

where $S_{z}$ is the set of all genotypes that belong to macrostate $z$. We apply the following classification rules.

\subsection*{$\texttt{No Regulation}$}

The $\texttt{No Regulation}$ macrostate
consists of all genotypes in which there is no regulation of any form
(no information transmitted from the signals to genes). This can happen
if both the TFs either do not sense any signal or do not bind well to
any binding sites.

\be x \in S_{\texttt{No Regulation}} \text{ if } \forall i \: \Big((\forall j \: k_{ij}\in\mathcal{W}) \text{ OR } (\sigma_i=00)\Big) \ee

\begin{figure}[H]
\centering{\includegraphics[scale=1]{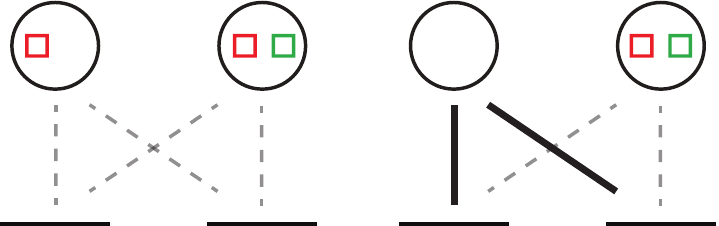}}
\caption{\textbf{Typical genotypes in $\texttt{No Regulation}$ macrostate.}
In the left genotype, even though both TFs sense some
signals, they do not bind well to either of the binding sites, hence
preventing any information transmission. In the right genotype one TF binds both the binding sites but does not sense any signal and the second TF
does not bind any binding site even though it senses both signals. This way or the other no information is transmitted between the signals and the genes.}
\end{figure}

\subsection*{$\texttt{Initial}$}

The $\texttt{Initial}$ macrostate consists of all genotypes in which
there is complete regulation with no form of specificity: both the
TFs sense both signals and bind both binding sites. This is the
typical initial state right after duplication.

\be x \in S_{\texttt{Initial}} \text{ if } \forall i \: \Big((\forall j \: k_{ij}\in\mathcal{S}) \text{ AND } (\sigma_i=11)\Big) \ee

\begin{figure}[H]
\centering{\includegraphics[scale=1]{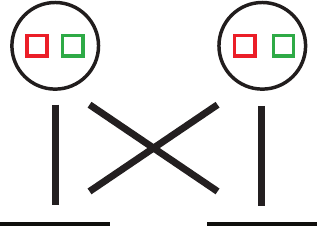}}
\caption{\textbf{ $\texttt{Initial}$ macrostate genotypes.} In
these genotypes, both TFs sense both signals and bind both binding
sites. 
}
\end{figure}

\subsection*{$\texttt{One TF Lost}$}

The $\texttt{One TF Lost}$ macrostate consists of all genotypes in
which one of the TFs is not involved in any regulation while the other
is involved in some regulatory activity (namely, one TF does not sense
any signal or does not bind well to any of the binding sites). This is equivalent to the
genotypes before duplication, except that there is a ``lost TF''.

\be x \in S_{\texttt{One TF Lost}} \text{ if } \Big|i: \: \Big((\forall j \: k_{ij}\in\mathcal{W}) \text{ OR } (\sigma_i=00)\Big)\Big|=1 \ee

\begin{figure}[H]
\centering{\includegraphics[scale=1]{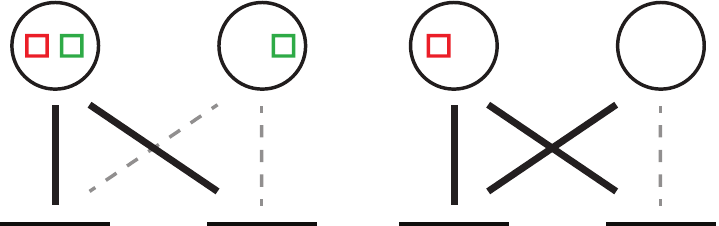}}

\caption{\textbf{Typical genotypes in  $\texttt{One TF Lost}$ macrostate.}
In the left genotype, only the first TF is involved in
regulation as it senses both signals and binds to both binding sites.
The second TF senses the green signal but does not bind any of the binding
sites, hence it is not involved in regulation and is ``lost''. In
the right genotype, again only the first TF is involved in regulation
as it senses the red signal and binds both binding sites. The second TF not involved in any regulation because it does not sense
any signal, although it binds the first binding site.}
\end{figure}

\subsection*{$\texttt{Specialize Both}$}

The $\texttt{Specialize Both}$ macrostate consists of all genotypes
in which there is correct specialization of TFs with respect to both
signal sensing and binding sites specificity. In these genotypes, one
TF senses only the first signal and binds only to the first binding
site, while the other TF senses only the second signal and binds only
to the second binding site.

\begin{align}
 & x\in S_{\texttt{Specialize Both}}\text{ if }\nonumber \\
 & (k_{11},k_{22}\in\mathcal{S}\text{ AND }k_{12},k_{21}\in\mathcal{W}\text{ AND }\sigma_{1}=10\text{ AND }\sigma_{2}=01)\nonumber \\
\text{ OR } & (k_{12},k_{21}\in\mathcal{S}\text{ AND }k_{11},k_{22}\in\mathcal{W}\text{ AND }\sigma_{1}=01\text{ AND }\sigma_{2}=10)
\end{align}

\begin{figure}[H]
\centering{\includegraphics[scale=1]{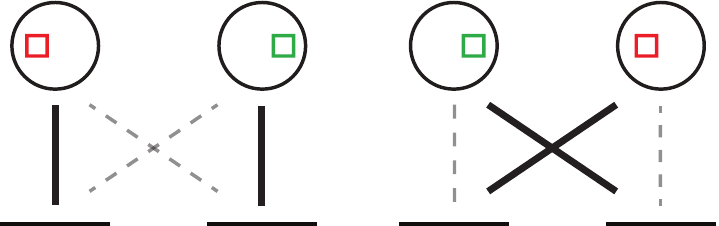}}

\caption{\textbf{ Genotypes in $\texttt{Specialize Both}$ macrostate.}
Both genotypes have specific paths from the signals to the genes.
In the left genotype, while the first TF senses the red signal and
binds the first (correct) binding site, the second TF senses the
green signal and binds the second (correct) binding site. Hence,
the first TF mediates the red signal to first gene pathway while the
second TF mediates the green signal to second gene pathway. In the
right genotype, the TFs exchange roles. The first TF mediates the
green signal to second gene pathway while the second TF mediates the
red signal to first gene pathway.}
\end{figure}

\subsection*{$\texttt{Specialize Binding}$}

In contrast, the $\texttt{Specialize Binding}$ macrostate consists
of all genotypes in which there is specialization of TFs with respect
to binding site specificities, but not with respect to the signal sensing domains.

\begin{align}
 & x\in S_{\texttt{Specialize Binding}}\text{ if }(\forall i\:\sigma_{i}\ne00)\:\text{AND}\nonumber \\
 & \Bigg(\text{\ensuremath{\Big(}}(k_{11},k_{22}\in\mathcal{S}\text{ AND }k_{12},k_{21}\in\mathcal{W})\text{\:AND\:}\neg(\sigma_{1}=10\text{ AND }\sigma_{2}=01)\Big)\nonumber \\
\text{OR\:} & \Big(\text{(\ensuremath{k_{12}}, \ensuremath{k_{21}\in\mathcal{S}\text{ AND }k_{11}},\ensuremath{k_{22}\in\mathcal{W}})\:AND\:}\neg(\sigma_{1}=01\text{ AND }\sigma_{2}=10)\Big)\Bigg)
\end{align}

\begin{figure}[H]
\centering{\includegraphics[scale=1]{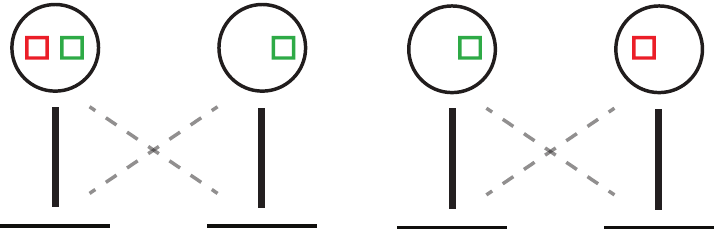}}

\caption{\textbf{Typical genotypes in $\texttt{Specialize Binding}$ macrostate.}
In both genotypes, the first TF binds
the first binding site and the second TF binds the second binding
site, but they have not correctly specialized in their signal sensing domains.
In the left genotype, while the second TF has specialized correctly
to sense only the green signal, the first TF still senses both the
signals. Hence, while the red signal pathway is established properly,
the green signal pathway is not - both genes are activated in the presence
of green signal. In the right genotype, the TFs have specialized in signal sensitivities, but opposite to the desired response pattern.
}
\end{figure}

\subsection*{$\texttt{Partial}$}

The $\texttt{Partial}$ macrostate consists of all genotypes which
do not belong in any of the other macrostates mentioned above. It contains a mixture of different regulatory architectures: both TFs regulate only one gene with
the other gene unregulated, one TF regulates both genes while the other
TF regulates only one gene or both TFs bind both binding sites but
at least one TF has specialized in signal sensing.

\begin{figure}[H]
\centering{\includegraphics[scale=1]{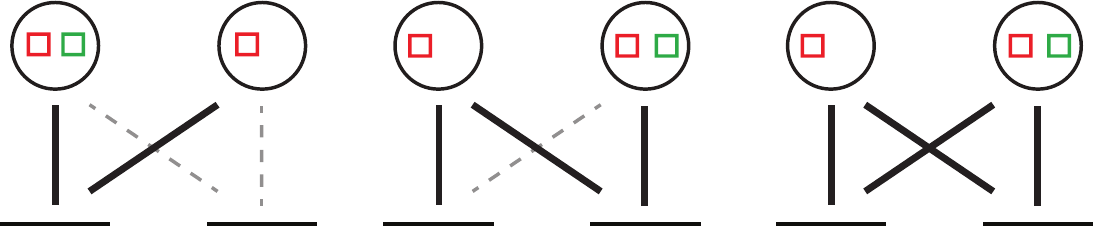}}

\caption{\textbf{ Typical genotypes in $\texttt{Partial}$ macrostate.} In the left genotype, both TFs regulate only the first
gene while the second gene is unregulated. In the middle genotype,
the first TF regulates both genes while the second TF regulates
only the second gene. In the right genotype, both TFs regulate both
genes but, unlike the $\texttt{Initial}$ macrostate, here the first TF does not mediate any information
from the green signal. }
\end{figure}

\subsubsection*{Role of $L$ in macrostate classification}

Keeping $\epsilon$ and $C_{0}$ constant while changing $L$ keeps
the threshold mismatch $k_{T}$ constant. Hence, the number of mismatches
$|\mathcal{S}|$ in the strong binding class remains the same while
the number of mismatches $|\mathcal{W}|$ in the weak binding class
increases. Hence, as $L$ increases, the number of genotypes in
all macrostates except $\texttt{{Initial}}$ increase. The volume of macrostates
with a larger number of weak mismatches increase more than the volume of macrostates
with a smaller number of weak mismatches. For instance, $\texttt{No Regulation}$
increases more than $\texttt{Specialize Binding}$. As $\texttt{One TF Lost}$
and $\texttt{Specialize Binding}$ have the same number of weak mismatches,
the ratio of the number of genotypes in them stays the same for different
$L$.

\section{Methods}
\label{sec:Methods}
\subsection{Markov chain formulation}

As explained in Section \ref{sec:Model}, we assume that the time between the emergence and fixation of a beneficial mutation is much shorter than the time until the emergence of the next beneficial mutation. Hence, by neglecting the times between emergence and fixation (or loss) of mutations the population can be captured at any time by a single genotype.
This so-called ``fixed state assumption''
lets us describe the state of the population as a probability distribution
over the possible genotypes, $P(\mathcal{D},t)$ or as a probability distribution over the possible reduced-genotypes, $P(\mathcal{G},t)$. This can be obtained
via a continuous-time discrete-space Markov chain defined over the genotype space $\mathcal{D}$ or the reduced-genotype space $\mathcal{G}=\{M,k_{ij},\sigma_{i}\}$.
The transition rate between $y$ and $x$, where either $x,y\in\mathcal{D}$ are genotypes, or $x,y\in\mathcal{G}$ are reduced-genotypes, is the rate of substitution~\cite{lassig_biophysics_2007}:
\begin{equation}
r_{xy}=2N\mu_{xy}\Phi_{y\rightarrow x}
\end{equation}
where $N$ is the population size, $\mu_{xy}$ is the mutation
rate from (reduced-) genotype $y$ to (reduced-) genotype $x$, and $\Phi_{y\rightarrow x}$ is the
probability of fixation of a single copy of $x$ in a population of $y$ (\eqref{eq:P_fix}). As the probability of fixation $\Phi_{y\rightarrow x}$ depends on $x$ and $y$ only via their fitness values $F(x)$ and $F(y)$, and $\mu_{xy}$ can be obtained analytically for reduced-genotypes, it is sufficient to consider the Markov chain on the space of reduced-genotypes $\mathcal{G}=\{M,k_{ij},\sigma_{i}\}$ rather than on the whole genotype space $\mathcal{D}$.
Each reduced-genotype $x=(M,k_{ij},\sigma_{i})$ can be realized by multiple genotypes (DNA sequences), whose number is given by $N_{\mbox{seq}}(k_{ij}|M)$ 
(\eqref{eq:Nseq} and \eqref{eq:P0}) below. Now, the
evolution of the probability distribution $P(\mathcal{G},t)$ is captured by

\begin{equation}
\label{eq:P_dyn}
\frac{\partial P(\mathcal{G},t)}{\partial t}=\mathbf{R}P(\mathcal{G},t),
\end{equation}
where $\mathbf{{R}}$ is the transition rate matrix of the underlying
Markov chain where each entry $r_{xy}$ denotes the rate of transition from $y$ to $x$.

\subsection{Steady state after duplication}

The probability distribution at steady state, $P_{SS}(\mathcal{G})=P(\mathcal{G},t\rightarrow\infty)$,
is the non-trivial solution of $\mathbf{R}P_{SS}(\mathcal{G})=0$.
It is also possible to obtain $P_{SS}(\mathcal{G})$ by invoking the set of detailed balance conditions, $r_{xy}P_{SS}(y)=r_{yx}P_{SS}(x)$, $\forall x,y$. This results
in an elegant expression

\begin{equation}
P_{SS}(\mathcal{G})=P_{0}(\mathcal{G})\exp(2NF(\mathcal{G}))\label{eq:SSdist},
\end{equation}

where $P_{0}$ is the neutral distribution of reduced-genotypes and $N$ is
the population size.

To calculate the neutral distribution $P_0$ of the reduced-genotypes, we begin by enumerating the number of possible BS sequences $j$ that have mismatch values ($k_{1j}$,$k_{2j}$) with respect to two TFs that match each other at $M$ out of $L$ consensus positions. This number equals:

\be
\label{eq:Nseq}
\begin{split}
&N_{\mbox{seq}}(k_1,k_2|M) = \sum_{j_0=j_0^{\mbox{\footnotesize{min}}}}^{j_0^{\mbox{\footnotesize{max}}}}
{M \choose j_0} 3^{M - j_0} {L-M \choose L-j_0-k_1} {j_0+k_1-M \choose L-j_0-k_2}2^{k_1+k_2+2j_0-L-M}\\
&j_0^{\mbox{\footnotesize{min}}}=\mbox{max}(\mbox{max}(0, M-\mbox{min}(k_1,k_2)),\lceil\frac{L+M-k_1-k_2}{2}\rceil )\\
&j_0^{\mbox{\footnotesize{max}}}=\mbox{min}(M,L-\mbox{max}(k_1,k_2))
\end{split}
\ee

\noindent where for brevity we write $k_1, k_2$ instead of $k_{1j}, k_{2j}$, and $\lceil x \rceil$ is the ceiling function, which maps $x$ onto the nearest integer larger than or equal to $x$. Now, the neutral distribution is (up to proportionality constant)
 \be
 \label{eq:P0}
 P_0(x) \sim N_{\mbox{seq}}(k_{11},k_{21}|M)N_{\mbox{seq}}(k_{12},k_{22}|M)\binom{L}{M}3^{L-M}.
 \ee

From Eq.~(\ref{eq:SSdist}) we obtain the steady state distribution
over the macrostate space. 
For every macrostate $z\in\mathcal{M}$ the probability to be in this macrostate at steady state equals the sum of probabilities of being in all reduced-genotypes $x$ that are assigned to that macrostate

\begin{equation}
Q_{SS}(z)={\displaystyle \sum_{x\in S_{z}}P_{SS}(x)}.
\end{equation}

\subsubsection*{Dominant macrostate}

We denote the the most probable macrostate at steady state by

\begin{equation}
z_{SS}^{*}:=\argmax_{z\in\mathcal{M}}\:Q_{SS}(z).
\end{equation}


\subsection{Evolutionary dynamics}

We obtain the evolutionary dynamics of $P(\mathcal{G},t)$ 
in units of generation time $t_g$ by numerically integrating the Markov chain in time-steps corresponding to one generation:

\begin{equation}
P(\mathcal{G},t+t_{g})=(\mathbf{I}+\mathbf{R}t_{g})P(\mathcal{G},t).
\end{equation}

We define $\mathbf{A}=\mathbf{I}+\mathbf{R}t_{g}$
as the transition probability matrix in this time-unit. From $P(\mathcal{G},t)$, we obtain the macrostate dynamics (\eqref{eq:P_dyn})
$Q(\mathcal{M},t)$. For every $z\in\mathcal{M}$,

\begin{equation}
Q(z,t)={\displaystyle \sum_{x\in S_{z}}P(x,t)}.
\end{equation}

\subsubsection*{Dominant macrostate}

To follow the macrostate dynamics in a more compact way, we refer to the most probable macrostate at each time-point $t$

\begin{equation}
z^{*}(t):=\argmax_{z\in\mathcal{M}}\:Q(z,t)
\end{equation}
as the dominant macrostate at that time.

\subsubsection*{Time to reach a particular macrostate}

We compute the mean first hitting time, $T_{S\leftarrow x}$,
to any subset of reduced-genotypes, $S$, from any other reduced-genotype $x$, by using the following recursive equation.

\begin{equation}
T_{S\leftarrow x}=t_{g}+{\displaystyle \sum_{y}a_{yx}T_{S\leftarrow y}},
\end{equation}

where $a_{yx}$ are elements of the transition probability matrix
$\mathbf{A}$. We consider subsets $S_{z}$ of genotypes that belong
to a particular macrostate $z$, and compute the mean first hitting
times, $T_{S_{z}\leftarrow x}$, to this macrostate. In particular,
we compute the mean first hitting times to $\texttt{Specialize Both}$, which we refer to as the ``time to specialization'', $\tau(x)$.

\subsubsection*{Dwell times}

For every macrostate $z$, we also compute the dwell time, $t^{dwell}(z)$,
which is the mean time to ``escape'' from that macrostate into any other macrostate $z'$. For every genotype $x$ in $S_{z}$, the mean time
to escape from $S_{z}$ is by definition  $T_{S_{z}'\leftarrow x}$,
the mean time taken to hit $S_{z}'=\mathcal{G}-S_{z}$, the complementary set of $S_z$.
We define the dwell time in macrostate $z$ as

\begin{equation}
t^{dwell}(z):=\langle T_{S_{z}'\leftarrow x}\rangle_{x\in S_{z}}
\end{equation}

\subsection{Stochastic simulations}
In addition to analytical solutions of the Markov chain formulation we also used stochastic simulations of TF and BS evolution to validate our analytical solution and also to test additional cases that were not analytically solvable, such as the case where each TF post-duplication regulates multiple genes.

\subsubsection{Gillespie Simulation - main model}

We use the Gillespie Stochastic Simulation Algorithm \cite{gillespie_general_1976} to track the evolutionary trajectories of the system.
Since we employ the fixed-state assumption, the time to fixation of each mutation is small compared
to the waiting time between mutations and we neglect it in the calculations. At each simulation run we obtain a temporal series, $s_{0},s_{1},s_{2},\dots$,
of genotypes (DNA sequences of TF consensus sequence and binding
sites, along with signal sensing alleles), and a corresponding sequence of times, $t_{0}=0,t_{1},t_{2},\dots$,
at which substitutions between consecutive genotypes occurred. Here, $s_{0}$ is the initial
DNA sequence with which we start the simulation. We construct $s_0$ by sampling a genotype from the steady state before duplication (with only 1 TF). For every $i$, from $t_{i}$ to $t_{i+1}$, the DNA sequence of the
system is $s_{i}$, from which there is a substitution event to $s_{i+1}$
at $t_{i+1}$. We obtain $s_{i+1}$ by appropriately sampling substitutions available from $s_{i}$, which can occur via TF consensus sequence mutations, or TF sensing domain mutations, or BS sequence mutations. We also draw $t_{i+1}-t_{i}$ (the waiting
time) from the appropriate exponential distribution in the Gillespie framework.
For each DNA sequence $s_i$, one can obtain the reduced-representation $(M,k_{ij},\sigma_{i})$.
From this, we obtain, for each simulation run $r$, the time trajectories of reduced-genotypes, $x_{r}(t)$, starting from $x_{r}(t=0)=x_{r0}$.
By running multiple times and computing the fractions of runs with
each reduced-genotype $x$ at each $t$, we obtain the dynamical trajectory of the probability distribution of reduced-genotypes, $P^{sim}(\mathcal{G},t)$,
and the steady state distribution, $P_{SS}^{sim}(\mathcal{G})$. Grouping the reduced-genotypes into macrostates, we also obtain the dynamical trajectory
of the probability distribution of macrostates, $Q^{sim}(\mathcal{M},t)$ and steady state distribution of macrostates, $Q_{SS}^{sim}(\mathcal{M})$.

The simulations enable us to compute non-trivial \textit{path-dependent} quantities relating to
an ensemble of trajectories $\{x_{r}(t)\}$, as well as to provide full distributions of quantities of interest. One such example is
the mean hitting time to some macrostate $z$, conditioned on not
hitting some other particular macrostate on the way. While it is possible in principle to compute such a path-dependent quantity exactly, in practice this requires too much numerical effort and Gillespie simulation becomes be the method of choice.

\subsubsection*{Time to specialization, dependent on pathway}

As explained in the main text, for a single trajectory (population),
there are two main paths from $\texttt{Initial}$ to $\texttt{Specialize Both}$,
each with a different dominant ``transient state''. One pathway
is fast and predominantly goes via genotypes in $\texttt{Partial}$ macrostate,
and the other is slow and predominantly via genotypes in $\texttt{One TF Lost}$ macrostate.
In each simulation run $r$, we calculate the time to specialization,
and also record the dominant transient state. By running many simulations,
we have a set of times to specialization that go via the fast pathway
of $\texttt{Partial}$ $\{\tau_{fast}\}$, and those via the slow
pathway of $\texttt{One TF Lost}$ $\{\tau_{slow}\}$. Using these,
we obtain the empirical distributions of $\tau_{slow}$ and $\tau_{fast}$, their means
($\bar{\tau}_{slow}=\langle\tau_{slow}\rangle$
and $\bar{\tau}_{fast}=\langle\tau_{fast}\rangle$); we also record the fraction of pathways proceeding via the slow and fast alternatives.

\subsubsection{Alternative model - fixed signal sensing domain}
\label{subsec:altSimul}
In the second model variant, as mentioned in \secref{sec:Model}, TFs are not equipped with an evolvable signal sensing domain $\sigma_i$. The active concentrations of the TFs, $C_i(m)$, in different environments $m$, are explicitly defined separately. In the stochastic simulation of this model variant, we therefore considered mutations only in the TF consensus sequence and the BS sequences. We also assumed a timescale separation, such that the TF consensus sequences evolve on a slower timescale compared to the BS sequences. We implement this by performing alternating rounds of one TF consensus sequence mutation and $50$ BS sequence mutations, resulting in $r_{TF}=0.02$. These two rounds together are considered a single time step of the simulation, which amounts to counting the number of TF consensus sequence mutations that have arisen.

As in the Gillespie simulation, we choose the starting point by sampling from the steady state before duplication with only 1 TF. The duplicate TF has the same binding preferences as the original TF but has different expression pattern $C_2(m)$ than the first $C_1(m)$. In each round, we calculate the fixation probability of the mutant using \eqref{eq:P_fix}, and compare a randomly drawn number between $0$ and $1$ to either fix them or not.

\begin{figure}[H]
    \centering
   \subfigure[]{
      \includegraphics[width=0.3\textwidth]{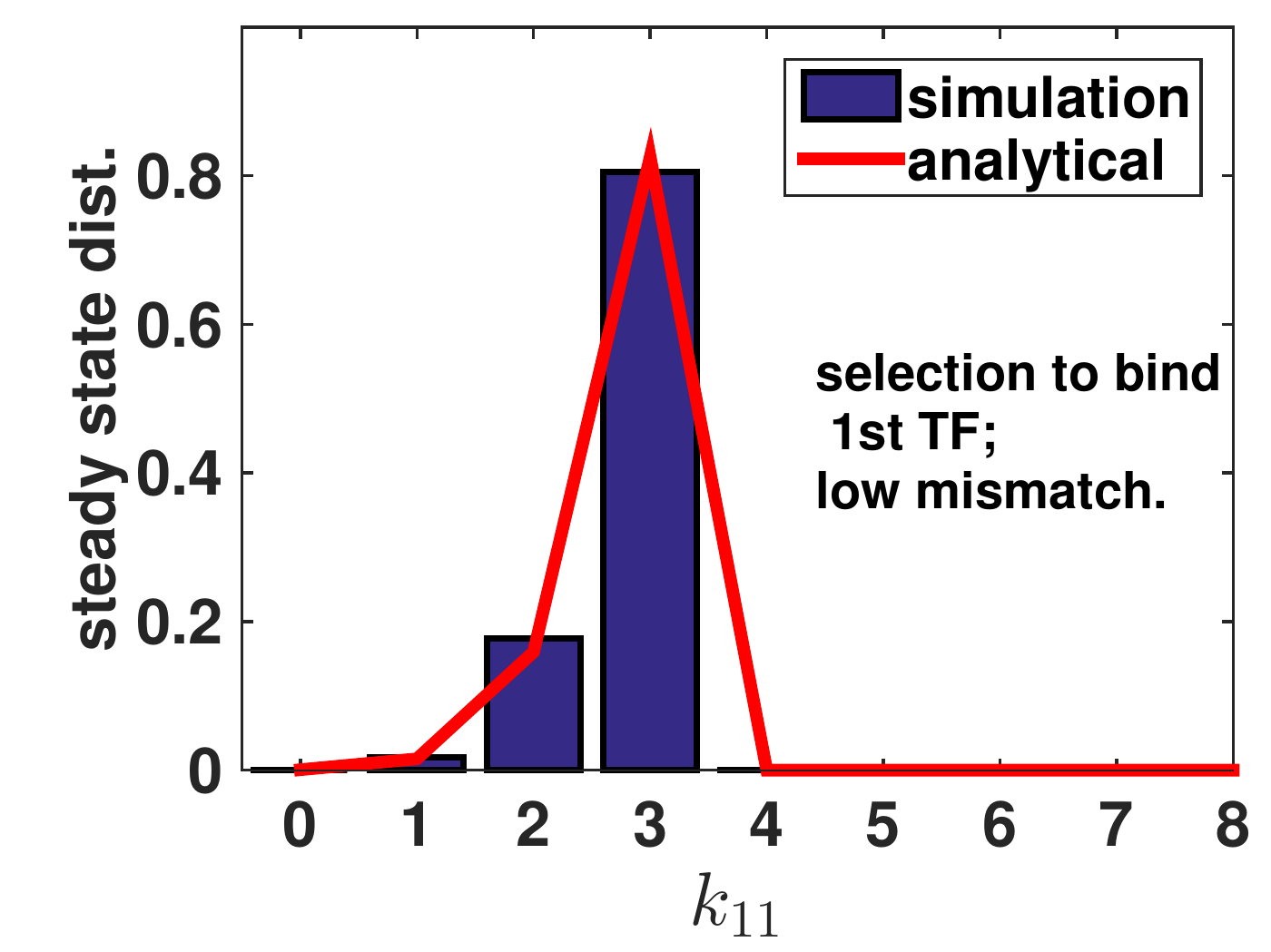}
        }
       \subfigure[]
{
        \includegraphics[width=0.3\textwidth]{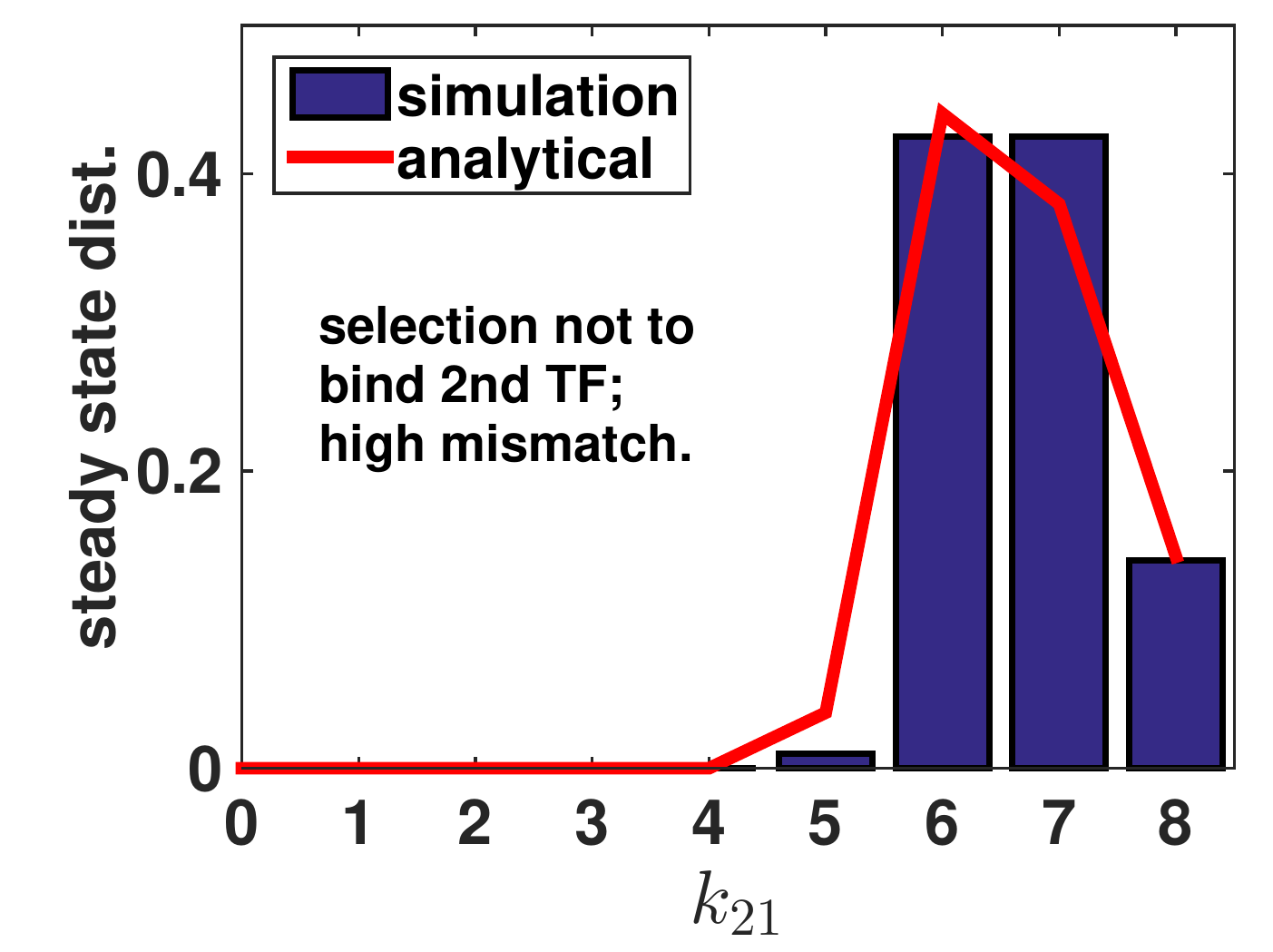}
    }
           \subfigure[]
{
        \includegraphics[width=0.3\textwidth]{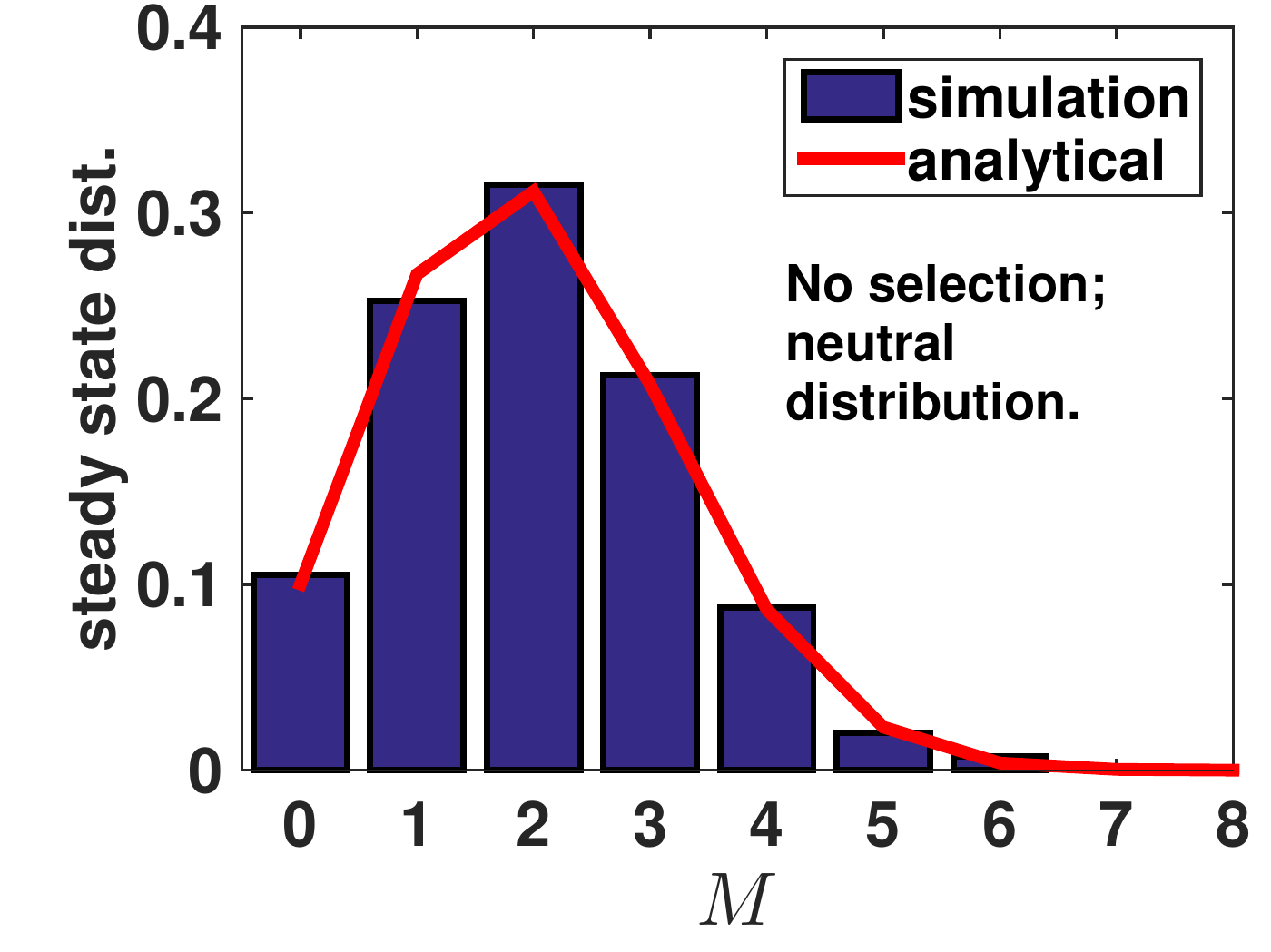}
    }
    \captionof{figure}
    {\label{fig:simu_validation} \textbf{Comparison between stochastic simulation and exact results.} Blue bars represent the statistics over 400 independent runs of the stochastic simulation. Red curve represents the analytical solution for the steady state distribution. We illustrate distributions of $k_{11}$ mismatch between first TF and first gene (a) where selection for the regulation of this gene incurs low mismatch; mismatch between second gene and TF $k_{21}$ where selection here results in high mismatch, such that this gene is NOT regulated by this TF. (c) illustrates the match distribution $M$ between the two TFs in the absence of selection, so that the Bernoulli distribution is obtained. Parameters: $L=8$,  $\epsilon=3$, $C_0=3.269\times10^5$.}
\end{figure}

In  \figref{fig:simu_validation} we compare between this stochastic simulation and the analytical solution for the steady state distributions of various mismatches the steady state distribution of the match between the two TFs.

\section{Steady state}
\label{sec:ss}
The steady state distribution (Eq. (3) in the main text) is a general result in Population Genetics, derived as a solution of the forward Kolmogorov Equation~\cite{gillespie_population_2004}. It is a product of two factors: the neutral distribution (entropy term) and the fitness weight of different genotypes (energy term).
The first factor, $P_0(\mathcal{G})$, is the neutral distribution (see below) which results from neutral processes only, such as mutation rates between different genotypes, assuming that all genotypes have equal fitness values.
If fitness values are unequal, the second factor, $\exp (2NF(\mathcal{G}))$, biases the probabilities of attaining different genotypes accordingly.
For a more comprehensive discussion and relation to statistical physics see Ref. ~\cite{sella_application_2005}.

\subsection{Distribution of $M$ for neutral and adaptive cases}
\begin{figure}[H]
    \centering
   \subfigure[]{
      \includegraphics[width=0.3\textwidth]{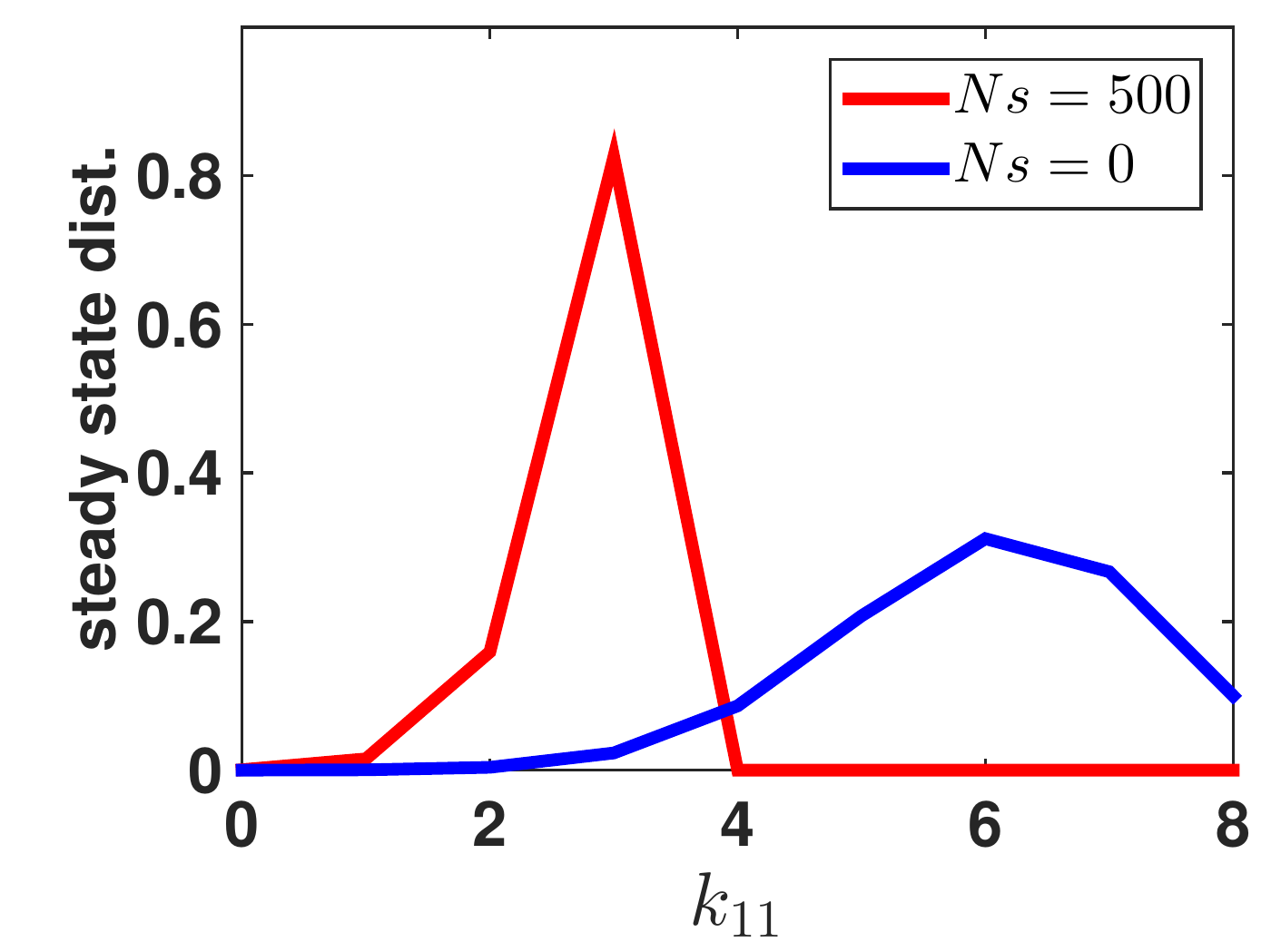}
        }
       \subfigure[]
{
        \includegraphics[width=0.3\textwidth]{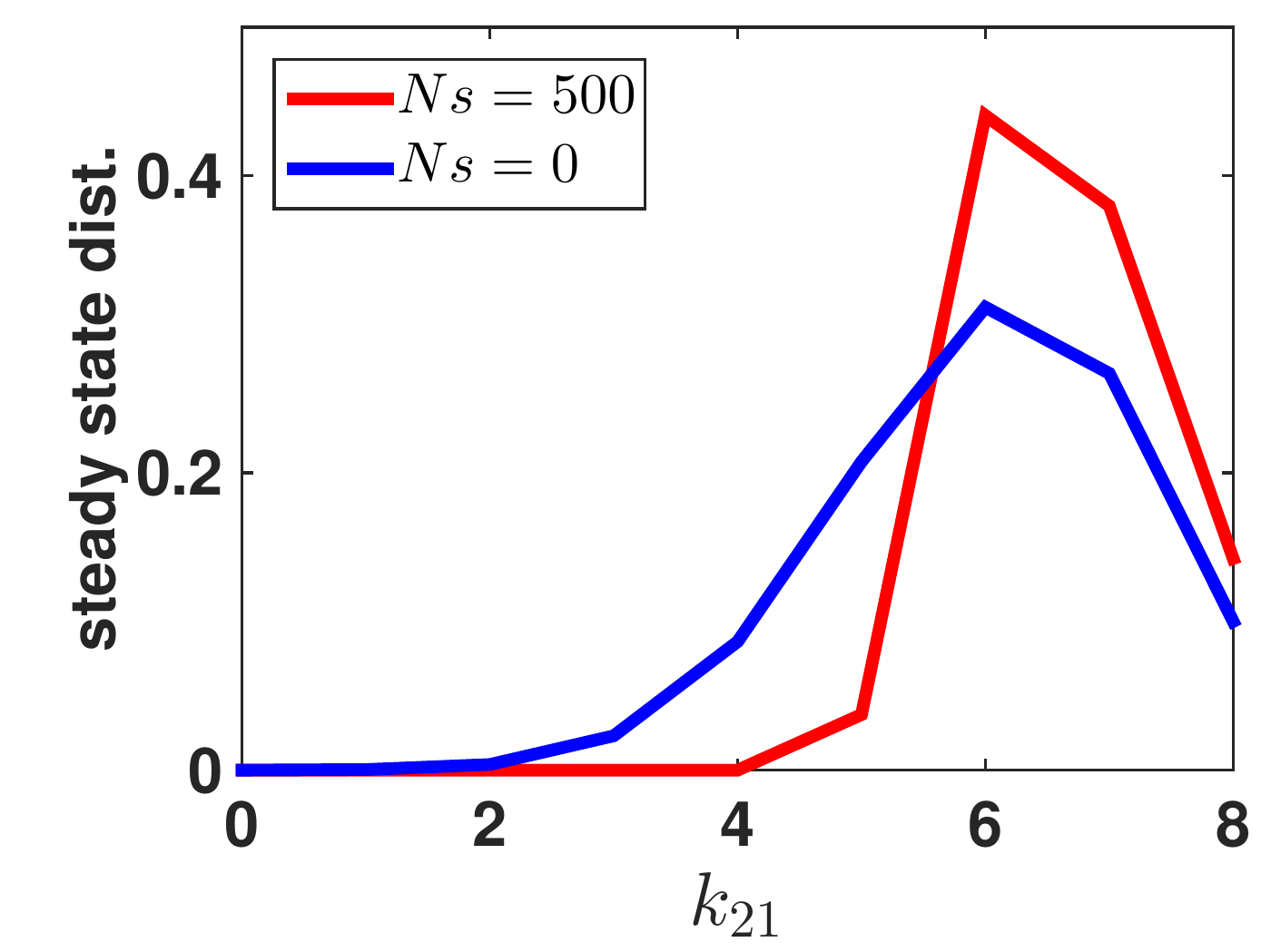}
    }
           \subfigure[]
{
        \includegraphics[width=0.3\textwidth]{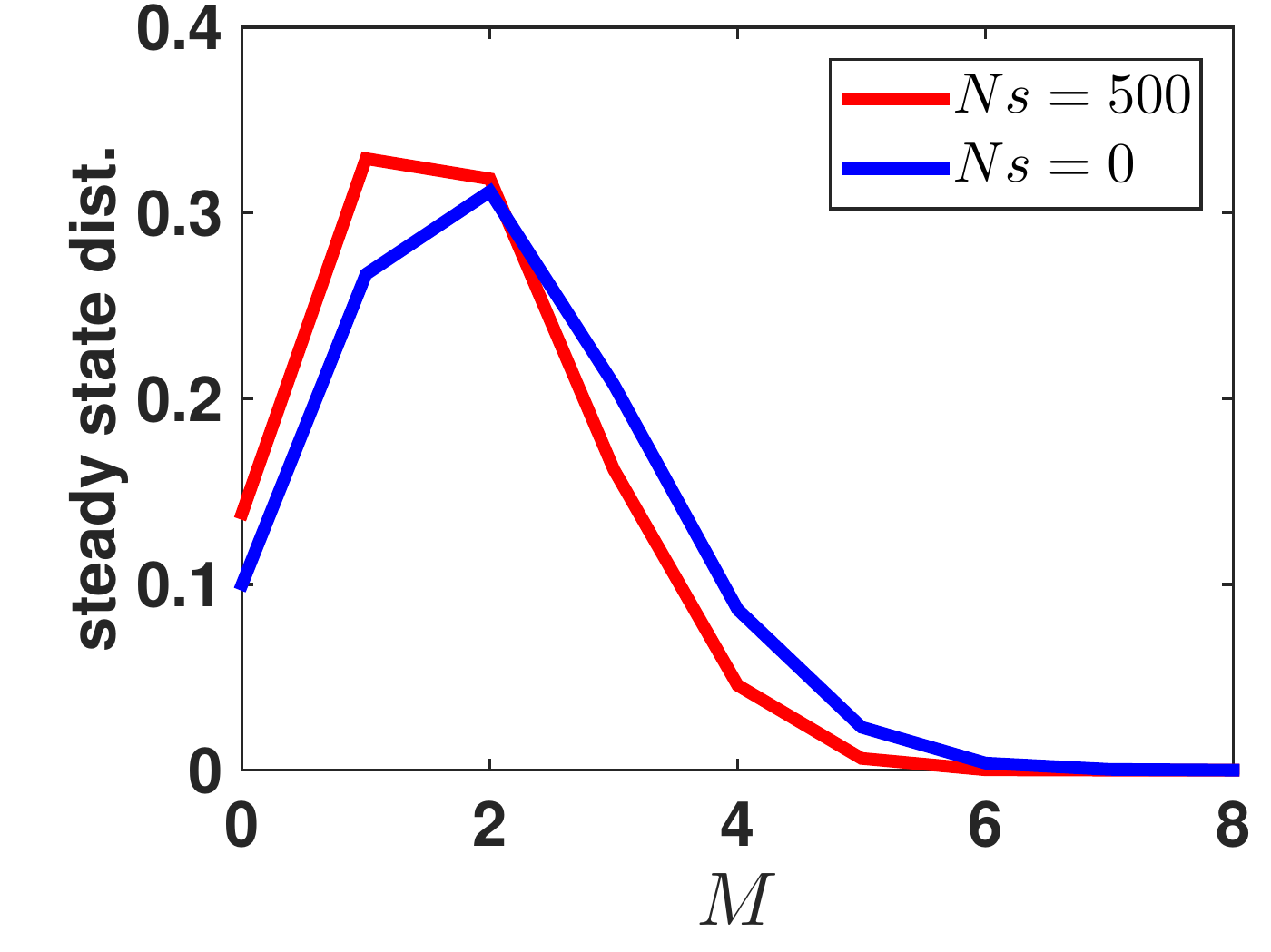}
    }
    \captionof{figure}
    {\label{fig:M_k_ss} \textbf{The steady state distributions of match, $M$, between TF consensus sequences when there is selection on the TFs to diverge is very similar to the neutral distribution. } We present analytically calculated steady state distribution of $k_{11}$ (a), $k_{21}$ (b) and $M$ (c) for $Ns=0$ (no selection, blue) and $Ns=500$ (strong selection, red).  The neutral distributions are always the Bernoulli distributions which here are peaked at $k=6$ and $M=2$. Selection to diverge biases the distribution to have a lower match than expected under neutrality. The difference between neutrality and selection becomes obvious only when looking at the distributions of $k$, the mismatches between BSs and TFs. Under neutrality the probability for match is low and the distribution is peaked around high mismatch values. When there is selection on the TF to remain functional it must preserve a low mismatch with at least one of the genes. Parameters: $L=8$,  $\epsilon=3$, $C_0=3.269\times10^5$.}
\end{figure}

In Fig. 3 of the main text  we compared the steady state distribution of $M$ (match between the two TFs) in the neutral case to the distribution of $M$  if selection to diverge applies. Parameters used were $L=5$, $Ns=25$, resulting in hardly distinguishable distributions. Here we repeat this calculation with different parameter values that emphasize the difference between these cases: a stronger selection $Ns=500$ and a longer binding site $L=8$. A stronger selection depletes the highest match values compared to the neutral (Bernoulli) distribution. Even under these more extreme values the difference between the two distributions is modest, as shown in \figref{fig:M_k_ss}. As a consequence, using distributions of $M$ as estimated from genomic data may provide insufficient statistical power to detect selection pressure on TFs to diverge.

\subsection{Probabilities of major macroscopic outcomes - losing a TF and specializing}
\begin{figure}[H]
\centering{\includegraphics{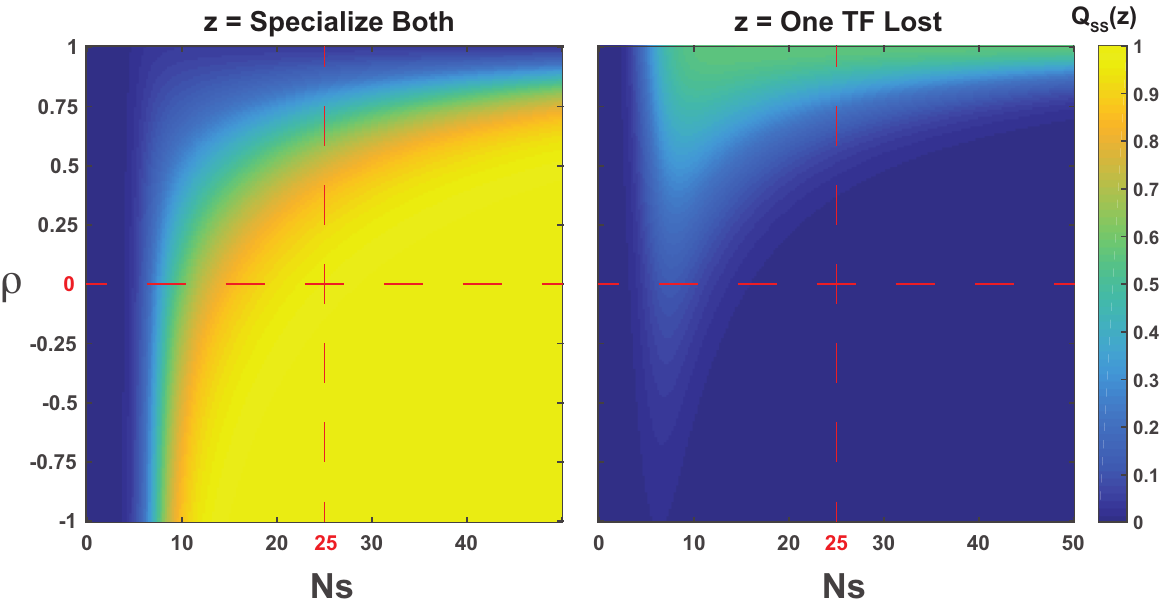}}

\caption{\textbf{Steady state probabilities of 'Specialize Both' (left) and 'One TF Lost' (right) macrostates for different values of selection intensity $Ns$ and correlation between the signals $\rho$.} The probability of either macrostate $Q_{SS}(z)$ is illustrated using a color code (blue = 0, yellow = 1). Intersection of the red dashed lines denotes the baseline parameters values.
\label{fig:spec_one}}
\end{figure}

In the main text we illustrate only the most probable macrostate for each parameter combination. Other macrostates are still possible, albeit with lower probability. Here we illustrate the probability to obtain either 'One TF Lost' or 'Specialize Both' macrostate at each parameter combination,
as described in Section \ref{sec:Methods}.  
As shown in Fig. \ref{fig:spec_one}, the
steady state probability of specialization $Q_{SS}(\texttt{Specialize Both})$
is high at large $Ns$ and intermediate $\rho$, and it decreases as selection strength decreases or signal correlation increases. The probability of having
$Q_{SS}(\texttt{One TF Lost})$ at steady state is significant only when selection
is not too weak and signals are highly correlated. Although for these parameter values it is the dominant macrostate its probability is only $\sim0.5$, such that other macrostates are not negligible. In contrast, for parameter values where 'Specialize Both' dominates its probability is close to 1.

\subsection{Asymmetric signal occurrence biases final outcomes}
\label{subsec:asymm}

\begin{figure}[H]
\centering{\includegraphics[scale=0.9]{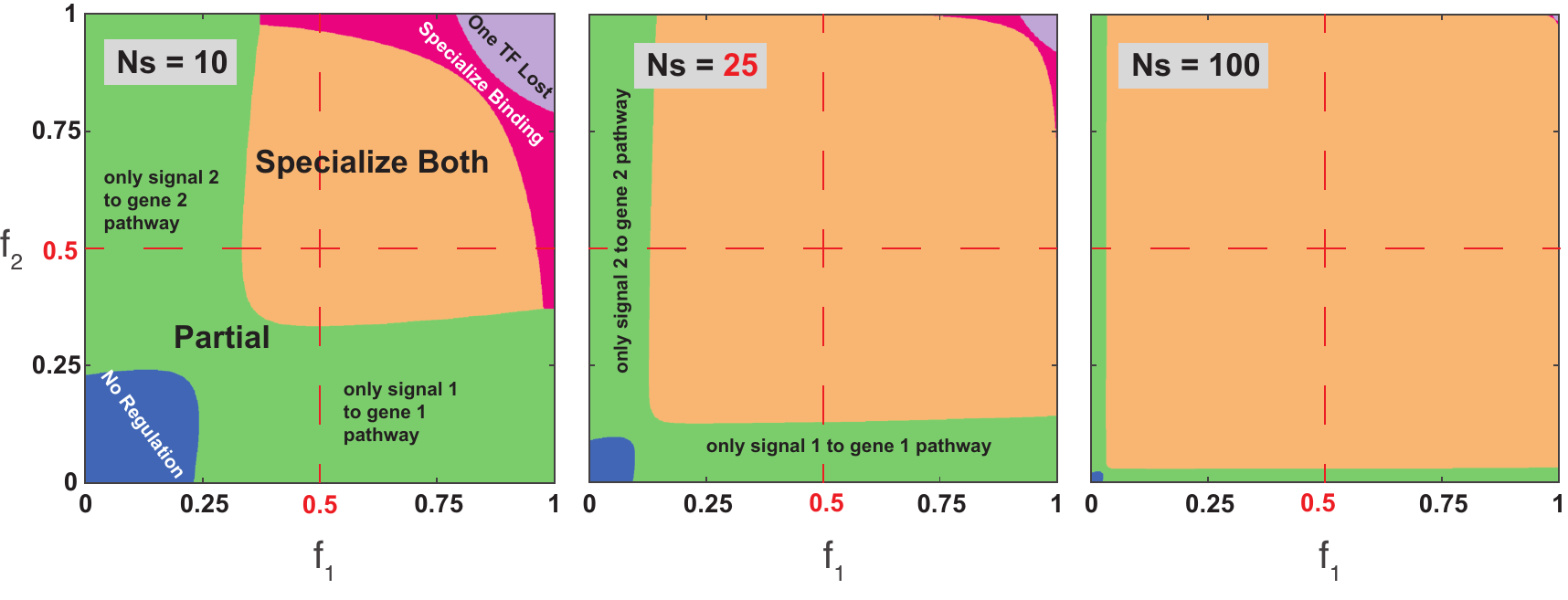}}

\caption{\textbf{Under medium to strong selection specialization occurs under a broad range of signal frequencies. Under weak selection specialization occurs only if signal frequencies are sufficiently high.} Phase plots of the most probable
macrostate at steady state as a function of signal frequencies
$f_{1}$ and $f_{2}$, at three different selection strengths $Ns=$ 10, 25, 100. The intersection between the red dashed lines, $f_1=f_2=0.5$, denotes the baseline parameters used anywhere else in this work.
\label{fig:asymm_envs}}
\end{figure}

In the main text, we assume symmetry between the occurrences of the two signals, namely their frequencies $f_1=f_2=0.5$, where $f_1=\alpha_{10}+\alpha_{11}$ is the frequency of the first signal, and  $f_2=\alpha_{01}+\alpha_{11}$ is the frequency of the second. Here we explore the effect of asymmetry in signal occurrence ($f_1 \neq f_2$) on the final evolutionary outcomes and in particular on the probability to fully specialize.
In \figref{fig:asymm_envs} we plot the most probable macrostate as a function of the signal frequencies $f_1$, $f_2$ for different values of selection intensities $Ns$. When both signals are rare, $f_1,f_2\ll 1$, \texttt{No Regulation} macrostate dominates, as selection on both pathways is weak. When one of the signals is frequent while
the other is rare, $f_1\gg f_2$, only the frequently used pathway is maintained, and the
dominant macrostate is $\texttt{Partial}$. Only when both signals are frequent and selection is not too weak, specialization occurs. Hence, a signal-gene pathway is maintained
only if it is required often enough, and the threshold for this (boundary
between $\texttt{Partial}$ and $\texttt{Specialize Both}$) depends on selection strength $Ns$.  As selection strength $Ns$ increases, this threshold moves to lower $f_1$ and $f_2$. As the frequencies of both signals increase, the dominant macrostate $\texttt{Specialize Both}$ is replaced by $\texttt{Specialize Binding}$, where sensing one signal is a good proxy for the other signal as well,
and later by $\texttt{One TF Lost}$ when one TF is sufficient to transduce both signals.

\section{Evolutionary dynamics}
\label{sec:dyn}
Evolutionary trajectories between the post-duplication unspecialized configuration ('Initial') to full specialization ('Specialize Both' macrostate) are multi-step processes that require several mutations and transiently pass through various macrostates. Here we describe the various trajectories for this functional transition.

\subsection{Evolutionary pathways - first model variant}
\begin{figure}[H]
\centering{\includegraphics[scale=0.6]{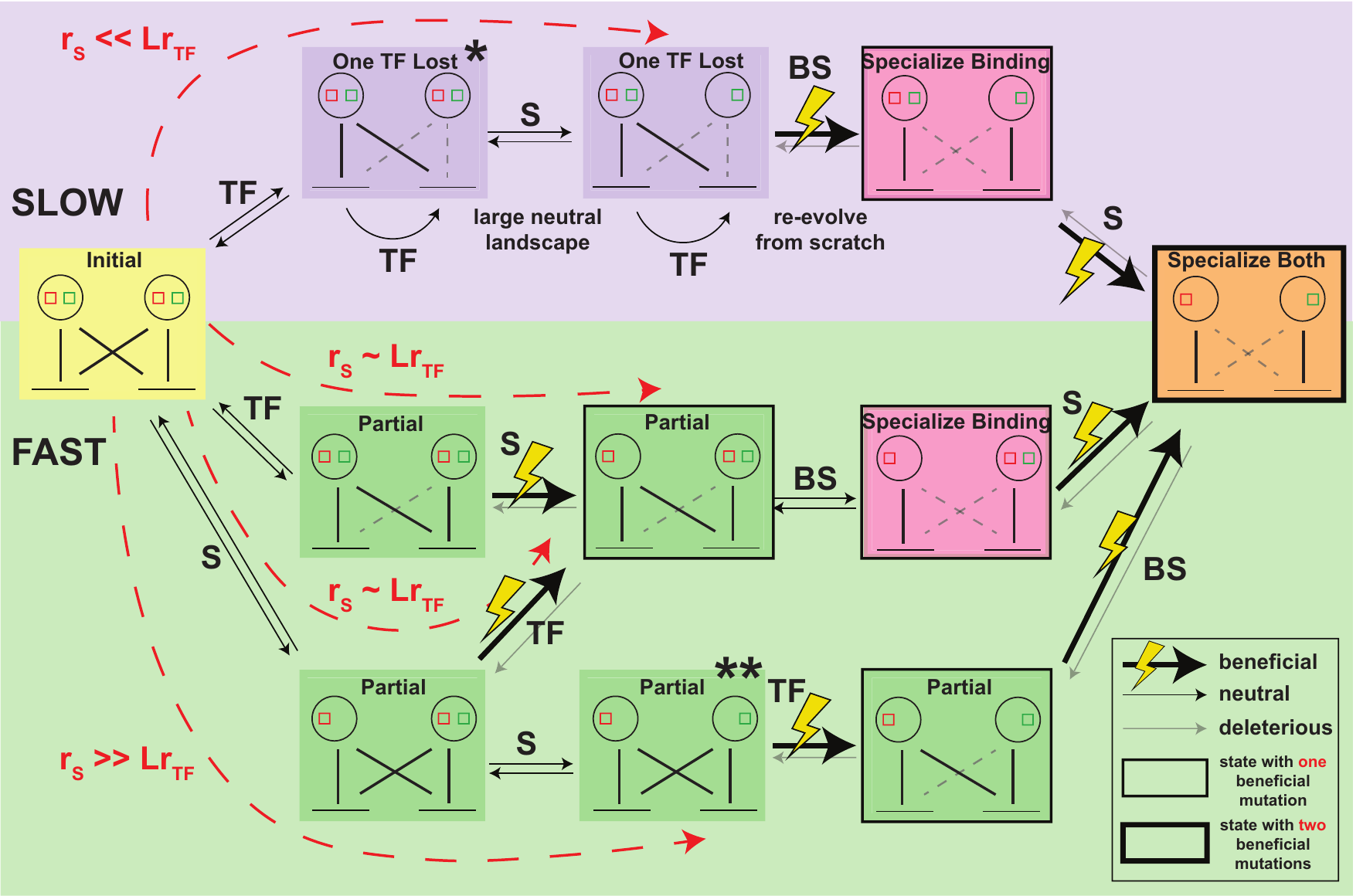}}

\caption{\textbf{Pathways to specialization differ in the order and
nature of mutations.} Here we detail the various mutations occurring along the different pathways to specialization. For each
mutation, we show the type of mutation (in text on the arrows): TF
consensus sequence mutation (TF), binding site sequence mutation
(BS), TF signal sensing domain mutation (S) and whether it is beneficial, (nearly) neutral or deleterious (style of the arrows). We also illustrate the macrostates along each pathway using the same color code in the background as in the main text. The number of beneficial mutations in each macrostate relative to the $\texttt{Initial}$ macrostate is depicted by box style (see legend). Text in red indicates the conditions on
mutation rates that favor the different pathways. Note that from
the $\texttt{One TF Lost}$ state marked with a star, the ``lost''
TF can actually take up new functions (by sensing and binding to signals
and binding sites other than those considered in our model), leading
to ``neo-functionalization''. Also, the $\texttt{Partial}$ state marked with two stars acts as the initial condition in the alternative model variant, with the TFs already specialize in signal sensing immediately post-duplication.\label{fig:dyn_pathway_details}}
\end{figure}

\begin{figure}[H]
\centering{\includegraphics[scale=0.8]{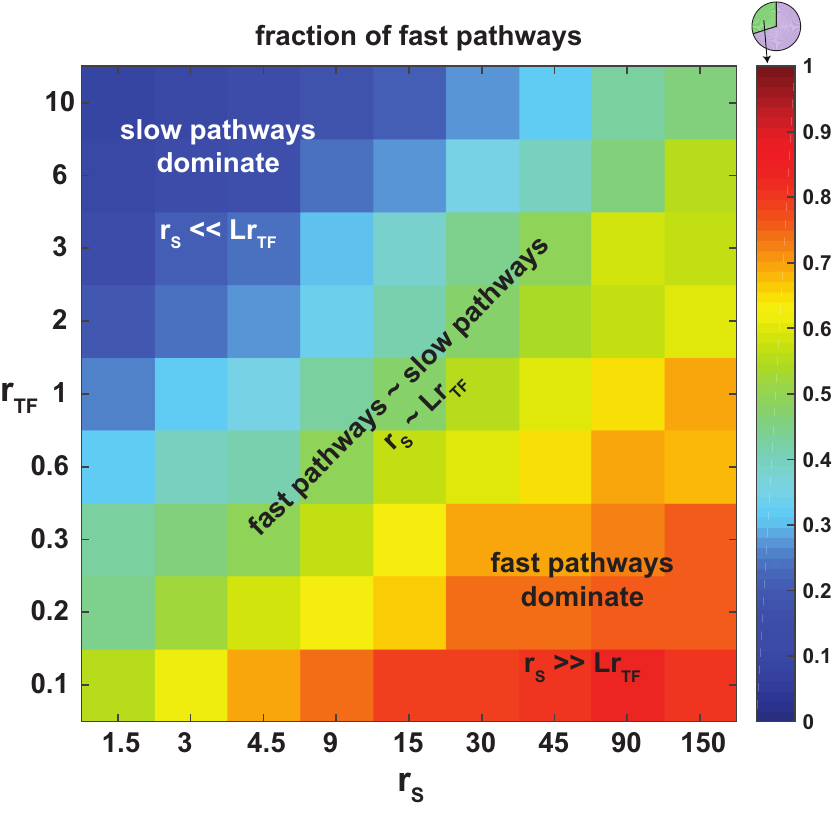}}

\caption{\textbf{The ratio between $r_{S}$ and $r_{TF}$ determines the dominant pathway.} We plot the fraction of fast $\texttt{Partial}$ pathways
as a function of $r_{S}$ (signal sensing domain mutation rate) and $r_{TF}$ (TF mutation rate). Other parameters remain at their
baseline values (see Section \ref{sec:Model}). Color code denotes the fraction of fast pathways (specialization is reached via 'Partial' intermediate state). \label{fig:dyn_pathway_muts}}
\end{figure}

In Fig. \ref{fig:dyn_pathway_details} we detail the different
pathways to specialization.
The pathways proceeding via $\texttt{One TF Lost}$ are slow compared to the pathways
proceeding via $\texttt{Partial}$ which are faster. The mutation initiating the process in all pathways is neutral and hence the ratio between $r_{S}$ (signal sensing domain mutations rate) and $r_{TF}$ (TF mutation rate) determines which pathway is more likely to occur - see Fig. \ref{fig:dyn_pathway_muts}.

Along the slow $\texttt{One TF Lost}$ pathway, typically, first a TF consensus sequence mutation occurs that weakens the binding of
one TF to both binding sites. Once binding is lost, further mutations cause
the TF consensus sequence to neutrally drift away.
Meanwhile, the lost TF gains a sensing mutation such that it senses
only one of the two signals. Next, a BS mutation in one of the binding sites flips its TF preference such that the system moves into $\texttt{Specialize Binding}$ macrostate.
This is a beneficial mutation as one of the signal-BS pathways becomes
specific. This involves evolving a TF-BS link essentially from scratch; the lost TF consensus sequence is a random number of mismatches
away from the binding site sequence, and the beneficial BS mutation
can occur only when the TF consensus sequence, by chance, becomes
close enough to the BS sequence. From $\texttt{Specialize Binding}$,
another beneficial sensing mutation leads the system to full specialization (BS and signal).

There are multiple routes in the $\texttt{Partial}$ pathway. In one
of the routes, first a neutral TF consensus sequence mutation occurs
such that the TF loses binding to only one of the two binding sites resulting in $\texttt{Partial}$ macrostate. This is different from the first
mutation in $\texttt{One TF Lost}$ pathway where the TF loses binding
to both binding sites. From here, a sensing domain mutation specializes
one of the signal-BS pathways, making this mutation beneficial. Further,
a neutral BS mutation brings the system to $\texttt{Specialize Binding}$,
from where a beneficial sensing domain mutation leads the system to
specialization.

In the second and third routes via the $\texttt{Partial}$ macrostate,
first a neutral sensing domain mutation occurs. Next, either a beneficial
TF consensus sequence mutation can bring the system onto the previous
route or if the sensing domain mutation rate is high, another neutral
sensing domain might occur first. From here, a beneficial TF consensus
sequence mutation and a beneficial BS mutation again lead to full specialization.

\subsection{Evolutionary pathways - second model variant}
\begin{figure}[H]
    \centering
   \subfigure[]{
      \includegraphics[width=0.4\textwidth]{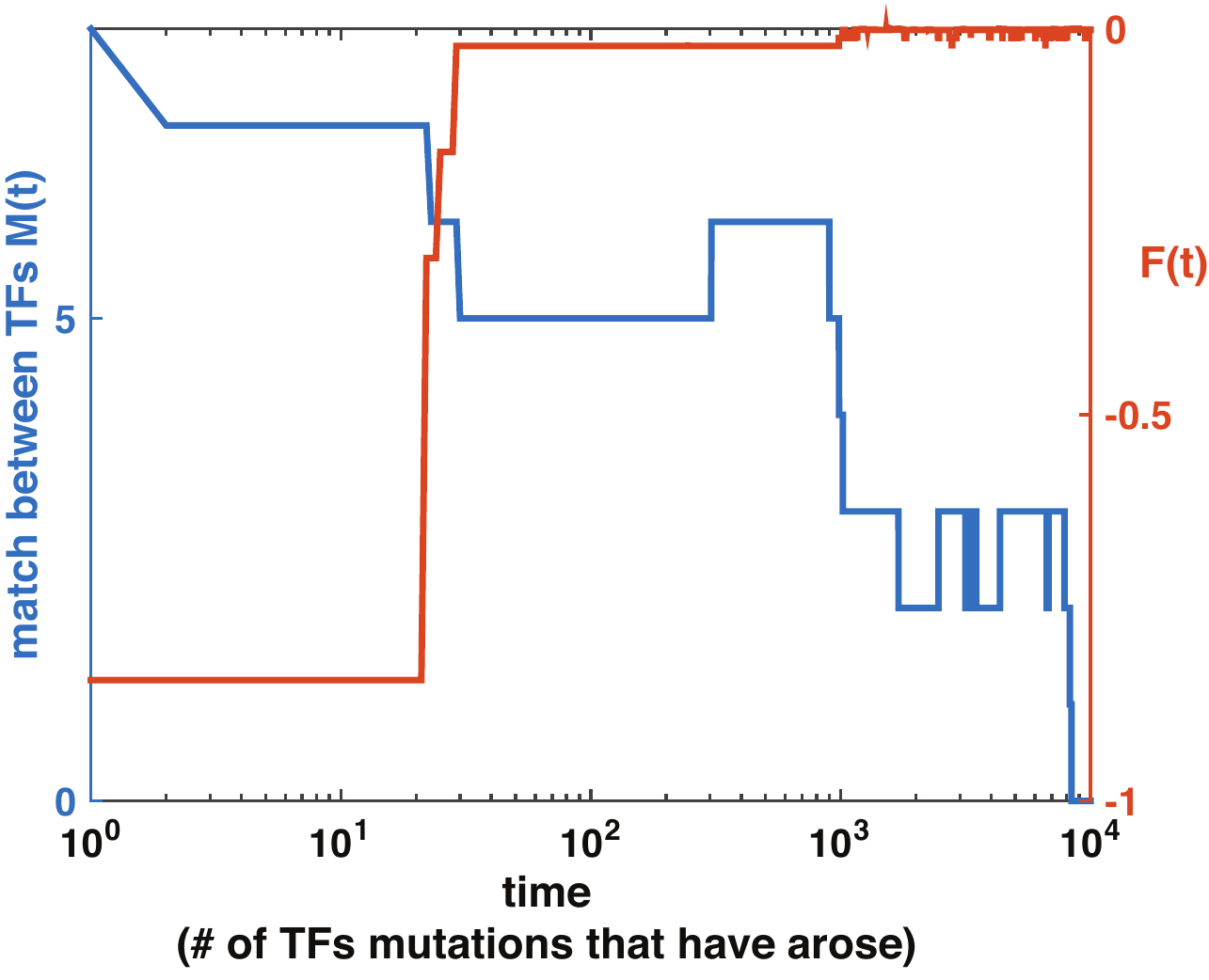}
        }
       \subfigure[]
{
        \includegraphics[width=0.48\textwidth]{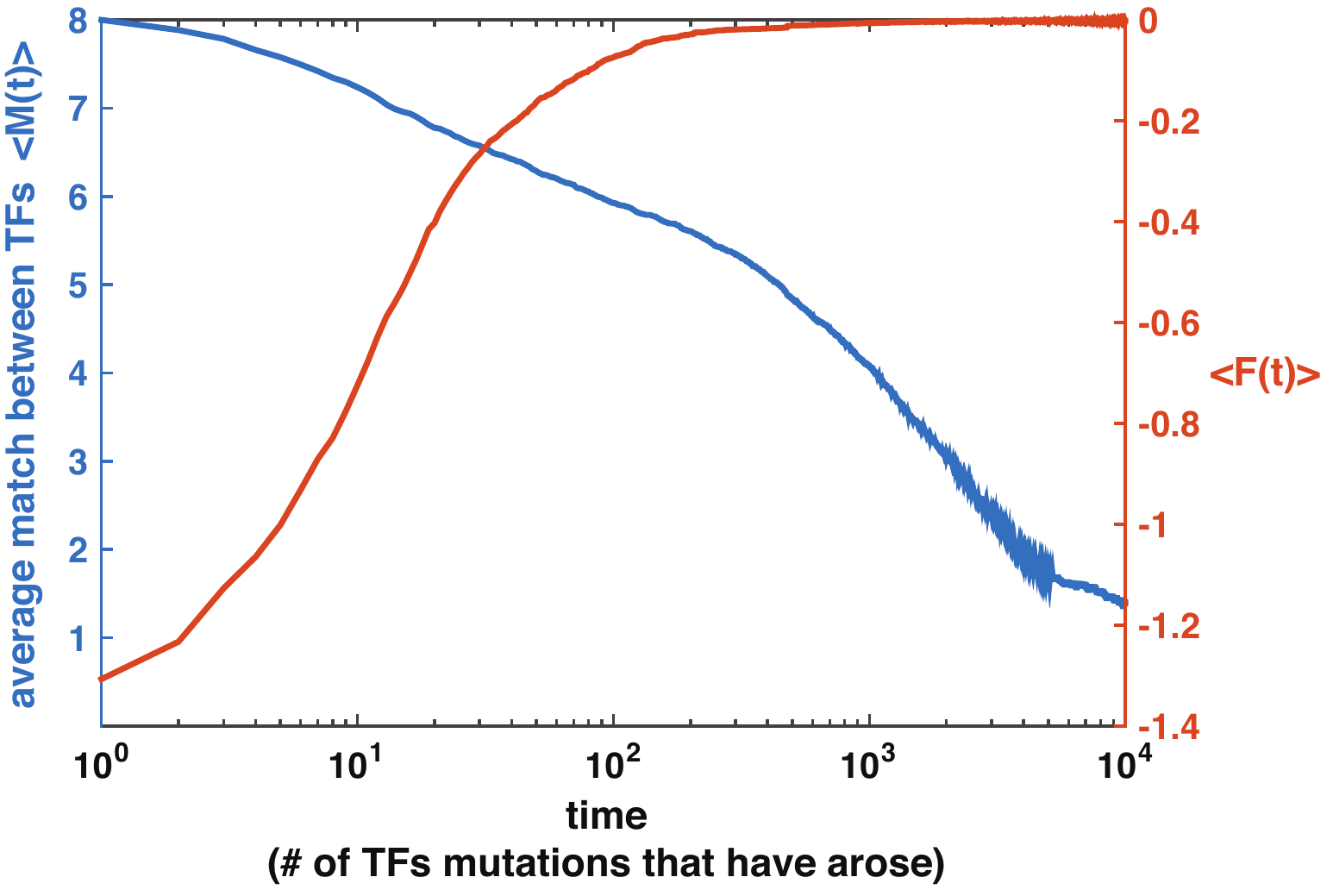}
    }
    \captionof{figure}
    {\label{fig:M_F_dynamics} \textbf{In the second model variant (TFs specialized in the signal sensitivities or expression times immediately post-duplication) a significant proportion of evolutionary time is spent in neutral evolution phase.} Selection only works in the beginning of the evolutionary trajectory to exert diversification, but a significant part of TF diversification occurs almost neutrally with only a modest fitness benefit involved. We illustrate dynamical trajectories of the match between TFs, $M$, and the fitness, $F$, obtained in stochastic simulations. (a) shows a single trajectory and (b) shows an average over 400 independent repeats of the simulation. Each time unit is a simulation iteration in which a mutation in one of TFs occurs, but does not necessarily fix (see \secref{subsec:altSimul}).
}
\end{figure}

The second model variant (see Section \ref{sec:Model}) assumes that immediately post-duplication, TFs are expressed at different times (or are already specialized with respect to their signal sensitivity), and that this is fixed for the rest of the evolutionary time. This $\texttt{Partial}$ macrostate is marked by two stars in the pathway schematic \figref{fig:dyn_pathway_details}. In this setting, selection to specialize starts with a phase of fast diversification where each pair of TF-BS mutates (in orchestrated manner) to diverge from the other. The fitness benefit in diversification is large at the beginning when the TFs are identical, but diminishes the more distinct they become. This is illustrated in \figref{fig:dyn_pathway_details} by the two TF and BS beneficial mutations that lead to specialization. After specialization, further TF diversification proceeds as a nearly neutral process, and hence occurs more slowly. These two phases, the fast adaptive one followed by the slow nearly-neutral one, are illustrated in \figref{fig:M_F_dynamics}.

\subsection{Time to specialization}

\begin{figure}[H]
\centering{\includegraphics{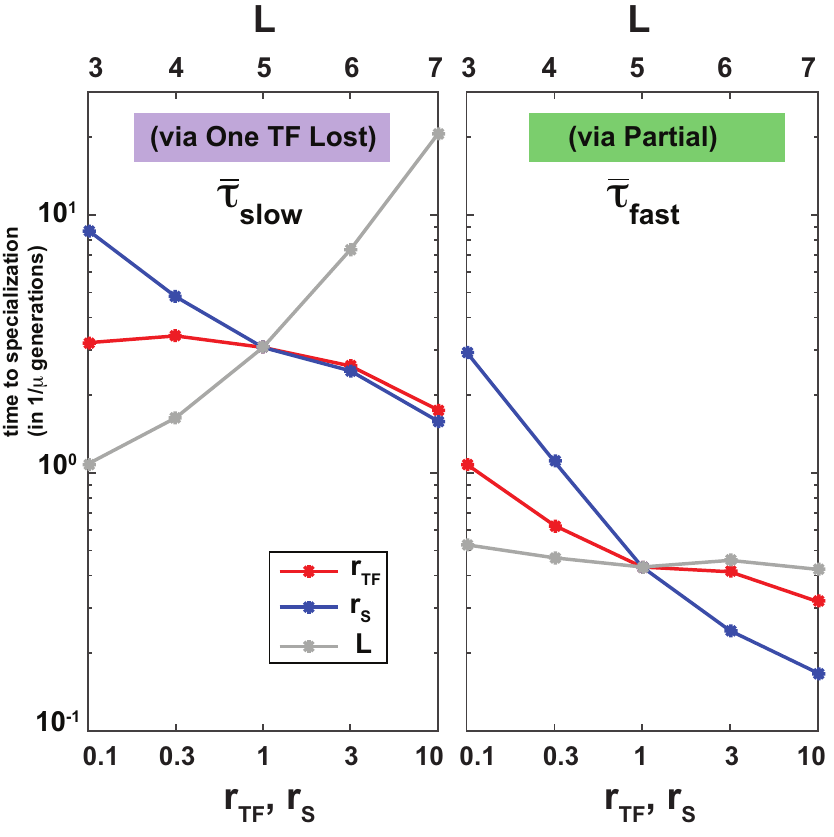}}

\caption{\textbf{Time to specialization via different pathways for different
parameters.} We plot the mean times to specialization, $\bar{\tau}_{slow}$
and $\bar{\tau}_{fast}$, via the slow (left panel) and the fast (right
panel) pathways, while varying $L$ (grey curve, top axis), $r_{TF}$ (red, bottom axis) and $r_{S}$ (blue, bottom axis) separately.
Other parameters remain at their baseline values. We find opposite dependence of the time to specialize on the binding site length $L$ in the distinct pathways. While for pathways going via 'One TF Lost' (left panel) time increases with $L$ due to increase in the sequence space, it mildly decreases with $L$ for pathways going via 'Partial'. For all pathways time decreases if mutation rates increase.
\label{fig:dyn_times}}
\end{figure}

In Fig. \ref{fig:dyn_times}, we plot the average time to specialization via slow and fast pathways for various values of $L$, $r_{TF}$ and $r_S$. The ratios of these times are
plotted in Fig. 4E of the main text. Increasing either mutation rate by changing $r_{TF}$ or $r_{S}$ speeds up specialization via both pathways because mutations occur faster. Increasing $L$ slows down the slow $\texttt{One TF Lost}$ pathway because of
an increase in size of the neutral landscape; strikingly, increasing $L$ does not lengthen the fast pathway through $\texttt{Partial}$ states. 

\section{Role of $\beta_X$, the relative fitness penalty on crosstalk interactions}
\label{sec:beta}
Transcription factors often bind weak secondary binding sites besides their primary target(s). This can lead to spurious activity of genes called crosstalk, i.e., deleterious activation of genes that does not happen via their primary regulatory pathway. For example, in our model a gene can be activated even if the signal to which it should respond is absent only because of (weak) binding of a transcription factor responding to another signal to its binding site. Previously, we studied the effect of crosstalk interference on gene regulation, and showed how it can place global constraints on the gene regulatory system~\cite{friedlander_intrinsic_2016}. Here, we explore the potential role of such crosstalk interactions in shaping the evolutionary trajectories of TF specialization.

The fitness of each reduced-genotype $x\in\mathcal{G}$ depends on the difference between the actual expression pattern the genotype generates and the ideal expression pattern as defined in \eqref{eq:F}.

\be F(x)=-s\displaystyle\sum_j\displaystyle \sum_m\alpha_{m}\beta_{jm}(p_{jm}-p_{jm}^*)^{2}. \ee

Here, $\beta_{jm}$ weigh the penalties on different deviations from the desired expression level $p_{jm}^{*}$. In a certain environment $m$ some genes should be active, $p_{jm}^{*}=1$, while others should remain inactive, $p_{jm}^{*}=0$. In our model, we allow for different penalties in either case. We penalize deviations from desired activity $p_{jm}^{*}=1$ by setting $\beta_{jm}=1$.  We consider deviations from desired inactivity $p_{jm}^{*}=0$ as less crucial and penalize them to a lesser extent $\beta_{jm}=\beta_{X}$, $\beta_{X}\in[0,1]$. At the two extremes, if $\beta_{X}=0$, no penalty on these crosstalk terms applies, while if $\beta_{X}=1$, penalties on all deviations are equally important.
In the main text, we used an intermediate value of $\beta_{X}=0.5$. In this section we explore the role of $\beta_{X}$ on the steady state distribution prior to and after TF duplication and on the evolutionary dynamics of specialization.

\subsection{Steady state before duplication}

\begin{figure}[H]
\centering{\includegraphics[scale=0.75]{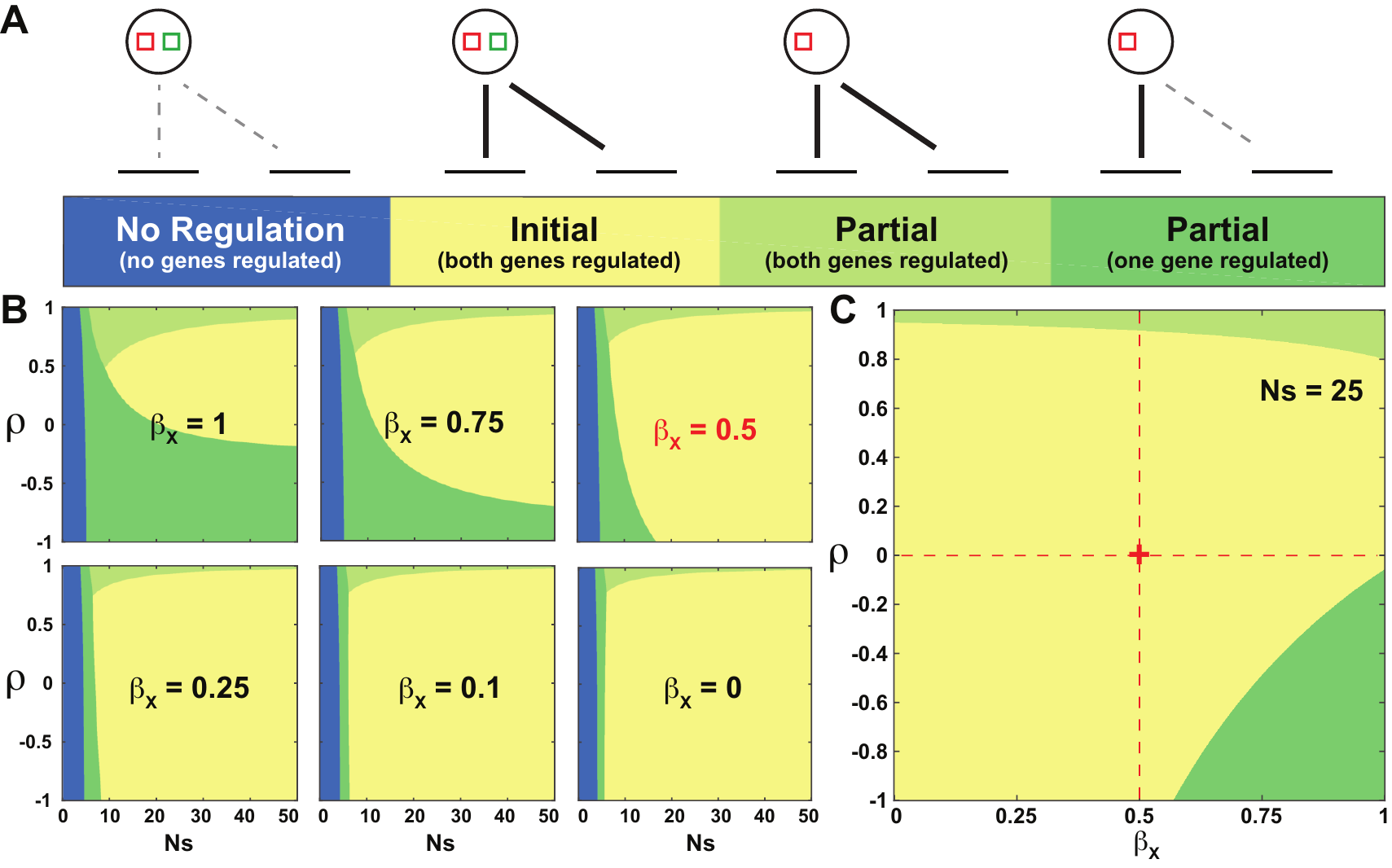}}

\caption{\textbf{Dominant macrostate at steady state before duplication depends on $\beta_{X}$ (crosstalk interaction penalty).}
(A) Illustration of the different macrostates when only a single TF exists. Macrostates before duplication are defined in terms of the macrostate they would result in, if a duplication occurred on those genotypes.
(B) Most probable macrostate at steady state before duplication, as
a function of selection strength, $Ns$, and the correlation between the two external signals, $\rho$, for different values of $\beta_{X}$,
the relative weight of fitness penalties corresponding to crosstalk
interactions. (C) The most probable macrostate at steady state before duplication,
as a function of $\beta_{X}$ and $\rho$ at $Ns=25$. \label{fig:beta_SSbef}}
\end{figure}

A steady state distribution is attained before duplication, when only a single TF regulates all genes. In \figref{fig:beta_SSbef} we illustrate the most probable macrostate prior to duplication for different values of cross-interaction penalties $\beta_X$. The macrostates possible before duplication are $\texttt{Initial}$ (both genes regulated), $\texttt{No Regulation}$ (none regulated) and some (but not all) variants of $\texttt{Partial}$ - see \figref{fig:beta_SSbef}A for illustration.
For $\beta_{X}\simeq 1$, the fitness penalty on mistakenly activating a gene is comparable to the fitness penalty on not fully inducing genes when needed, resulting in network configurations in which only one of the two genes is regulated (corresponding to $\texttt{Partial}$ macrostate immediately after duplication for most $\rho<0$). This is because, while configurations with only one gene regulated have one functional interaction and no crosstalk interactions, configurations with both genes regulated have two functional interactions and two crosstalk interactions. As $\beta_{X}$ decreases, the selection against crosstalk interactions becomes weaker, resulting in configurations in which both genes are regulated ($\texttt{Initial}$
macrostate immediately after duplication) even when $\rho<0$.

\subsection{Steady state after duplication}

\begin{figure}[H]
\centering{\includegraphics[scale=0.75]{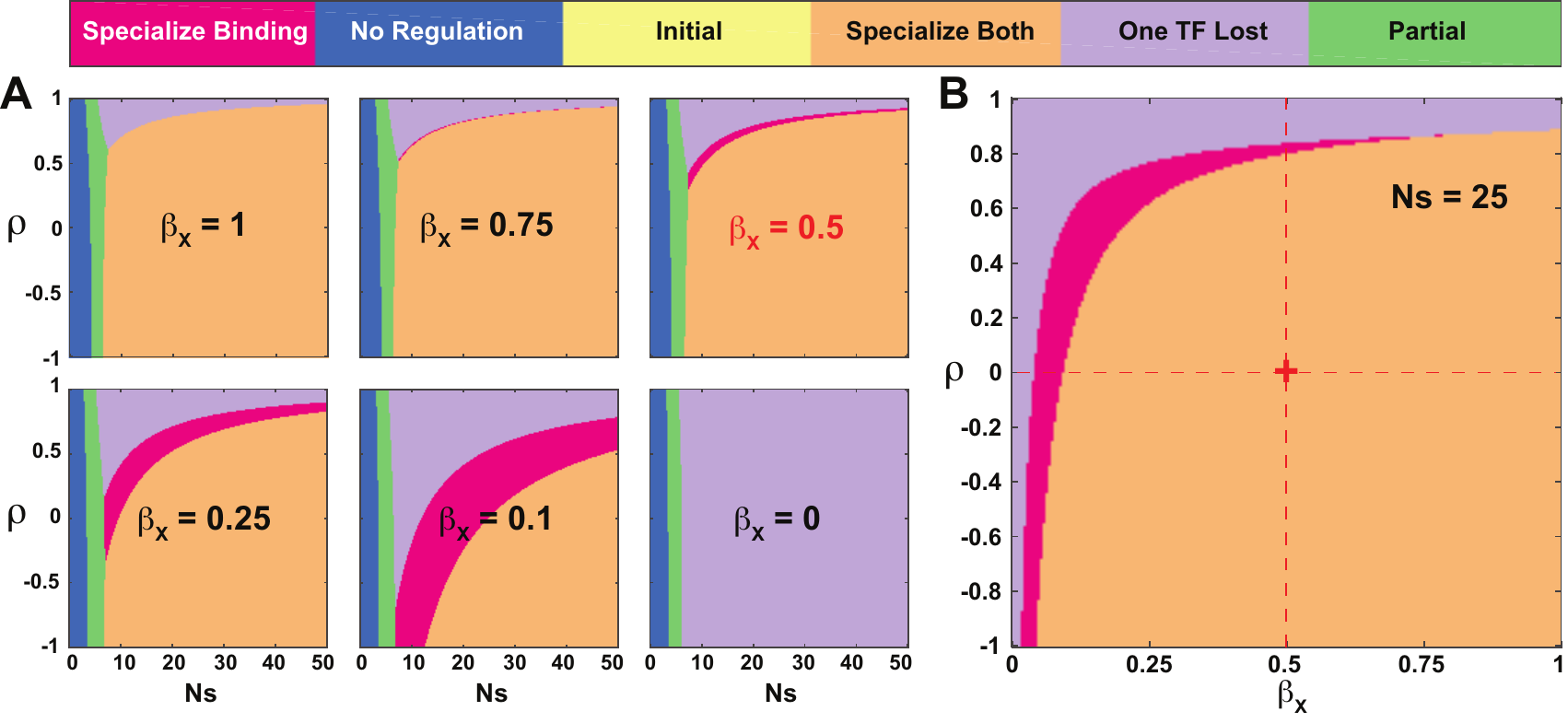}}

\caption{\textbf{Dependence of steady state after duplication on $\beta_{X}$, the fitness penalty on cross-interactions.}
(A) The most probable macrostate at steady state after duplication,
as a function of selection strength, $Ns$, and the correlation between the two external signals, $\rho$, is plotted for six different values
of $\beta_{X}$. (B) The most probable macrostate at steady state
after duplication, as a function of $\beta_{X}$ and $\rho$ at $Ns=25$. An increase in $\beta_X$ has a a similar effect to an increase in selection intensity on all interactions by varying $Ns$.
\label{fig:beta_SSaft}}
\end{figure}

We proceed to observe the effect of varying $\beta_X$ on the steady state after duplication, analogous to Fig. 3C of the main text where we assumed $\beta_X=0.5$. In Fig. \ref{fig:beta_SSaft}, we show the phase plot of the most
probable outcome of duplication at steady state for different values of $\beta_{X}$.
The qualitative features of this phase plot are invariant to changes in $\beta_X$, as long as $\beta_X > 0$. For $\rho$ not too close to $1$,
we obtain transitions from $\texttt{No Regulation}$ to $\texttt{Partial}$ and to $\texttt{Specialize Both}$  as $Ns$ increases. For large enough
$Ns$, as $\rho$ increases, there is a shift from $\texttt{Specialize Both}$
to $\texttt{One TF Lost}$, via $\texttt{Specialize Binding}$, the
width of which increases as $\beta_{X}$ decreases. This is because there is reduced selection pressure on avoiding crosstalk interactions as
$\beta_{X}$ decreases. For small $\beta_{X}$, as $\rho$ increases,
it is sufficient that one of the TFs senses both signals
while the TFs are still specialized in binding. As $\rho$ increases even
further, it is sufficient to have one TF mediating both pathways, marking the shift to the $\texttt{One TF Lost}$ macrostate. These transitions occur very prominently for very small $\beta_{X}\approx0$, where
$\texttt{One TF Lost}$ is the most probable outcome for all $\rho$ values. Many models of duplication do not consider crosstalk interactions in their fitness function, and hence deal with the case of $\beta_X=0$, making it important for comparison to our results.

\subsection{Evolutionary dynamics}

\begin{figure}[H]
\centering{\includegraphics[scale=0.8]{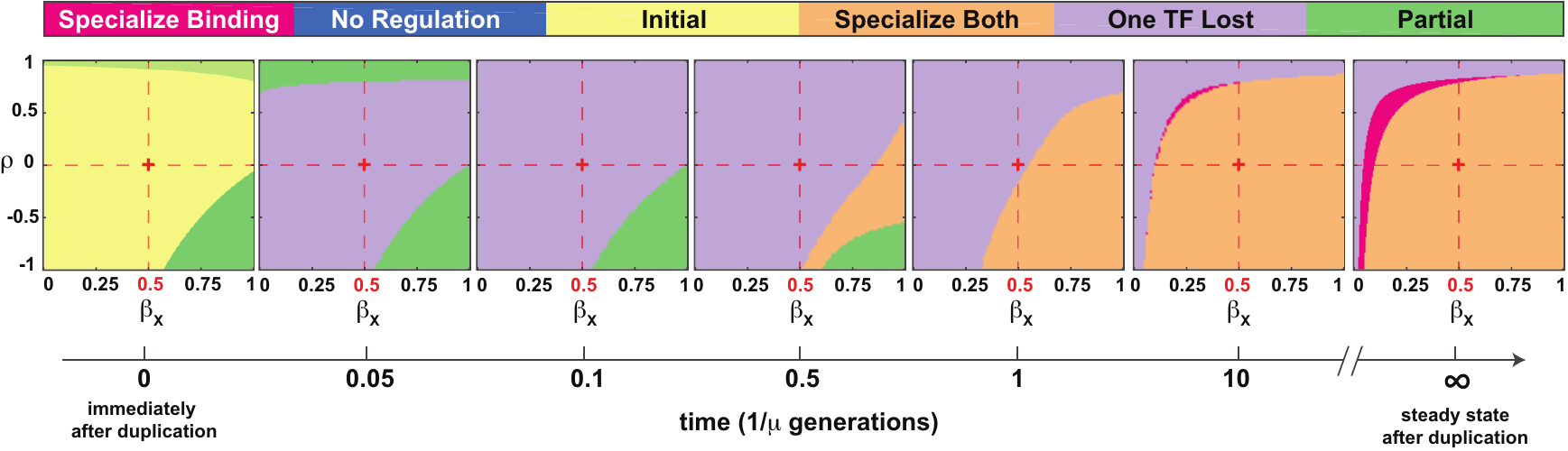}}

\caption{\textbf{Snapshots of the most probable macrostate at different time-points
post-duplication.} The most probable macrostate as a function of signal
correlation, $\rho$, and $\beta_{X}$, the relative weight of fitness
penalties corresponding to crosstalk errors, for $Ns=25$. The left-most phase plot corresponds
to the time-point immediately after duplication, and the right-most
phase plot corresponds to the steady state after duplication. For
other parameters, the baseline values have been used. $\beta_{X}=1$
corresponds to equal-magnitude selection strengths on functional as well
as crosstalk interactions; $\beta_{X}=0$ corresponds to no selection
against crosstalk interactions. In the main text, we choose $\beta_{X}=0.5$
as the baseline parameter value. \label{fig:beta_MPMS_snapshots}}
\end{figure}

To understand how $\beta_{X}$ affects the evolutionary dynamics of
specialization, we first obtained the dynamics of the most probable
macrostate as a function of $\rho$ and $\beta_{X}$ for fixed selection intensity $Ns=25$ (baseline
parameters). In Fig. \ref{fig:beta_MPMS_snapshots}, we plot a few
snapshots of the phase diagram of the most probable macrostate at
different time-points after duplication, starting from $t=0$ (immediately
after duplication), to $t=\infty$ (steady state after duplication). Specialization
is faster for smaller $\rho$ because the fitness benefit of eliminating crosstalk interactions is larger. Likewise, specialization is faster for larger $\beta_{X}$ as the selection strength against crosstalk interactions is higher.
A huge region of the $(\beta_{X},\rho)$ plane corresponding to small
$\beta_{X}$ or large $\rho$, most of which starts at $\texttt{Initial}$ and specializes via the slow pathway of $\texttt{One TF Lost}$.

\begin{figure}[H]
\centering{\includegraphics[scale=1]{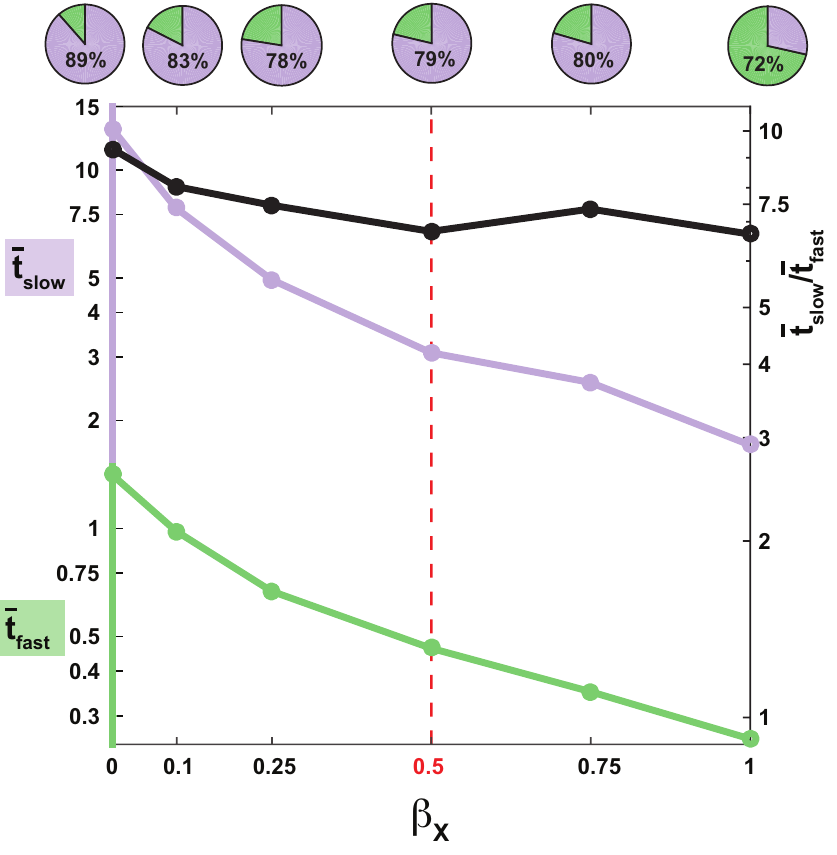}}

\caption{\textbf{How do the slow and fast pathways to specialization depend
on $\beta_{X}$?}
For large $\beta_X$, the time to specialization shortens for all pathways and the fraction of trajectories to specialization taken via fast pathways (through $\texttt{Partial}$ macrostate) increases. Pie charts illustrate the fraction of slow (lavender) and fast (green) trajectories for different values of $\beta_X$. The black line (right y-axis) shows the ratio between average specialization times, which does not significantly change with $\beta_X$.
For other parameters, the baseline values were used.  $\beta_{X}=1$
corresponds to equal-magnitude selection strengths on functional as well
as crosstalk interactions; $\beta_{X}=0$ corresponds to no selection
against crosstalk interactions. In the main text, we choose $\beta_{X}=0.5$
as the baseline parameter value.
\label{fig:beta_tslowfast}}
\end{figure}

Next we sought to understand which pathways are taken towards specialization for different $\beta_{X}$ by running many repeats of simulations at each $\beta_{X}$. For each $\beta_{X}$, we found the most probable genotype at steady state before duplication and ran many repeats of the simulation starting from that genotype. In Fig.  \ref{fig:beta_tslowfast}, we explore the dependence on $\beta_{X}$ of fraction of the two pathways to specialization (slow via $\texttt{One TF Lost}$ and fast via $\texttt{Partial}$), and also the corresponding times to specialization. First of all, specialization becomes quicker as $\beta_{X}$ increases from $0$ to $1$. This is because stronger selection against the crosstalk interactions eliminates them faster. Secondly, the relative speed of the fast pathway (compared to the slow pathway) depends only very weakly on $\beta_{X}$. Thirdly, about $80\%$ of trajectories follow the slow pathway, and this depends only very weakly on $\beta_{X}$, till $\beta_{X}=0.75$. In contrast, for $\beta_{X}=1$, the fast pathways via $\texttt{Partial}$ become predominant. This occurs because the steady state before duplication (which acts as the initial condition for the trajectories) flips from $\texttt{Initial}$ to $\texttt{Partial}$.

\section{Comparison between biophysically-realistic model and simple models}
\label{sec:biallelic}
\begin{figure}[H]
\centering{\includegraphics[scale=0.6]{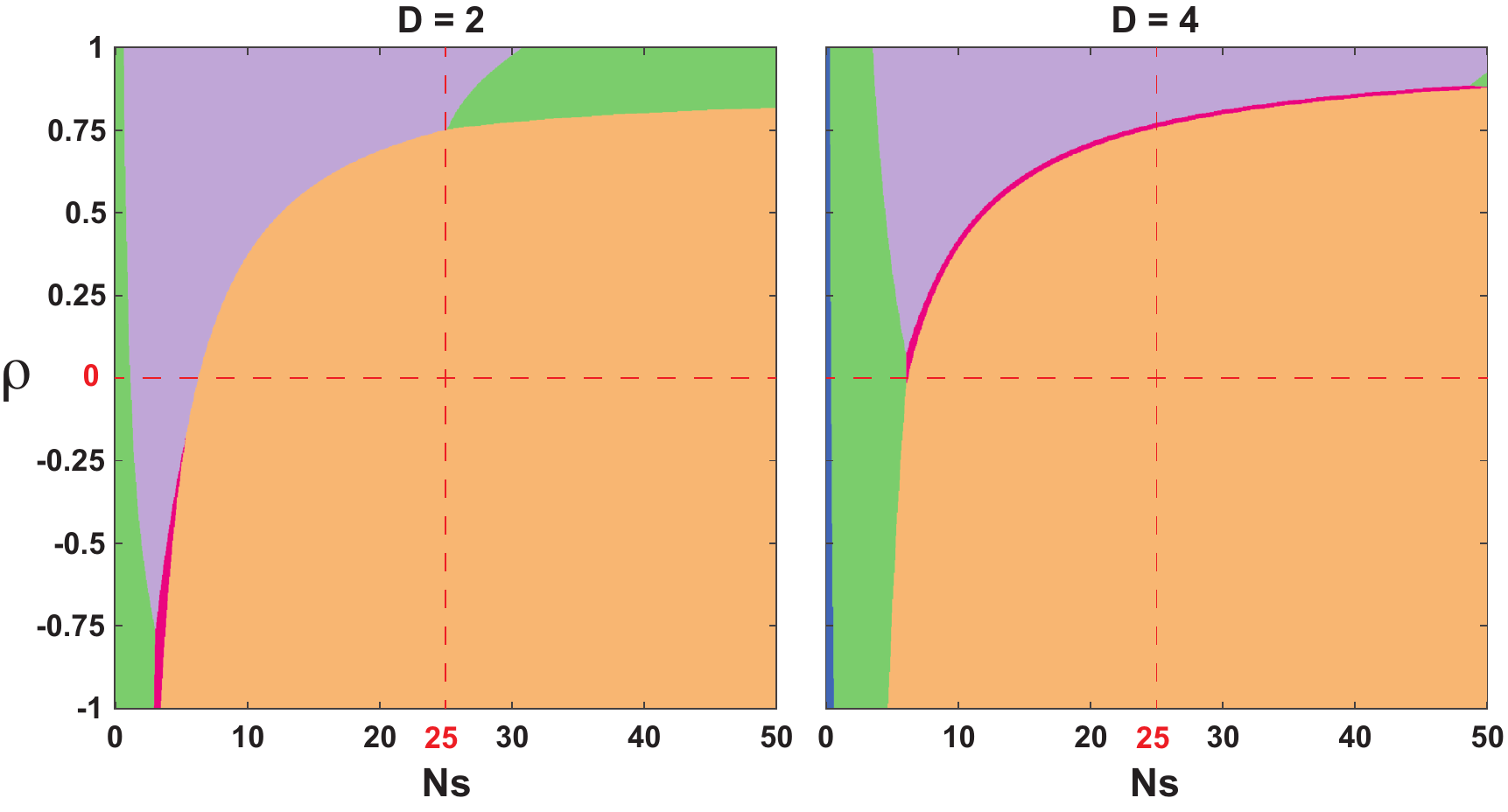}}

\caption{\textbf{Dominant macrostate at steady state for biallelic-like models.}
Here we plot the dominant macrostate at steady state  as a function of $Ns$ and $\rho$ for biallelic-like models with alphabet size $D=2$ (left panel) and $D=4$ (right panel). Color code used to indicate different macrostates is the same as in the main text. 
\label{fig:biall_SS}}
\end{figure}

Gene duplication literature often studies models with a small number of discrete alleles, for example, binary alleles informing whether TF-BS binding occurs. Throughout this work we employ a different approach by including a biophysical description of TF/DNA interactions. Consequently, a large number of different genotypes can often realize each functional architecture (macrostate), capturing naturally the important effects of neutral processes (mutational entropy). Our framework reduces to biallelic models at $L=1$ and alphabet size $D=2$ (and multiallelic version with $D=4$), so we can directly study the relationship between the results for a biophysically realistic fitness landscape and various common simplifications. We refer to these simpler models with $L=1$ here as the biallelic-like model. The biallelic-like model cannot reproduce some of the results obtained with the biophysically-realistic model of the main text. In particular, certain important macrostates do not exist in the biallelic-like model. We also find an opposite dependence on time to specialization for the different pathways ($\texttt{One TF Lost}$ vs. $\texttt{Partial}$).
In Fig.~\ref{fig:biall_SS} we plot the dominant macrostate at steady state for two values of $D$. For $D=4$ (right panel of the figure), many qualitative features
are retained from the more realistic main text model: for instance,
the change from $\texttt{No Regulation}$ to $\texttt{Partial}$ to
$\texttt{Specialize Both}$ as $Ns$ increases, and the change from
$\texttt{Specialize Both}$ to $\texttt{Specialize Binding}$ to $\texttt{One TF Lost}$
as $\rho$ increases. For $D=2$, we have
$\texttt{Partial}$ macrostate dominating at $Ns=0$, because its entropy is larger than that of the $\texttt{No Regulation}$ macrostate. Also, at large $Ns$ and large $\rho$, $\texttt{Partial}$
dominates via the genotypes in which all TF-BS links are strong but
the signal sensing domain is not specialized.

\begin{figure}[H]
\centering{\includegraphics{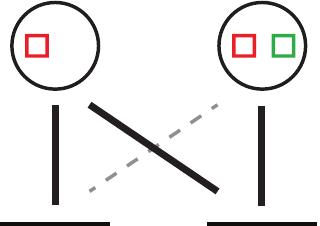}}

\caption{\textbf{This type of $\texttt{Partial}$ macrostate is absent in biallelic-like
models.} In biallelic-like models, strong TF-BS link means an exact match between TF and BS. Hence, the description of \textbf{$\texttt{Partial}$} states of the
kind shown here is impossible. \label{fig:biall_partial}}
\end{figure}

\begin{figure}[H]
\centering{\includegraphics{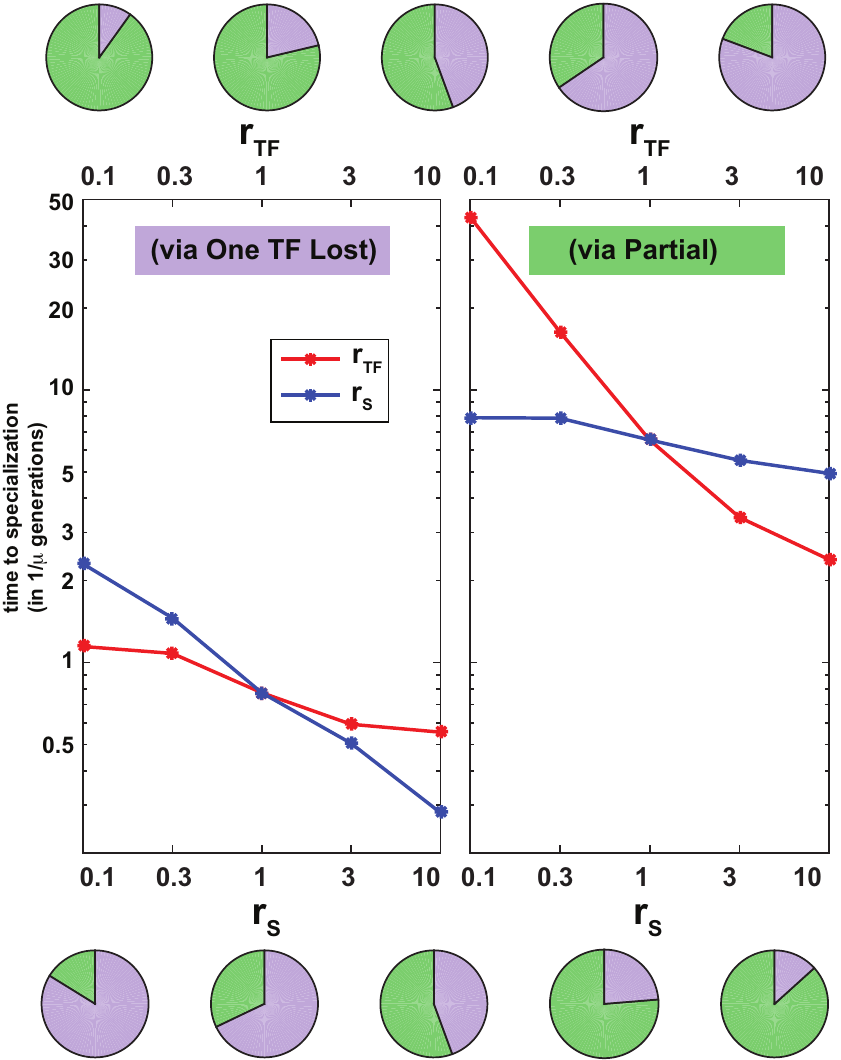}}

\caption{\textbf{Biallelic-like models reverse the relation between different pathways to specialization: $\texttt{Partial}$ pathways are the slow ones and $\texttt{One TF Lost}$ pathways  are faster, in contrast to the full model studied in the main text.} We plot the times to specialization
via $\texttt{One TF Lost}$ (left panel) and via $\texttt{Partial}$
(right panel), at $Ns=100$, while changing $r_{TF}$ (red curve) and $r_{S}$ (blue curve) separately, keeping the other parameters at their baseline values
in each case. We also show the fraction of these pathways as pie charts (upper pie charts refer to different $r_{TF}$ values; lower ones to  different $r_S$ values).
\label{fig:biall_times}}
\end{figure}

Certain variants of $\texttt{Partial}$ that exist in the general model do not exist in the biallelic-like model, as shown in Fig.~\ref{fig:biall_partial}. These states have intermediate fitness and they arise in the fast $\texttt{Partial}$
pathway of the main text model, where they form a bridge between the
$\texttt{Initial}$ and the $\texttt{Specialize Both}$ macrostates. Hence,
in biallelic models, fast $\texttt{Partial}$ pathways do not exist
and instead, passing through $\texttt{Partial}$ entails either losing
a BS or specializing very fast in the signal sensing domain. These
states have low fitness in the biallelic-like model and hence $\texttt{Partial}$
pathway is actually slow. This is plotted in Fig.~\ref{fig:biall_times}.

In summary, biallelic-like models and the biophysically realistic model share a few similarities but also differ in certain important aspects. Biallelic-like models, while being very simplistic, still capture a few key qualitative features of the steady state distribution, for example, the transitions of dominant macrostates along the $\rho$ and $Ns$ axes. On the other hand, biallelic-like models paint a completely different picture of evolutionary dynamics and timescales. Because they do not consider intermediate-fitness $\texttt{Partial}$ states, unlike in the biophysically realistic model, time to specialization through $\texttt{Partial}$ becomes slower than through $\texttt{One TF Lost}$.

\section{Multiple genes regulated by each TF post-duplication}
\label{sec:MultipleGenes}
\subsection{Steady state after duplication}

Transcription factors often regulate multiple downstream genes, rather than one gene post-duplication, as we considered so far. Here we generalize our analysis to account for a general number of  genes, $n_G$. The steady state distribution in the general case is
\be
P(M,\{k_{ij}\},\{\sigma_i\}) = P_0(M,\{k_{ij}\})P_0(\{\sigma_i\})\exp(2NF),
\ee
where $P_0$ is the neutral distribution and $F$ is the fitness of the reduced-genotype. First, we need to account for the neutral distribution $P_0$ (entropic factor). This is straightforward, because for given TF consensus sequences, the probability that a particular binding site $j$ has mismatch values $(k_{1j},k_{2j})$ is independent of the state of other binding sites.  Thus, we can simply factor out the probabilities for different genes:
\be
\label{eq:factor}
P_0(M,\{k_{ij}\},\{\sigma_i\})= P_0(\{\sigma_i\})P_0(M)\prod_j P_0(k_{1j},k_{2j}|M),
\ee

\noindent where $j$ enumerates the genes.

\noindent Second, we need to take care of the adaptive (energy) factor $\exp(2NF)$ in the general case. Because $F=\sum_j F_j$ is linear in terms of contributions $F_j$ from each gene $j$, $\exp(2NF)$ factorizes into $\prod_j \exp(2NF_j)$. Hence, we have

\be
P(M,\{k_{ij}\},\{\sigma_i\}) = P_0(M)P_0(\{\sigma_i\})\prod_j P_0(k_{1j},k_{2j}|M)\exp(2NF_j).
\ee

\noindent Now, for $\langle M \rangle$, we have,
\be
\begin{split}
\langle M \rangle  & = \displaystyle \sum_{\{k_{ij}\},M,\{\sigma_i\}} M P(M,\{k_{ij}\},\{\sigma_i\}) \\
		            & = \displaystyle \sum_{\{\sigma_i\}} P_0(\{\sigma_i\}) \displaystyle \sum_{M} M P_0(M)\prod_j \displaystyle \sum_{k_{1j},k_{2j}} P_0(k_{1j},k_{2j}|M)\exp(2NF_j) \\
		            & = \displaystyle \sum_{\{\sigma_i\}} P_0(\{\sigma_i\}) \displaystyle \sum_{M} M P_0(M) \prod_j  \langle \exp(2NF_j) \rangle_{P_{0}(\{k_{ij}\}|M)}. \\
\end{split}
\ee

$\langle \exp(2NF_j) \rangle_{P_{0}(\{k_{ij}\}|M)}$ can be calculated for each gene $j$ separately. We consider $n_G$ downstream genes split into two sets of size $a$ and $b$ ($n_G = a+b$), such that $a$ genes should respond to the first signal and $b$ genes respond to the second signal. We write this as $a+b$ schematically in the figures. For the main model, we had $a=b=1$.

\begin{figure}[H]
    \centering
   \subfigure[]{
      \includegraphics[width=0.3\textwidth]{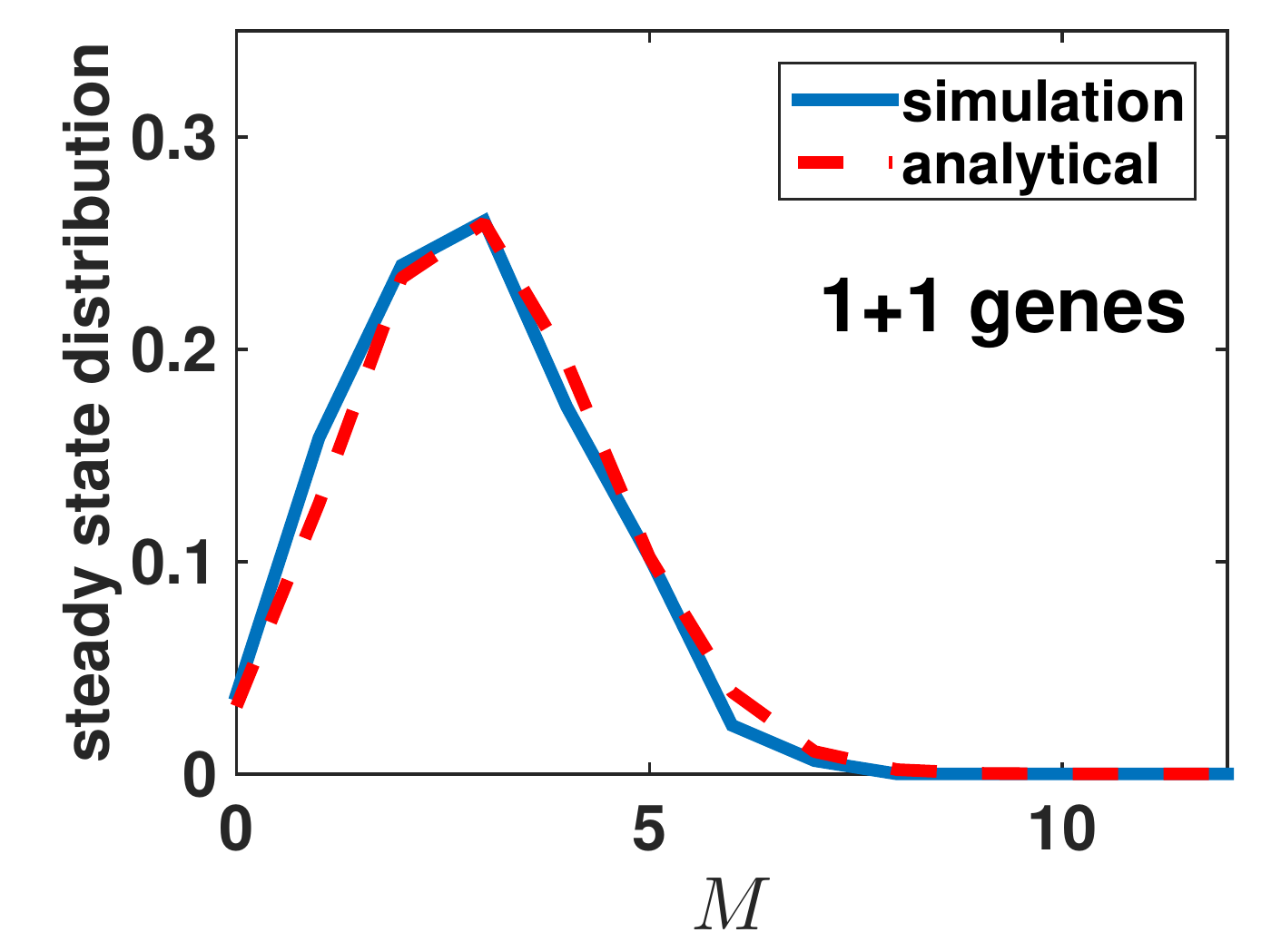}
        }
       \subfigure[]
{
        \includegraphics[width=0.3\textwidth]{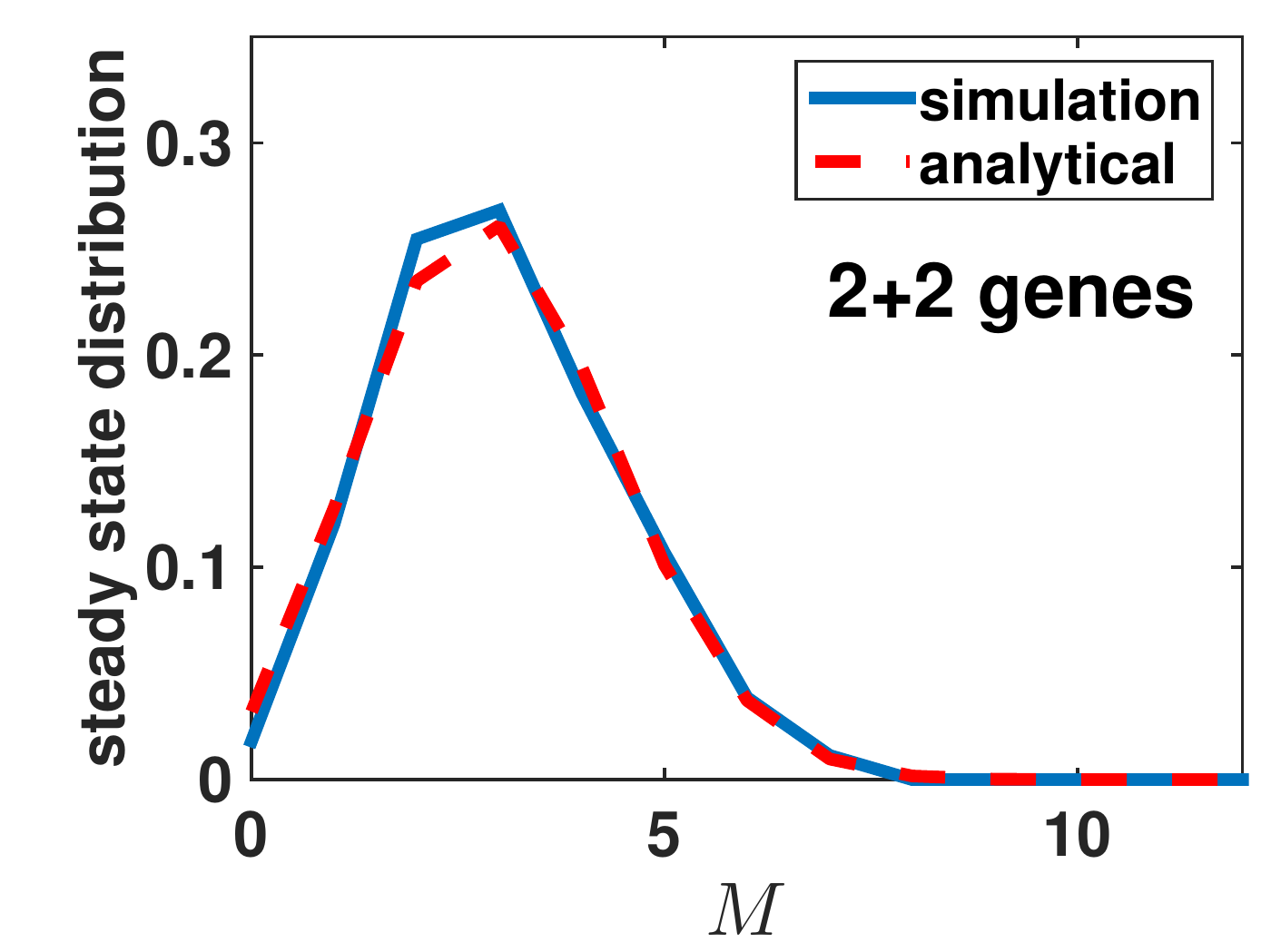}
    }
           \subfigure[]
{
        \includegraphics[width=0.3\textwidth]{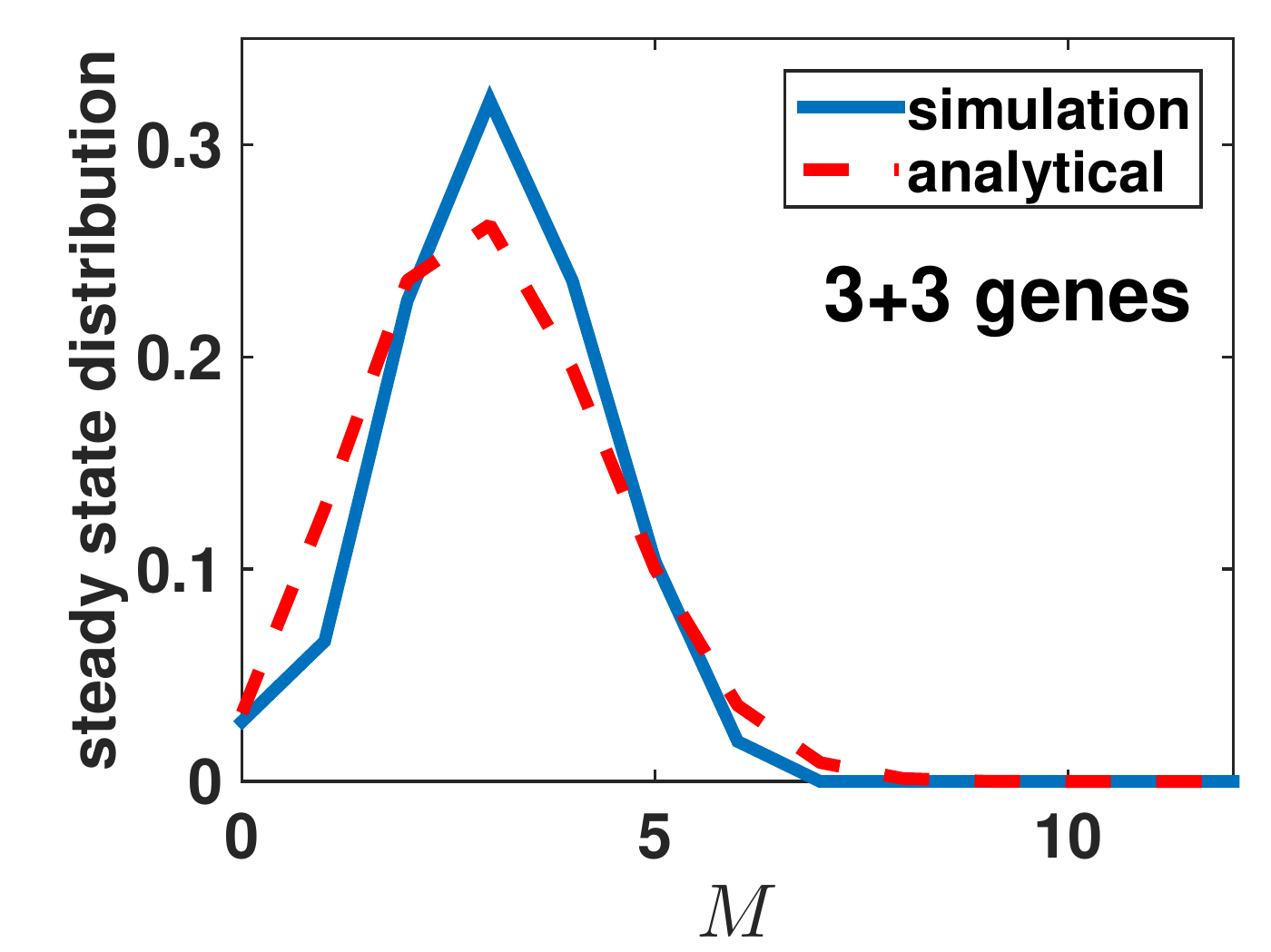}
    }
    \captionof{figure}
    {\label{fig:M_ss_dif_gene_num} The steady state distribution of $M$, the match between TF consensus sequences, is independent of the number of downstream genes regulated by these TFs. We present the analytically calculated steady state distribution and stochastic simulation results for $a+b=$1+1, 2+2 and 3+3 downstream genes.  Simulation steady state is the distribution obtained after 50,000 generations (1+1, 2+2 genes) or 150,000 generations (3+3 genes). Parameters: $L=12$, $Ns=500$, $\epsilon=3$, $C_0=3.269\times 10^5$, $r_{TF}=0.02$ (TF mutation rate is 50 times lower than the BS mutation rate).}
\end{figure}

We find that the steady state distribution of $M$, the match between the two transcription factors, is independent of the number of downstream genes - see \figref{fig:M_ss_dif_gene_num}.

\subsection{Evolutionary dynamics}

\subsubsection{Frustration of fitness landscape}
\begin{figure}[H]
    \centering
   \subfigure[]{
      \includegraphics[width=0.3\textwidth]{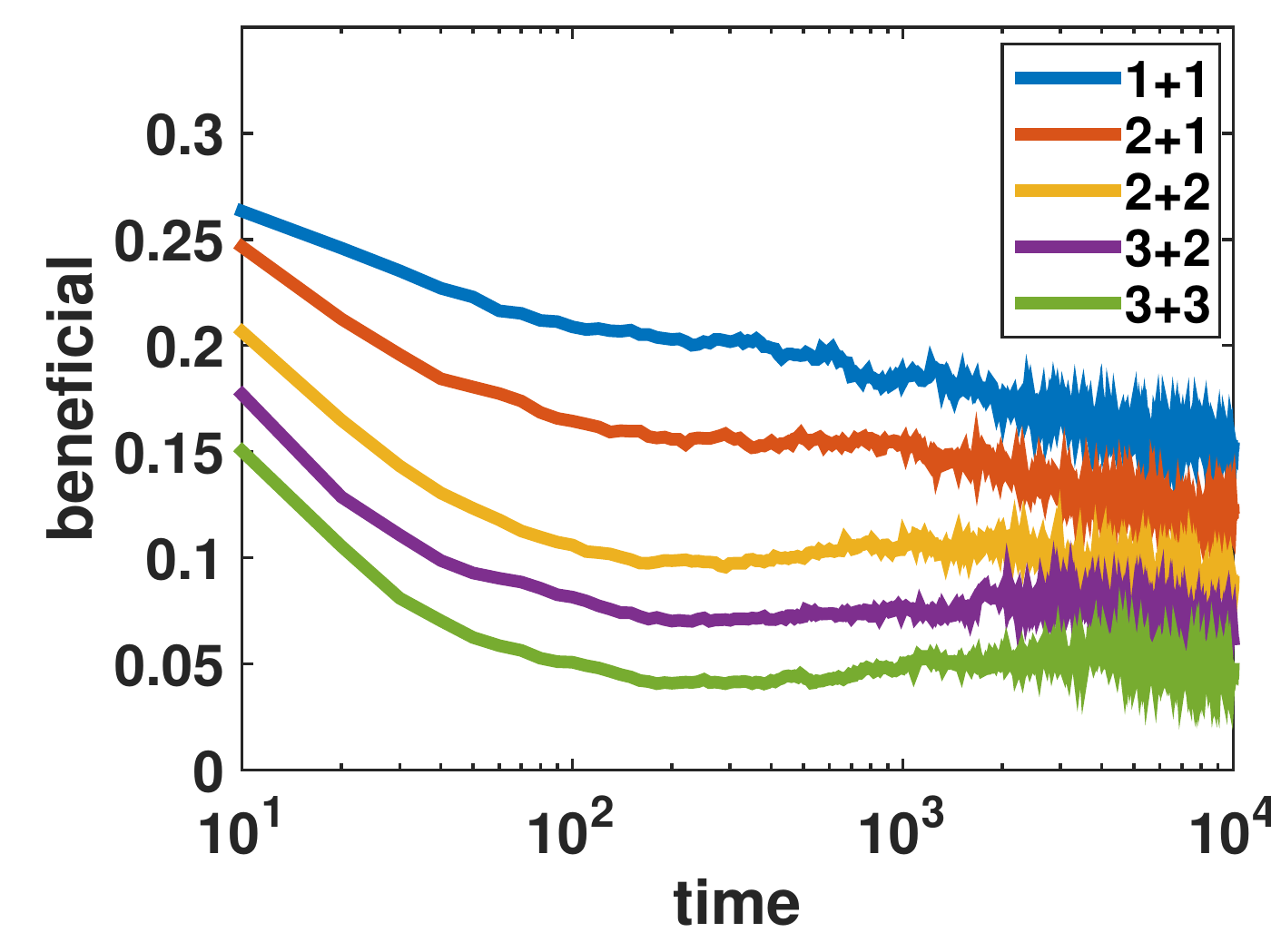}
        }
       \subfigure[]
{
        \includegraphics[width=0.3\textwidth]{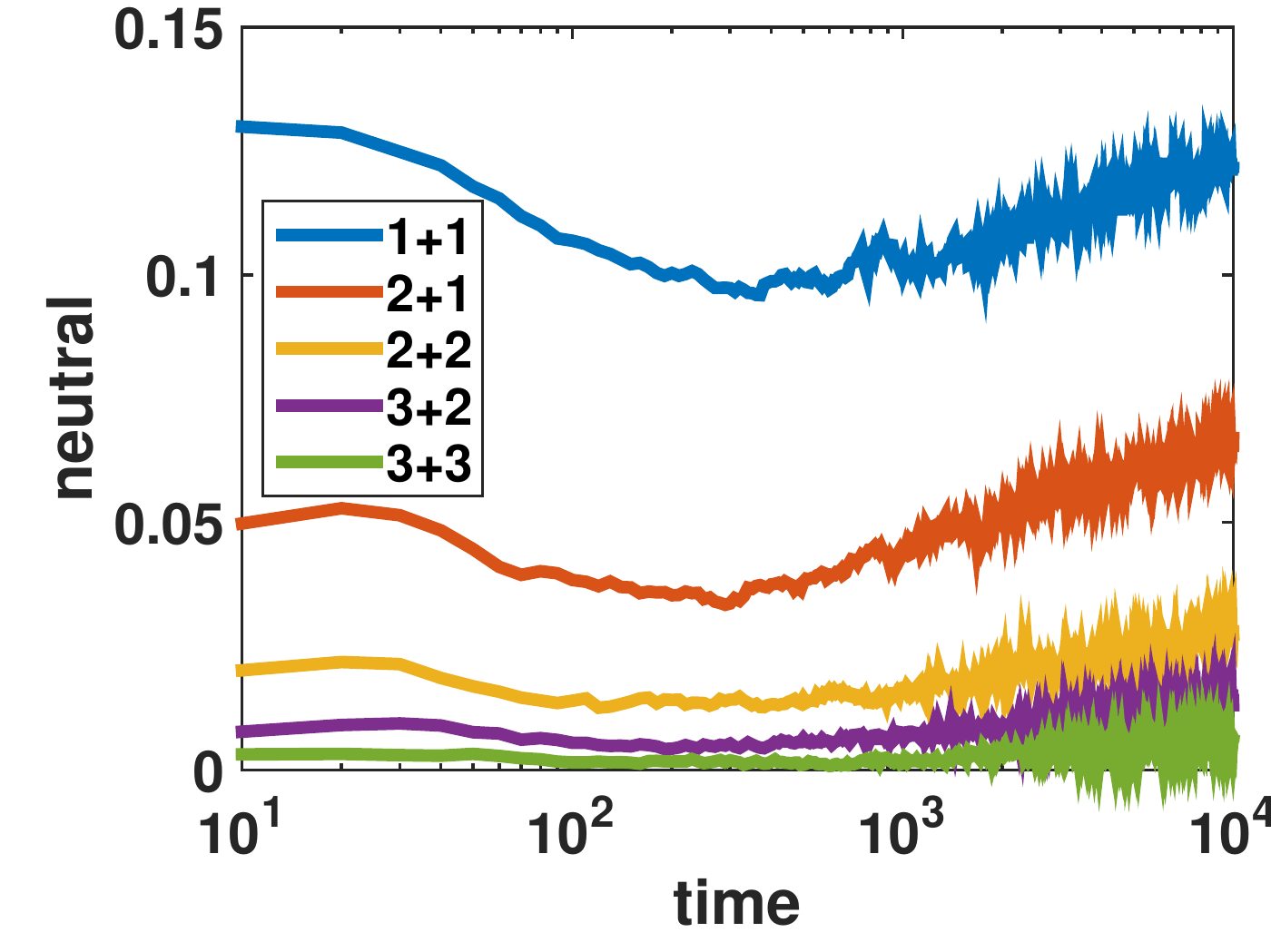}
    }
           \subfigure[]
{
        \includegraphics[width=0.3\textwidth]{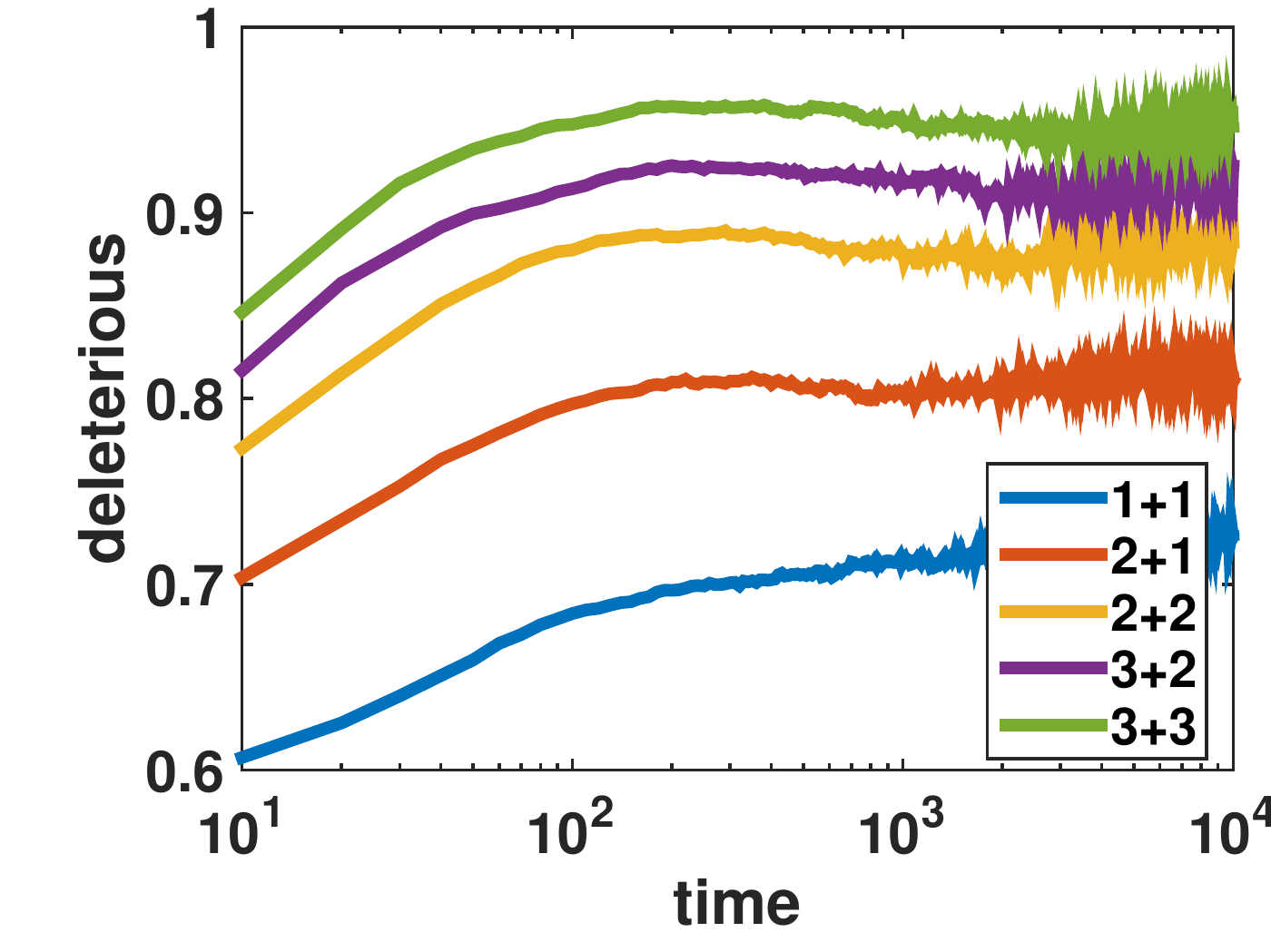}
    }
    \caption[]
    {\label{fig:TF_mut}  {\bf The fitness landscape becomes more frustrated when $n_G>2$ (i.e., when each TF post-duplication regulates more than $1$ gene).}
  At every time point in the stochastic simulation we analyze all possible TF consensus sequence mutations and classify them according to their effect on fitness as beneficial (a) neutral (b) or deleterious (c).  
  With increasing number of downstream genes, $n_G = a+b$, regulated by each TF (different curve colors, see legend), the fractions of beneficial and neutral mutations decrease and the fraction of deleterious mutations increases. This is because TFs become more constrained as $n_G$ increases, resulting in fewer potential mutations that are beneficial or neutral.}
\end{figure}

Each TF needs to simultaneously regulate a subset of the genes while avoiding regulation of the remaining ones. This increasing number of constraints, relative to the $n_G=2$ case, incurs a diminishing number of feasible evolutionary trajectories. The fitness change due to a TF consensus sequence mutation is assessed according to its effect on the binding affinities of this TF with all existing genes. Hence, for each TF, as $n_G$ increases, the number of constraints also increases. This limits the number of possible substitutions a TF can access via fewer beneficial and neutral mutations. In contrast, for each binding site, the number of constraints does not change because it is only constrained by the two TFs and not by other binding sites.  
To demonstrate how extra constraints arising for $n_G>2$ genes affect evolutionary trajectories, we classified in \figref{fig:TF_mut} the effects of all TF mutations on fitness for various numbers of downstream genes $a+b$.

\begin{figure}[H]
\begin{center}
\includegraphics[width=0.8\linewidth]{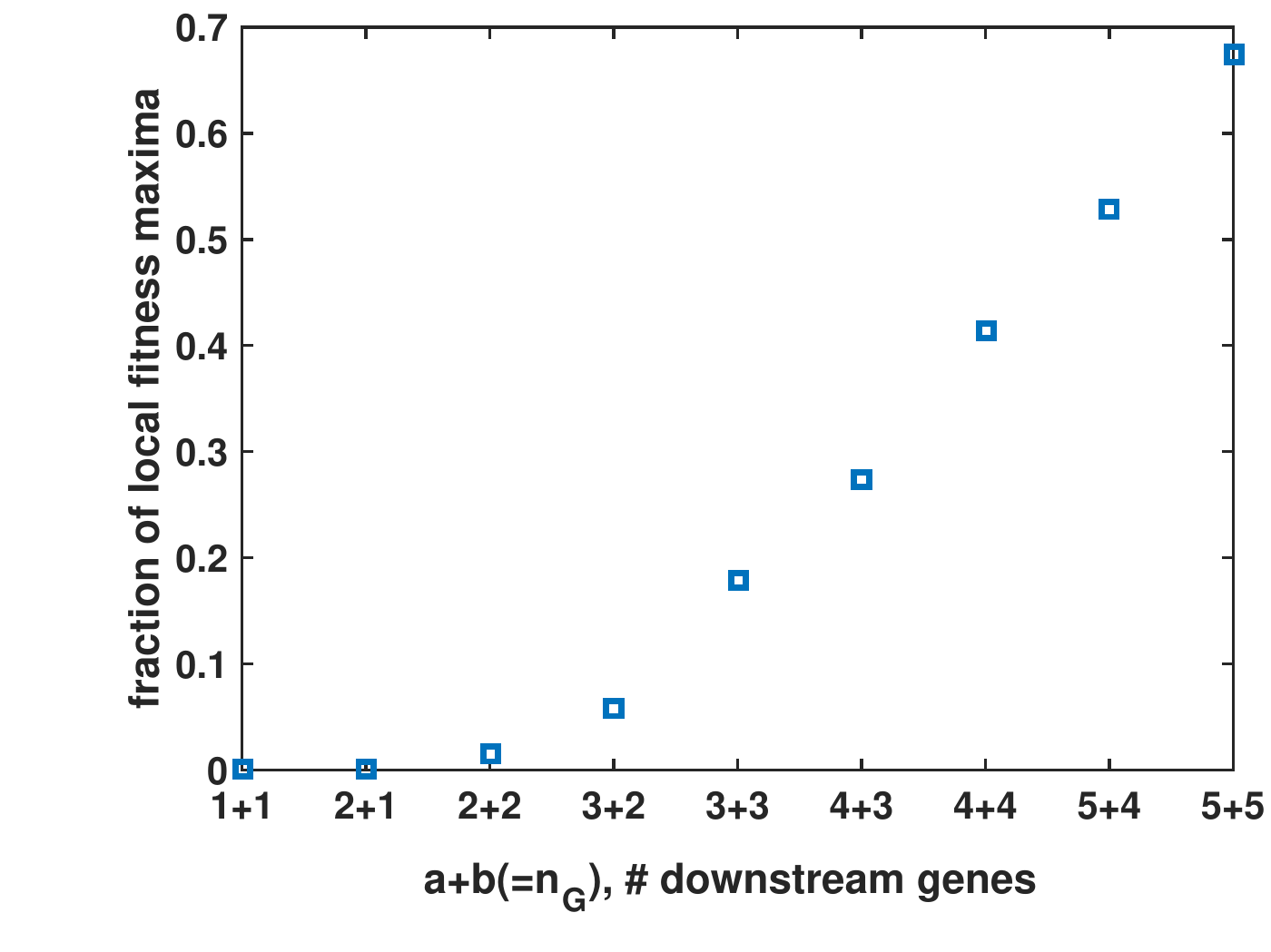}
  \caption[]
 { \label{fig:fitness_peaks}
 {\bf The adaptive landscape of TFs becomes more rugged the more genes they regulate. }
We classify all possible TF mutations according to their fitness effect as beneficial, deleterious or neutral. If at a certain time point all mutations of both TF are strictly deleterious, this indicates a local fitness peak. A way out of such a peak, if there is one, proceeds by means of BS mutation(s), following which the TF can evolve further. The figure illustrates simulation-based statistics of the fraction of time points in which such fitness peaks are encountered for different $n_G$, split (un)equally, $n_G = a+b$, between the TFs (indicated on x-axis). Clearly, the more genes a TF needs to regulate, the more constrained it is, and the fewer are the trajectories it can take. The fraction of local fitness maxima depicted in the plot were obtained by sampling the fitness landscape along typical evolutionary trajectories, and hence does not reflect the entire fitness landscape. Each point is an average over 160,000 points (400 independent simulation repeats, 4000 time points sampled at a uniform interval between $t$=6000-10,000 when the dynamics is already nearly neutral (see \secref{subsec:altSimul} for details). Parameters: $L=8$, $Ns=100$, $C_0=3.269\times 10^5$, $\epsilon=3$, $\beta_X=1$.
}
\end{center}
 \end{figure}

With increasing numbers of downstream genes, evolutionary trajectories are more often stuck in local fitness peaks. We demonstrate this effect in \figref{fig:fitness_peaks}, where we classified at each time point in the simulation all possible TF mutations, and determined that a particular point is a fitness peak if all possible TF mutations from that point are strictly deleterious. Evolution can still continue thanks to the binding sites mutations which are much less constrained.

\subsubsection{Evolutionary pathways}

\begin{figure}[H]
\centering{\includegraphics[scale=0.8]{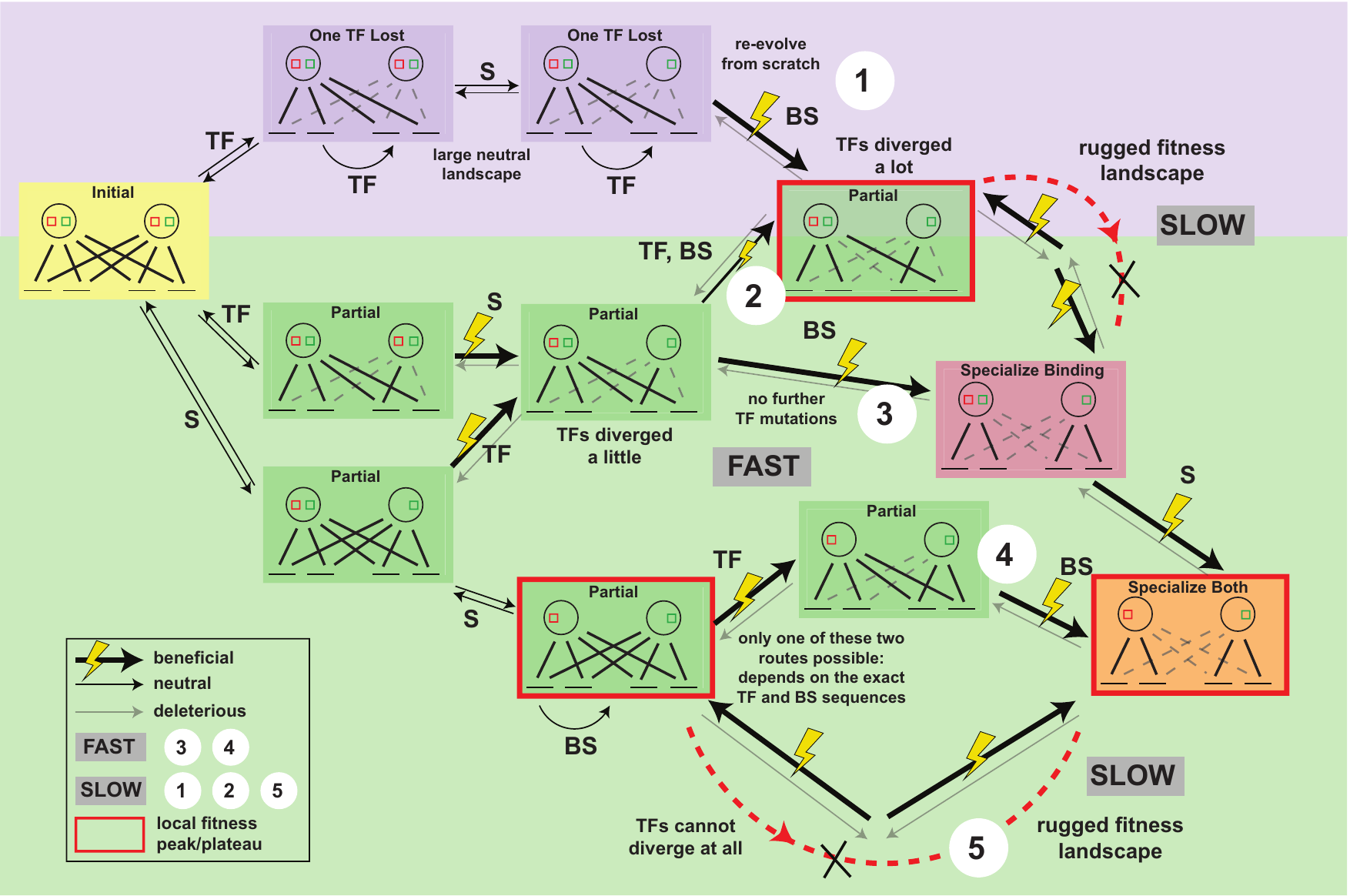}}

\caption{\textbf{Different pathways to specialization vary in the order and
nature of mutations, and might have to cross a rugged fitness landscape for $n_G>2$.}
Here we show in detail the various mutations that occur along the
different pathways (marked with numbers inside white circles) to specialization.
For each mutation, we show the type of mutation (text on the arrows):
TF consensus sequence mutation (TF) or binding site sequence mutation
(BS), TF signal sensing domain mutation (S) and whether it is beneficial
or (nearly) neutral or deleterious (style of the arrows, see legend). We also depict the macrostates along each pathway
graphically, and mark local fitness peaks/plateaus with red boxes.
In red dotted curved lines, we denote parts of the pathways which
involve a fitness valley and hence, are very difficult to cross. Routes
not involving any fitness valleys (numbered $3$ and $4$) are fast,
while those involving a fitness valley (numbered $1,2$ and $5$)
are slow.
\label{fig:mult_pathways}}
\end{figure}

The pathways to specialization in the case of multiple regulated genes are more complex than those described in Section \ref{sec:dyn} for $n_G=2$.  The
primary difference is that for $n_G>2$ some pathways involve fitness valley crossings, where there is a chance of being stuck on local fitness
peaks/plateaus. Hence, these paths take longer times to specialize.
The following are the main pathways that are depicted in Fig. \ref{fig:mult_pathways}.
The first proceeds via $\texttt{One TF Lost}$ macrostate while the other pathways proceed only via $\texttt{Partial}$ configurations.
\begin{enumerate}
\item The first pathway involves the $\texttt{One TF Lost}$ macrostate, where as before one TF does not bind to any binding site. Evolving a TF-BS link to this TF entails a random walk on a neutral landscape and essentially involves regulatory evolution from scratch. After gaining a TF-BS link from a BS mutation, the system ends up on a local fitness
plateau (marked with a red box in Fig. \ref{fig:mult_pathways}) in the $\texttt{Partial}$ state. This is because the ``lost'' TF (second
TF in the figure) has considerably diverged from the first TF yet has specialized only for some, but not all, of the genes associated with the green signal, but not for all of them.
All of the TFs and BSs are constrained to maintain match beyond some minimal level.

Hence specialization can only occur if one of the strong TF-BS links weakens. Such weakening decreases the fitness, and hence incurs crossing a fitness valley. This pathway is consequently very slow.
\item The remaining pathways do not involve $\texttt{One TF Lost}$ macrostate and
go only via $\texttt{Partial}$ macrostate. In the second pathway, first, a TF
consensus sequence mutation and a signal sensing mutation (either
can occur first) lead the system to a $\texttt{Partial}$ state with
some of the signal-BS pathways specialized. Then, an additional TF
consensus sequence mutation pushes the TFs further apart. This, together
with BS mutations, brings the system to the local fitness plateau
(in the $\texttt{Partial}$ macrostate) described in the previous
pathway. This pathway is also slow, because of the fitness valley crossing described above.
\item In the third pathway also, first, a TF consensus sequence mutation and
a signal sensing mutation (either can occur first) lead the system
to a $\texttt{Partial}$ state with some of the signal-BS pathways
specialized. From here, no additional TF consensus sequence mutations
occur that push the TFs away. Hence, there are paths for the BSs to
realign their binding preferences (to the other TF) such that fitness
is always maintained and does not involving crossing any fitness valleys.
Hence, this pathway is fast.
\item In the fourth and the fifth pathways, the first two mutations are
signal sensing mutations that specialize the TFs' signal sensing domains.
From here, a TF mutation and subsequent BS mutations can specialize
without going through fitness valleys. Hence, this is a fast pathway.
For a given genotype (specifying the TF and BS sequences), this fourth
pathway is either possible or not. If it is not possible, then the
only resort is the fifth pathway.
\item The fifth pathway comes into play when the fourth pathway is not possible.
This happens when any TF mutation loses some signal-BS pathways, hence
dropping the fitness considerably. The TFs cannot diverge at all,
and this involves crossing a fitness valley. Hence, this is a slow
pathway.
\end{enumerate}

\subsubsection{Time to specialization}

\begin{figure}[H]
\centering{\includegraphics[scale=0.6]{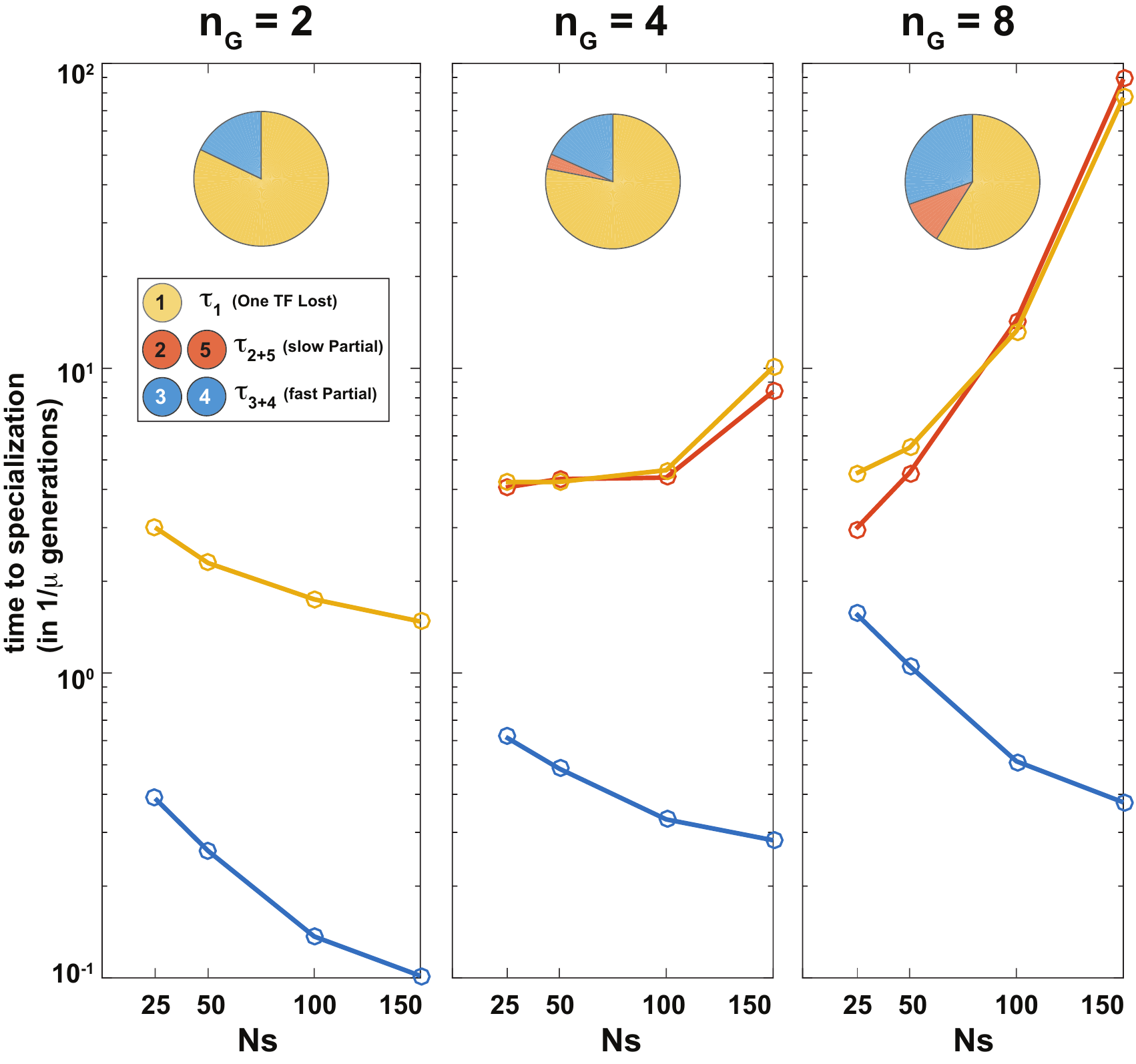}}

\caption{\textbf{Times to specialization via different pathways for various numbers of downstream genes.} Shown are the
times to specialization via different pathways as a function of $Ns$
for different values of $n_{G}$. We plot the times for the slow $\texttt{One TF Lost}$
pathway (numbered $1$, yellow), the slow $\texttt{Partial}$ pathway (numbered
$2$ and $5$, red), and the fast $\texttt{Partial}$ pathway (numbered
$3$ and $4$, blue). Plotted as pie charts also are the fraction of various pathways for different $n_G$ values as pie charts; these fractions depend only very weakly on $Ns$. In general, the higher the $n_G$, the larger the fraction of fast trajectories (3 and 4) and the longer the time needed to specialize. Pathways whose time lengths with $Ns$, which are the slow $\texttt{Partial}$ pathway (red) and the $\texttt{One TF Lost}$ pathway (yellow) for $n_G>2$, involve crossing fitness barriers.  \label{fig:mult_times}}
\end{figure}

By running simulations, we calculate the time to specialization for different values of $n_{G}>2$ (total number of downstream genes) via the different pathways described in the previous section. Specifically, we calculate the time to specialization, $\tau_{1}$,
via the $\texttt{One TF Lost}$ pathway (pathway $1$), $\tau_{3+4}$,
via the fast $\texttt{Partial}$ pathways (pathways $3$ and $4$),
and, $\tau_{2+5}$, via the slow $\texttt{Partial}$ pathways (pathways
$2$ and $5$). We also calculate the fractions of these pathways.
These are shown in Fig. \ref{fig:mult_times}. The slow $\texttt{Partial}$
pathway (numbered $2$ and $5$) is absent for $n_G=2$. The fast
$\texttt{Partial}$ pathway (numbered $3$ and $4$) does not involve
crossing any fitness valleys, and hence the time to specialization
via this pathway decreases with increasing $Ns$ for all $n_G$.
The time to specialization via the slow $\texttt{One TF Lost}$ pathway
(numbered $1$) decreases with increasing $Ns$ for $n_G=2$, and
so does not involve crossing fitness valleys. For $n_G>2$, the time to specialization via both the slow $\texttt{One TF Lost}$
pathway and the slow $\texttt{Partial}$ pathway increases as $Ns$
increases. Both these pathways for $n_G>2$ involve crossing fitness
valleys. With increasing $n_G$, the fractions of the fast $\texttt{Partial}$
pathway and slow $\texttt{Partial}$ pathway increase at the expense
of the slow $\texttt{One TF Lost}$ pathway.

\section{Promiscuity-promoting mutations}
\label{sec:Promiscuity}
So far we considered the "mismatch-energy model" for TF-BS specificity, where each position in the TF and the binding site contributed equally to the total binding energy, depending on whether the position has a mismatch between the TF consensus sequence and the BS sequence. Let the TF consensus sequence be $s^{*}$ and the binding site sequence
be $s$, both of length $L$. In general, we have
\begin{equation}
E={\displaystyle {\displaystyle \sum_{i}}E_{i}}
\end{equation}
where $i$ runs over all the positions of the binding site. For each
specific position $i$, the contribution is $E_{i}=0$ if $s_{i}=s_{i}^{*}$ (match)
and $E_{i}=\epsilon$ if $s_{i}\ne s_{i}^{*}$ (mismatch).

\begin{figure}[H]
\centering{\includegraphics{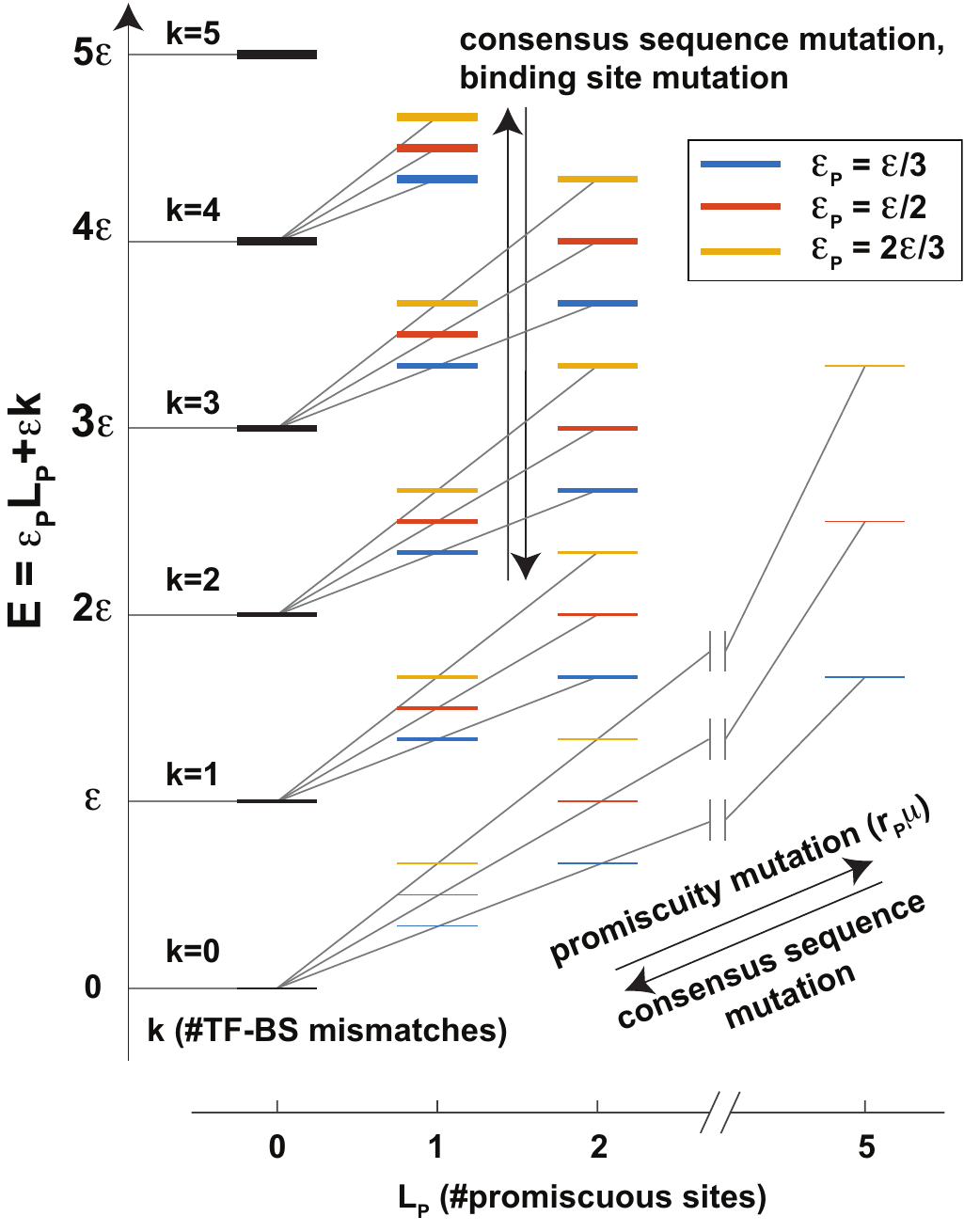}}

\caption{\textbf{Total TF-DNA binding energies depend on number of mismatches as well as on the number of promiscuous TF positions. } We plot the different energy levels depicting the TF-BS binding energy, $E=\epsilon_{P}L_{P}+\epsilon k$, for TFs with varying number of  promiscuous positions $L_{P}$ and $k$ mismatches between the TF and BS in the remaining $L-L_{P}$ specific positions. Note that lower $E$ corresponds to tighter TF-BS binding. We illustrate this for three different values  of $\epsilon_{P}$, the energy contribution per
promiscuous position (different colors). Increasing line thickness of the energy levels represents higher mismatch values $k$. While promiscuity-promoting mutations increase
$L_{P}$ by converting a specific position to a promiscuous one, regular TF mutations that hit a promiscuous position can convert it to be specific and decrease $L_{P}$.
\label{fig:energy_ladder}}
\end{figure}

Experiments on TF-BS specificity, however, suggest that some TF (and binding site) positions dominate while others only have minor energetic contributions. In this section we study a simple generalization of the mismatch-energy model, where we allow for two levels of contribution: some positions are specific (favor a unique nucleotide) and have large energetic contribution while others are non-specific or promiscuous (all nucleotides are equally favorable) and have a smaller energetic contribution. For each specific position $i$, the contribution $E_i$ is, as in the mismatch-energy model, $\epsilon$ if there is mismatch between the TF consensus sequence and the BS sequence in that position, and $0$ if there is a match. On the other hand, for each promiscuous position $i$, the contribution is $E_{i}=\epsilon_{P}$ (typically $0\le\epsilon_P \le \epsilon$), independent
of $s_{i}$. Hence, for a TF with $L_{P}<L$ promiscuous positions in total, and $k$ mismatches in the remaining $L-L_{P}$ specific positions, the total binding energy would be $E=\epsilon_{P}L_{P}+k\epsilon$. The different possible energy levels for specific and promiscuous TFs are illustrated in Fig. \ref{fig:energy_ladder}.

We also introduce an additional type of mutation, called ``promiscuity-promoting'' mutation, that occurs at rate $r_P \mu$. As illustrated in Fig. 5A of the main text, these mutations convert a specific TF position in the consensus sequence to a promiscuous one. A promiscuous position can return to be specific again if it is hit by a consensus TF mutation (regular TF mutations we considered until now, happening at rate $r_{TF}\mu$).

Promiscuity entails a cost in terms of TF-BS binding. To elucidate this cost, we consider the dependency of the free (dimensionless) concentration, $C_0$, of a TF, on the binding preferences of the TF. For a TF with no promiscuous positions, $C_0$ can be calculated in the chemical potential framework as

\be
C_0(L_P=0) = \dfrac{C}{GS(\epsilon,L) + \displaystyle \sum_n \exp(-E_n)},
\ee

where $C$ is the copy number of the TF, $G$ is the number of sites on the DNA where the TF can bind in a sequence-specific manner, $n$ enumerates other possible energy configurations of the TF that are sequence-independent (residing in the free solution, or nonspecific binding to DNA), and $E_n$ is the free energy in configuration $n$. $S(\epsilon,L)=\langle e^{-\epsilon k} \rangle_{P(k)}$ 
is the similarity between binding sites defined in \cite{friedlander_intrinsic_2016}, with $GS(\epsilon,L)$ acting as the Boltzmann factor for all possible specific binding configurations. This term captures the sequestration of TFs on the DNA due to spurious binding. Assuming that the DNA sequence is random,
$P(k)\sim B(L,3/4)$ is the Binomial distribution for the number of mismatches that a random DNA sequence has with a given TF consensus sequence.

\begin{figure}[H]
\centering{\includegraphics{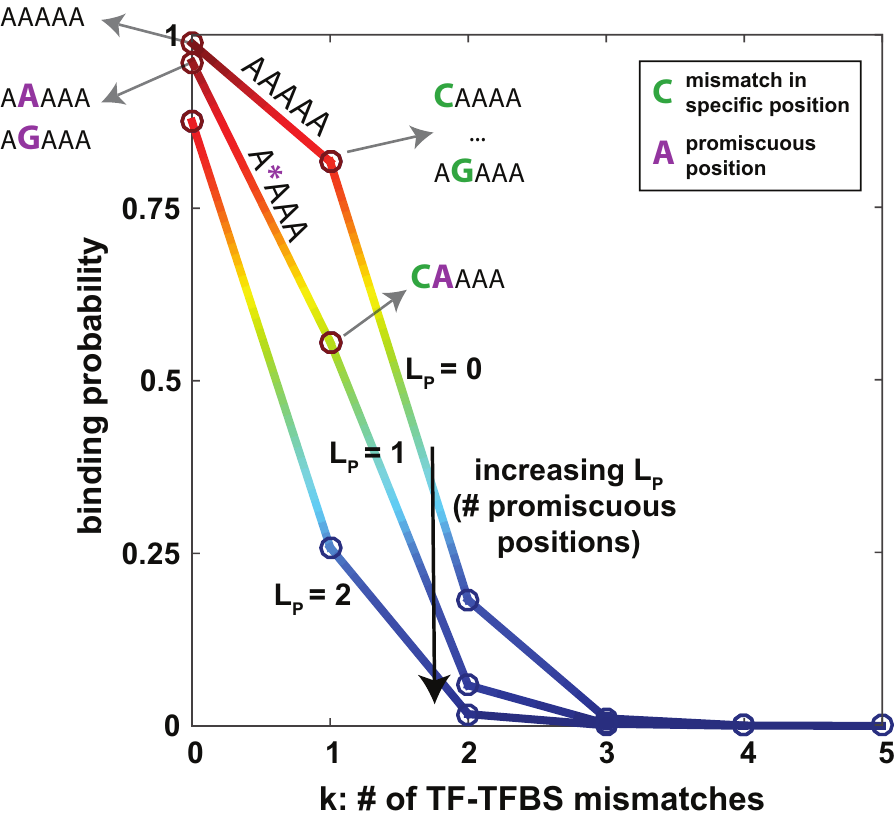}}

\caption{\textbf{Binding probability of the TF to DNA decreases the more promiscuous it is.} The TF-BS binding
probability is plotted as a function of the number of TF-BS mismatches $k$ among the $L-L_P$ specific positions for different values of $L_{P}$, the number of promiscuous positions in the TF. 
We list, as an example, different sequences that are consistent with given $(L_P,k)$. \label{fig:prom_pon}}
\end{figure}

\noindent For a promiscuous TF with $L_P$ promiscuous positions, we have,

\be
\begin{split}
C_0(L_P) & = \dfrac{C}{Ge^{-\epsilon_P L_P}S(\epsilon,L-L_P) + \displaystyle \sum_n \exp(-E_n)} \\
& = C_0(L_P=0) \dfrac{GS(\epsilon,L) + \displaystyle \sum_n \exp(-E_n)}{Ge^{-\epsilon_P L_P}S(\epsilon,L-L_P) + \displaystyle \sum_n \exp(-E_n)} \\
& = C_0(L_P=0) \dfrac{1+A}{e^{-\epsilon_P L_P} \frac{S(\epsilon,L-L_P)}{S(\epsilon,L_P)} + A},
\end{split}
\ee
where $A =  \frac{\sum_n \exp(-E_n)}{GS(\epsilon,L)}$ is an effective parameter that captures the relative contribution of the Boltzmann factor corresponding to spurious specific binding on the DNA, compared with all other Boltzmann factors. We have assumed that $A=0.1$ is fixed in our calculations, and the results we present are fairly robust to the value of $A$. The probability that a binding site is bound by a TF with $L_P>0$ promiscuous positions and $k$ mismatches  with respect to the binding site in the remaining $L-L_P$ positions, assuming no other TF type is present, is

\be
p = \dfrac{C_0(L_P)e^{-\epsilon k - \epsilon_P L_P}}{1+C_0(L_P)e^{-\epsilon k - \epsilon_P L_P}}.
\ee
This probability is plotted in \figref{fig:prom_pon} for various $k$ and $L_p$ values. While $C_0(L_P)$ can be greater or lesser than $C_0(L_P=0)$ depending on the value of $\epsilon_P$, we have $C_0(L_P)e^{-\epsilon_P L_P}<C_0(L_P=0)$. Hence, as the number of promiscuous positions, $L_P$, in the TF increases, the binding probability decreases.

For instance, consider a TF with consensus sequence $AAAAA$ (see Fig. \ref{fig:prom_pon}).
This TF is specific for $A$'s in all five positions of the binding site sequence.
Each mismatch in the binding site sequence (green positions in the sequences in Fig. \ref{fig:prom_pon}) with respect to $AAAAA$ decreases the binding affinity, and thereby decreases the binding probability. Now consider a promiscuous TF with consensus sequence $A*AAA$, where $*$ denotes a promiscuous position. The second position, independent of the bp in the BS sequence (purple positions in the sequences in Fig. \ref{fig:prom_pon}), decreases the binding affinity, but by a lesser amount than a specific position mismatch (green positions). Hence, the binding probabilities of the promiscuous TF to $AAAAA$, $AGAAA$, $ATAAA$ or $ACAAA$ are equal, and higher than the binding probability of the specific TF to $CAAAA$ or $AGAAA$ or other single-mismatch BS sequences.

\subsection{Steady state after duplication}

\begin{figure}[H]
\centering{\includegraphics[scale=0.7]{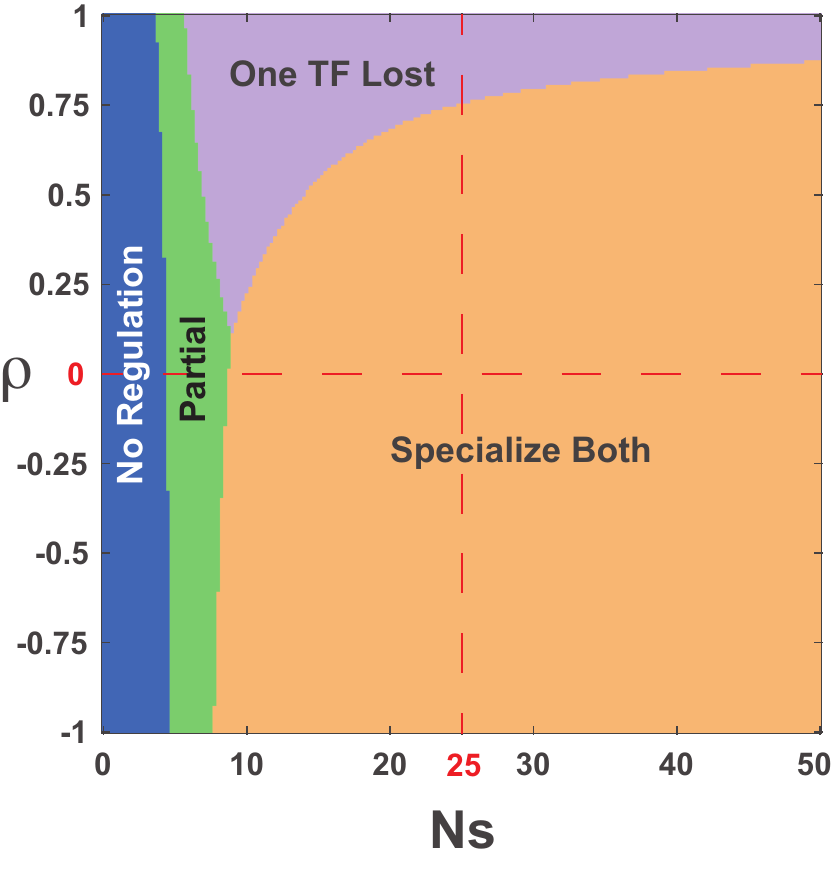}}

\caption{\textbf{Most probable macrostate in the presense of promiscuity-promoting
mutations.} We plot the most probable macrostate at steady state,
$z_{SS}^{*}$, for different $\rho$ and $Ns$, for $n_G=2$ and relative mutation rate
$r_{P}=3$, keeping other parameters at their baseline values. We choose $r_P=3$ so that at each position, a specific bp has equal effective mutation rate towards a promiscuous state or another specific bp. \label{fig:prom_macro_SS}}
\end{figure}

In the presence of promiscuity-promoting mutations, we obtain
the steady state distribution over the genotypic space analytically,
from which we obtain the dominant macrostate at steady state for different
$\rho$ and $Ns$ values (Fig. \ref{fig:prom_macro_SS}). The inclusion of promiscuity-promoting mutations does not significantly change the dominant macrostate
phase plot except for a slight increase in the range of $\texttt{One TF Lost}$ macrostate.

\begin{figure}[H]
\centering{\includegraphics[scale=0.7]{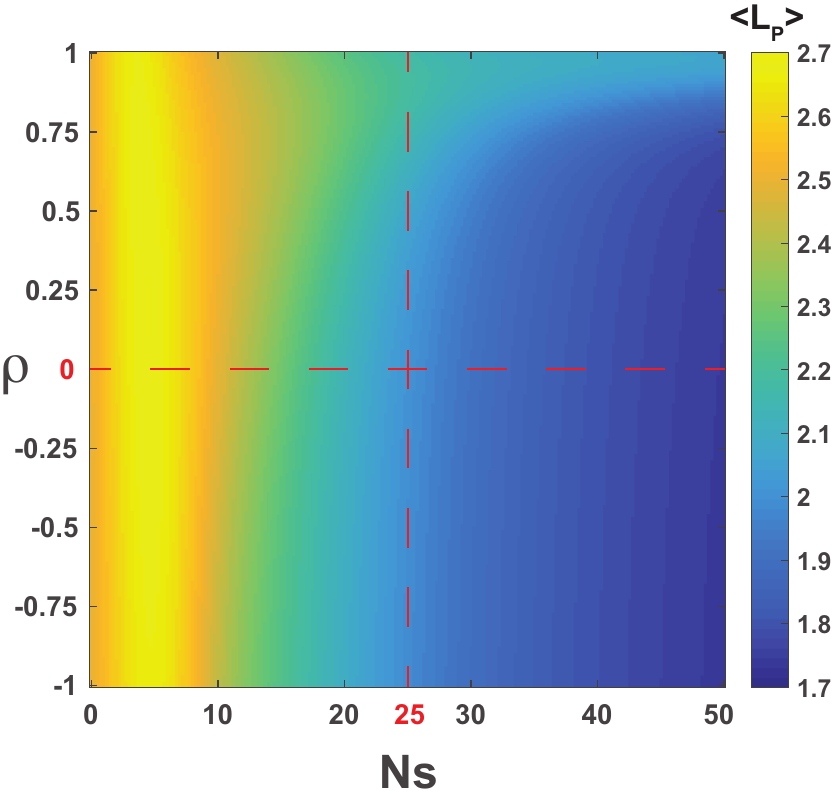}}

\caption{\textbf{Mean number of promiscuous TF positions at steady state decreases with selection intensity.} We plot the mean number of promiscuous positions at steady state, $\langle L_{P}\rangle$ (out of $L=5$),
for different values of signal correlation $\rho$ and selection strength $Ns$.  Steady state values of $\langle L_{P}\rangle$ are within a relatively small range. As selection strength increases, $\langle L_{P}\rangle$ decreases, yet still remains above zero.
Parameter values: $n_G=2$, $r_{P}=3$; other parameters are at their baseline values.
\label{fig:prom_LP_SS}}
\end{figure}

We also plot the mean number of promiscuous positions at steady state
in Fig. \ref{fig:prom_LP_SS}. This number decreases with selection intensity, because promiscuous positions decrease the TF binding probability (see Fig. \ref{fig:prom_pon}) making them less favorable once specialization has occurred.

\subsection{Evolutionary dynamics}

\subsubsection{Time to specialization}

\begin{figure}[H]
\centering{\includegraphics[scale=0.6]{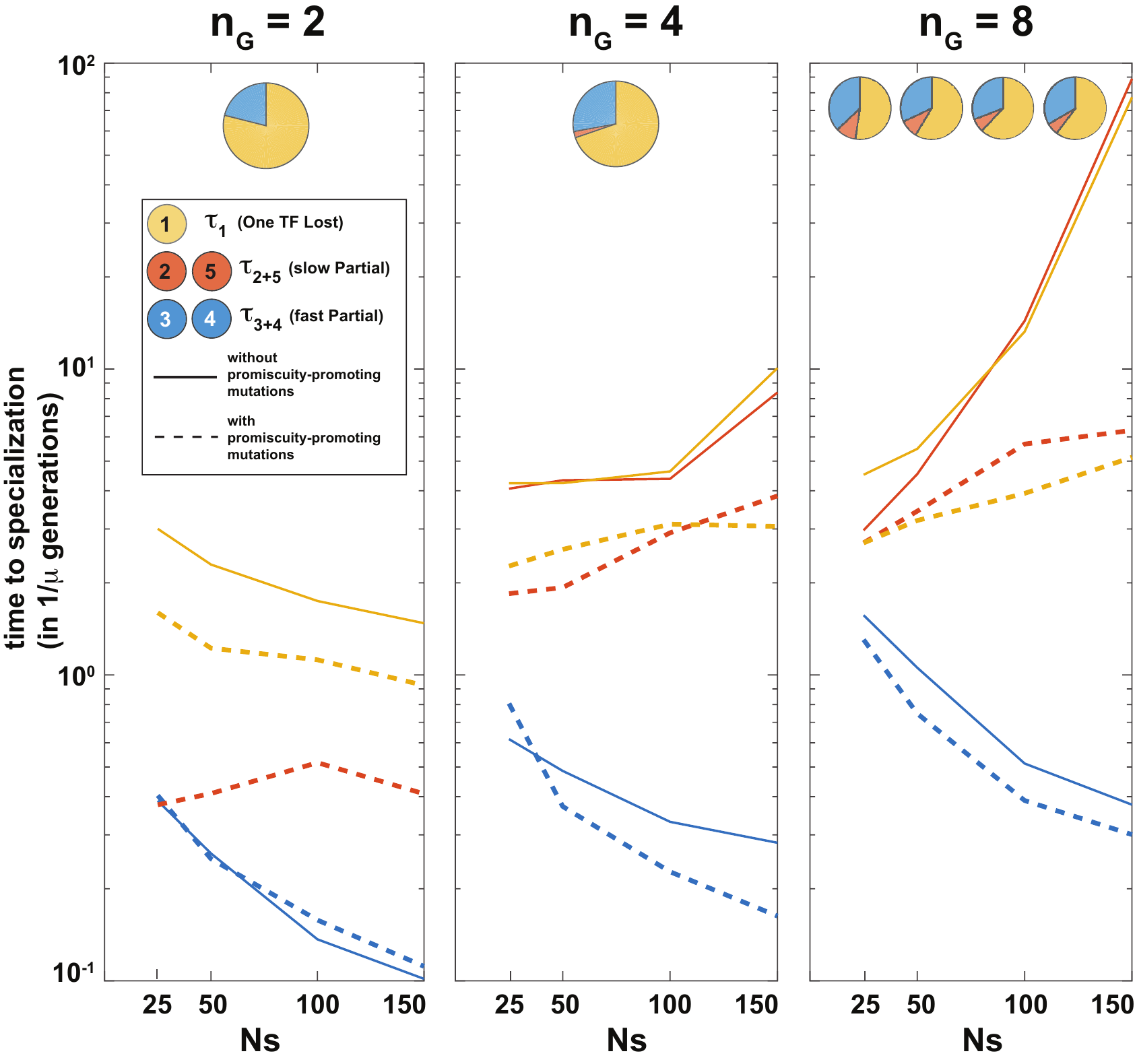}}

\caption{\textbf{Promiscuity-promoting mutations accelerate specialization.} We plot the times to specialization via different
pathways that are depicted in Fig. \ref{fig:mult_pathways}, as a function of $Ns$ for different values of $n_G$ (the number of downstream genes per TF), in the absence (solid lines) and presence (dotted lines) of promiscuity-promoting mutations. Specialization times are shown for the slow $\texttt{One TF Lost}$
pathway (numbered $1$, yellow), the slow $\texttt{Partial}$ pathway (numbered $2$ and $5$, red), and the fast $\texttt{Partial}$ pathway (numbered
$3$ and $4$, blue). In general, promiscuity-promoting mutations shorten evolutionary specialization times. This effect is particularly marked for the slow pathways ($\texttt{One TF Lost}$ and slow $\texttt{Partial}$) and for large numbers of downstream genes $n_G$.
The pie charts illustrate the fraction of the various pathways at each $n_G$ value. For $n_G=8$, we plot the pie
charts for the different $Ns$ values marked on the x-axis.
\label{fig:prom_times}}
\end{figure}

In general, promiscuity-promoting mutations accelerate specialization, as shown in Fig. \ref{fig:prom_times}. The speedup of the fast
$\texttt{Partial}$ pathway ($3$ and $4$) is not very large, but
the speedup of the slow $\texttt{Partial}$ ($2$ and $5$) and the
slow $\texttt{One TF Lost}$ ($1$) pathways  is considerable, an effect that increases with increasing $Ns$ (see Fig. \ref{fig:mult_pathways} for details of the pathways). Promiscuity-promoting mutations act by converting deleterious BS mutations into neutral or beneficial
ones. By that they effectively lower or even remove fitness barriers. This effect is more significant with a large number of downstream genes, where more constraints on TF evolution exist. The fraction of different pathways does not change much if promiscuity-promoting
mutations are present. Note that as a function of $Ns$, the fraction
of fast $\texttt{Partial}$ pathways does not change considerably, but the
fraction of slow $\texttt{Partial}$ pathways decreases while increasing
the fraction of slow $\texttt{One TF Lost}$ pathways.

\subsubsection{Typical trajectory}
\begin{figure}[H]
\centering{\includegraphics{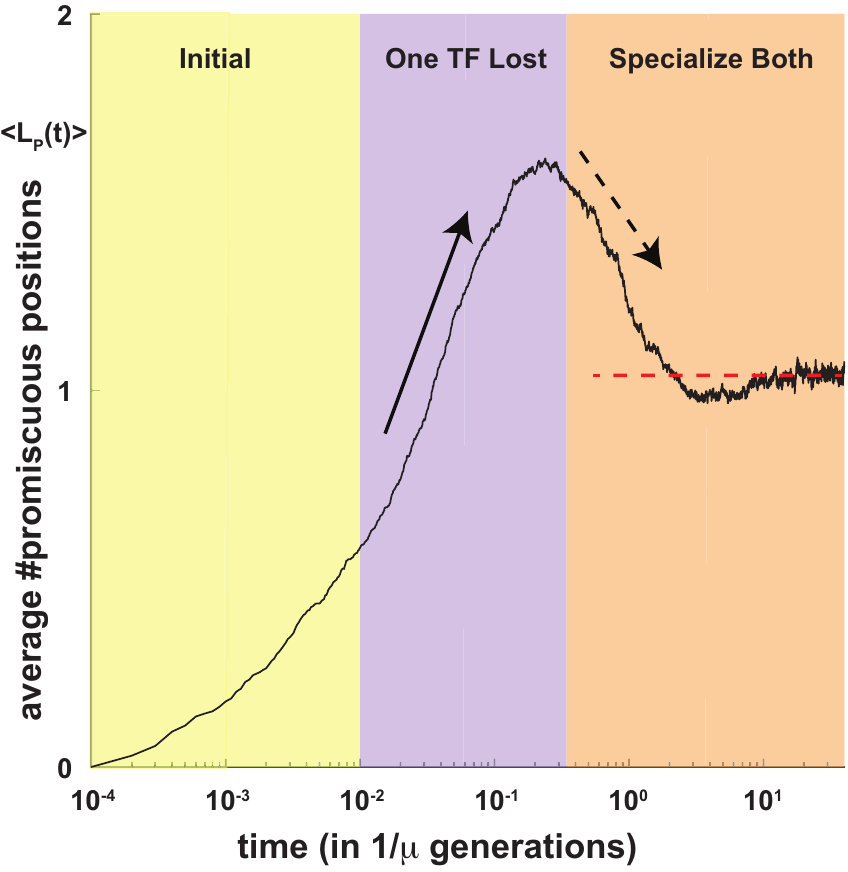}}

\caption{\textbf{Number of promiscuous positions transiently peaks during adaptation and relaxes after
specialization to an intermediate steady state value.} We plot the average number
of promiscuous positions $\langle L_{P}(t)\rangle$ as a function
of time for $L=5,n_G=4,Ns=250$ and $r_{P}=10$; other parameters are at baseline values. Solid black arrow indicates the increase
in the number of promiscuous positions in the transient $\texttt{One TF Lost}$
phase, while the dotted black arrow indicates their decrease after
specializing. The red dotted line indicates the steady state value of $\langle L_{P}\rangle$.
\label{fig:prom_dyn_traj}}
\end{figure}

Promiscuity-promoting mutations play different roles in different phases of the evolutionary trajectory. While after specialization they are less favorable (because they lower binding affinity and potentially destabilize the specialized state), during adaptation they can facilitate fitness valley crossing.
In Fig. \ref{fig:prom_dyn_traj}, we plot the trajectory of the average
number of promiscuous TF positions as a function of time. Starting with no promiscious positions
in the $\texttt{Initial}$ state, the number of promiscuous positions
increases during  the transient $\texttt{One TF Lost}$ state, and then decreases to reach its steady state value after reaching the
$\texttt{Specialize Both}$ state. The speedup of evolution is mainly during the transient
$\texttt{One TF Lost}$ phase, where the number of promiscuous positions
peaks.

\clearpage

\end{document}